\documentclass[useAMS,usenatbib,usegraphicx]{mn2e}

\usepackage{amssymb}  
\usepackage{hyperref}

\hypersetup{
    colorlinks=true,        
    citecolor=black,        
    linkcolor=blue,         
    urlcolor=blue           
}

\def \aap{A\&A}
\def \aaps{A\&AS}

\def \aj{AJ}

\def \apj{ApJ}
\def \apjl{ApJ}
\def \apjs{ApJS}
\def \araa{ARA\&A}

\def \mnras{MNRAS}

\def \nat{Nat}

\usepackage{color}
\usepackage{graphicx}

\newcommand{\AlII}{\hbox{{\rm Al}{\sc \,ii}}}

\newcommand{\CII}{\hbox{{\rm C}{\sc \,ii}}}

\newcommand{\CIV}{\hbox{{\rm C}{\sc \,iv}}}
\newcommand{\CaII}{\hbox{{\rm Ca}{\sc \,ii}}}

\newcommand{\FeII}{\hbox{{\rm Fe}{\sc \,ii}}}
\newcommand{\HI}{\hbox{{\rm H}{\sc \,i}}}

\newcommand{\Ha}{\hbox{{\rm H}$\alpha$}}
\newcommand{\Hb}{\hbox{{\rm H}$\beta$}}
\newcommand{\Hg}{\hbox{{\rm H}$\gamma$}}
\newcommand{\Hd}{\hbox{{\rm H}$\delta$}}
\newcommand{\MgI}{\hbox{{\rm Mg}{\sc \,i}}}
\newcommand{\MgII}{\hbox{{\rm Mg}{\sc \,ii}}}
\newcommand{\NI}{\hbox{{\rm N}{\sc \,i}}}
\newcommand{\NII}{\hbox{{\rm N}{\sc \,ii}}}

\newcommand{\OI}{\hbox{{\rm O}{\sc \,i}}}
\newcommand{\OII}{\hbox{{\rm O}{\sc \,ii}}}
\newcommand{\OIII}{\hbox{{\rm O}{\sc \,iii}}}
\newcommand{\OVI}{\hbox{{\rm O}{\sc \,vi}}}
\newcommand{\SiII}{\hbox{{\rm Si}{\sc \,ii}}}
\newcommand{\SiIII}{\hbox{{\rm Si}{\sc \,iii}}}
\newcommand{\SiIV}{\hbox{{\rm Si}{\sc \,iv}}}

\newcommand{\lya}{\mbox{Ly$\alpha$}}
\newcommand{\lyb}{\mbox{Ly$\beta$}}
\newcommand{\lyg}{\mbox{Ly$\gamma$}}

\newcommand{\kms}{\mbox{km~s$^{-1}$}}
\newcommand{\cmm}{\mbox{cm$^{-2}$}}

\newcommand{\kmsmpc}{\mbox{km~s$^{-1}$~Mpc$^{-1}$}}
\newcommand{\ergs}{\mbox{erg~s$^{-1}$}}
\newcommand{\ergscmm}{\mbox{erg~s$^{-1}$~cm$^{-2}$}}
\newcommand{\msunyr}{\mbox{M$_{\sun}$~yr$^{-1}$}}
\newcommand{\msun}{\mbox{M$_{\sun}$}}

\newcommand{\zem}{\mbox{$z_{\rm em}$}}

\newcommand{\NHI}{\hbox{$N_{\rm HI}$}}

\newcommand{\gimic}       {\mbox{\textsc{Gimic}}}
\newcommand{\gadget}      {\mbox{\textsc{Gadget3}}}
\newcommand{\subfind}     {\mbox{\textsc{Subfind}}}

\newcommand{\hMpc}        {\mbox{$h^{-1}$~Mpc}}
\newcommand{\hMsun}       {\mbox{$h^{-1}$~M$_{\odot}$}}

\title[Galaxies and the IGM across three quasar sightlines]{Galaxies
  at Redshift $\sim0.5$ Around Three Closely Spaced Quasar Sightlines}

\author[N. H. M. Crighton et al.]{Neil H. M. Crighton$^1$\thanks{E-mail: \href{mailto:neil.crighton@durham.ac.uk}{neil.crighton@durham.ac.uk}
}, 
Simon L. Morris$^1$, 
Jill Bechtold$^2$, 
Robert A. Crain$^3$, \newauthor
Buell T. Jannuzi$^4$, 
Allen Shone$^1$  \&
Tom Theuns$^{1,5}$ \\
\\
  $^1$Department of Physics, University of Durham, South Road, Durham, DH1 3LE, UK.  \\
  $^2$Department of Astronomy, University of Arizona, Tucson, AZ 85721, USA.\\
  $^3$Centre for Astrophysics and Supercomputing, Swinburne University of Technology, Mail H39, PO Box 218, Victoria 3122, Australia\\
  $^4$National Optical Astronomy Observatory, 950 N. Cherry Ave., Tucson, AZ 85719, USA \\
 $^5$Universiteit Antwerpen, Campus Groenenborger, Groenenborgerlaan
171, B-2020 Antwerpen, Belgium
}
\begin{document}

\date{Accepted 2009 December 15. Received 2009 December 14; in
  original form 2009 October 11}

\pagerange{\pageref{firstpage}--\pageref{lastpage}} \pubyear{2009}

\maketitle

\label{firstpage}

\begin{abstract}
  We examine the relationship between galaxies and the intergalactic
  medium at $z < 1$ using a group of three closely spaced background
  QSOs with $z_{\rm em} \approx 1$ observed with the \textsl{Hubble
    Space Telescope}.  Using a new grouping algorithm, we identify
  groups of galaxies and absorbers across the three QSO sightlines
  that may be physically linked. There is an excess number of such
  groups compared to the number we expect from a random distribution
  of absorbers at a confidence level of $99.9$\%.  The same search is
  performed with mock spectra generated using a hydrodynamic
  simulation, and we find the vast majority of such groups arise in
  dense regions of the simulation. We find that at $z<0.5$, groups in
  the simulation generally trace the large-scale filamentary structure
  as seen in the projected 2-d distribution of the \HI\ column density
  in a $\sim 30h^{-1}$~Mpc region.  We discover a probable sub-damped
  Lyman-$\alpha$ system at $z=0.557$ showing strong, low-ionisation
  metal absorption lines. Previous analyses of absorption across the
  three sightlines attributed these metal lines to \HI. We show that
  even when the new line identifications are taken into account,
  evidence remains for planar structures with scales of $\sim 1$~Mpc
  absorbing across the three sightlines.  We identify a galaxy at
  $z=0.2272$ with associated metal absorption in two sightlines, each
  200~kpc away. By constraining the star formation history of the
  galaxy, we show the gas causing this metal absorption may have been
  enriched and ejected by the galaxy during a burst of star formation
  2 Gyr ago.
\end{abstract}


\section{Introduction}
\label{sec:introduction}

Ever since QSO absorption lines were found to be produced by
intergalactic gas unassociated with the background QSO, researchers
have been speculating on their connection to galaxies
\citep[e.g.][]{bahcall_absorption_1969}. By measuring galaxy positions close
to QSO sightlines, many groups have attempted to identify galaxies
responsible for absorption lines, with varying degrees of success. It
is particularly timely to study this relationship given the importance
that gaseous inflows and outflows are now believed to have in
regulating galaxy star formation rates. The relationship between QSO
absorbers and galaxies can reveal important clues as to how the
intergalactic medium (IGM) and galaxies interact.

It is well established that strong metal line absorbers seen in QSO
sightlines are closely associated with galaxies. This connection was
first seen between a \CaII\ doublet and galaxy at $z=0.53$
\citep{boksenberg_existence_1978}, and subsequently confirmed for a
number of galaxies using \MgII\ absorption \citep{Bergeron91}. \MgII\
has a rest wavelength of $\sim2800$~\AA\ and an easily identified
doublet, thus it is straightforward to detect such absorbers
associated with low-redshift galaxies using optical spectra. At $z<1$,
strong \MgII\ systems (with rest equivalent width $>0.3$~\AA) are
generally accompanied by a galaxy within a few tens of physical kpc of
the QSO sightline \citep{Bergeron91, Steidel94b, Churchill00b,
  Kacprzak08}.  Studies of galaxies near these absorbers have shown
that the covering factor of the halos causing the absorption must be
less than unity, and there may be a different mechanisms associated
with very high equivalent width \MgII\ systems (galactic winds?) and
mid- to low-equivalent width \MgII\ systems (infall?).  With UV
spectrographs available on the \textsl{Hubble Space Telescope}
(\textsl{HST}), shorter rest wavelength metal lines and the \HI\ \lya\
transition can also be observed at low redshift. At $z<1$, strong
\CIV\ doublet absorption is generally found within $50-100$ physical
kpc of a galaxy \citep{Chen01}, and its clustering strength with
galaxies is similar to that of galaxy-galaxy clustering
\citep{Morris06}. \citet{Adelberger05} used a large sample of Lyman
break selected galaxies at $z \approx 2.5$ around high-redshift QSOs
to show that this relationship between \CIV\ and galaxies also holds
at higher redshifts. Damped \lya\ absorbers (DLAs), predominantly
neutral \HI\ absorbers with column densities \NHI~$>
10^{20.3}$~absorbers~\cmm, are also found to be closely associated
with a wide range of galaxy types \citep{Rao03, Chen05}.  These
results have led to the picture of `halos' of absorbing gas around
galaxies that cause the observed metal lines. A halo of gas around
each galaxy gives rise to \MgII\ absorption up to a radius of a few
tens of kpc, and \CIV\ absorption up to larger radii. Evidence for
this picture has been assembled on a statistical basis from many
single QSO sightlines close to galaxies, and there are no direct
detections of a halo of metal-enriched gas around a galaxy using
multiple nearby sightlines. The detailed geometry of such gas will
depend on how it was placed there -- perhaps by winds induced by
star-formation in the galaxy \citep[e.g.][]{Theuns02}.

The relationship between galaxies and the more tenuous IGM as
represented by the \lya\ forest (\HI\ absorption with \NHI~$ <
10^{16}$ \cmm) is even less clear. The first \textsl{Hubble Space
  Telescope} observations of the low-redshift \lya\ forest showed many
more lines than expected from an extrapolation of the high-redshift
number counts \citep{Morris91, Bahcall91}. This trend was subsequently
confirmed by \citet{Weymann98} using a larger number of \textsl{HST}
UV spectra.  A combination of the reduced ionizing background
radiation at low redshifts and thus an increase in the fraction of
neutral hydrogen \citep{Theuns_loz_evol_98, Dave99}, and structure
evolution \citep{Scott02}, is believed to explain the unexpectedly
high \HI\ line density.  This large line density opened the
possibility of comparing the low redshift IGM and galaxy
distributions.

The first results towards QSO sightline 3C~273 showed little
correlation between galaxy and \HI\ except for the strongest \NHI\
absorbers \citep{Morris93}, a conclusion that has been supported in
more recent studies \citep{Penton02, Chen05corr, Morris06,
  Wilman07}. An anti-correlation between absorber rest equivalent
width (EW) and the impact parameter of nearby galaxies was reported in
\citet{Lanzetta95}, \citet{Chen98} and \citet{Chen01b} for lines with
rest EW $> 0.3$~\AA.  This was initially interpreted as evidence that
a significant fraction of \lya\ absorbers could be attributed to
gaseous halos associated with galaxies. However, other similar studies
did not find such a strong relationship \citep{Stocke95, LeBrun96,
  Tripp98}, and concluded that an anti-correlation was only
significant for the \HI\ absorbers with large rest EWs ($>0.24$~\AA).

The current accepted picture of the low redshift \lya\ forest has been
heavily influenced by the results of N-body smoothed particle
hydrodynamics (SPH) simulations \citep[e.g.][]{Theuns_sph_sim_98,
  Dave99}. These predict dense, condensed gas very close to galaxies,
shocked gas further out, and diffuse gas far from galaxies. The gas
density increases with proximity to galaxies, explaining the
correlation of equivalent width with impact parameter. There is no
clean differentiation between gas that is associated with galaxies and
the diffuse IGM.

Closely spaced QSO sightlines, either due to a fortunate asterism of
background QSOs or a single lensed QSO, can be used to search for
correlations in \HI\ and metal absorption across the sightlines
\citep[e.g.][]{Bechtold94a, Dinshaw95}.  Here `closely spaced' means
separations on the expected size scale of \lya\ absorbers; up to a few
100 kpc based on SPH simulations and theoretical arguments
\citep{Schaye01}. Lensed QSO sightlines show that spectra of the
\lya\ forest are very similar over 10s of kpc at $z=2.3$
\citep{Smette95} and non-lensed close sightlines show correlations
exist even over 100s of kpc. SPH simulations of the \lya\ forest are
consistent with such correlation lengths, although they are
interpreted as a coherence length in the IGM rather than the size of
individual `clouds'. Indeed, \citet{Rauch95} have shown there are not
enough available baryons in the universe to explain the observed
incidence of \lya\ absorbers if they are spherical structures with
such large radii.

Here we analyse the distribution of galaxies in a field containing
three closely spaced QSO sightlines - towards LBQS~0107-024A,
LBQS~0107-024B and LBQS~0107-0232. This QSO group has been studied
extensively, and showed some of the first evidence for very large
scale correlations in \lya\ forest absorption \citep{Dinshaw95}. Using
multiple QSO sightlines opens the possibility of directly
constraining the geometry of the absorbing gas, both around individual
galaxies close to the sightlines and in large scale structures traced
by galaxies.

At what distances and velocity scales do we expect to see associations
between gas and galaxies? Absorbers could plausibly be associated with
galaxies on the scale of a galactic halo, galaxy group, galaxy
cluster, or large scale structure.  Velocity dispersions for early
type galaxies are typically $\lesssim 400$~\kms\ \citep{Sheth03}, and
their rotational velocities are $\lesssim 150$~\kms\ \citep{Franx91}.
Rotational velocities in spirals are usually less than $\sim
400$~\kms\ \citep{Ramella89}. \citet{Eke04} find groups of galaxies in
the 2df galaxy redshift survey have median velocity dispersions of
$\sim260$~\kms. Large galaxy clusters at $z<0.15$ show velocity
dispersions of $\sim800$~\kms\ \citep{Fadda96}. While the
intra-cluster medium is expected to be too hot for any \HI\ gas to
exist, galaxies with associated cold gas and large relative
velocity dispersions may be present in the cluster. \citet{Colberg05}
find dark matter filaments strung between galaxy cluster `knots' in
$\Lambda$CDM simulations that have a typical radius of 2~Mpc and
length of $5-10$~Mpc.

Winds that expel gas from galaxies could introduce a further velocity
offset between galaxies and any associated absorbers. At $z\sim3$, the
velocity of \lya\ emission seen in Lyman-break selected galaxies has a
large offset ($\sim600$~\kms) with respect to the position of nebular
lines, interpreted as being partially due to winds
\citep[e.g.][]{Adelberger05}. Low redshift starburst galaxies can show
similarly large wind velocities \citep{Heckman90}. In our analysis we
generally consider three velocity cutoffs for association between
absorbers or between absorbers and galaxies: 200, 500 and 1000~\kms.

We intend to analyse galaxies around multiple QSO sightlines to
address the questions: Are absorbers more likely to be found near
galaxies?  Are groups of galaxies more or less likely to be associated
with absorbers than single galaxies?  Do large scale structures seen
in absorption across multiple sightlines coincide with large scale
structures seen in the galaxy distribution? Can we put any constraints
on the geometry of the absorbing gas associated with galaxies using
the multiple sightlines?  Do we see evidence that galaxies are linked
with metal-line systems in one or more sightlines?

In this paper we identify candidate structures at $z=0.501$ and
$z=0.535$ comprised of both galaxies and \HI\ absorption spanning the
three sightlines. Similar structures we identify in an SPH simulation
are likely to arise inside filamentary large-scale structure in high
density regions of the universe.  We also analyse a bright galaxy at
$z=0.2272$ with associated metal absorption in two sightlines.

We use a cosmology with $\Omega_{\rm m}=0.3$, $\Omega_{\Lambda}=0.7$
and $H_0=70$~\kmsmpc, where $\Omega_{\rm m}$ and $\Omega_{\Lambda}$
are the ratios of the matter density and cosmological constant energy
density to the critical density, respectively. We give distances
scaled by the parameter $h_{70}$, where $H_0 \equiv
70h_{70}$~\kmsmpc. All the distances given are physical (proper)
distances unless stated otherwise. We convert between velocity,
redshift and wavelength differences using:
\begin{equation}
  \frac{\Delta v}{c} = \frac{\Delta \lambda}{\lambda} = \frac{\Delta z}{1+z}\, ,
\end{equation}
where $c$ is the speed of light, and $\lambda$ and $z$ are the mean
wavelength and redshift where the difference is being measured. These
relations assume $v \ll c$, which is true for the velocity differences
of $\lesssim1500$~\kms\ we consider.

We calculate the impact parameter of a galaxy from a QSO sightline by
first calculating the comoving line of sight distance $D_{\rm C}$ for
a galaxy with redshift $z$ using
\begin{equation}
  D_{\rm C} = \frac{c}{H_0} \int^z_0 {dz' \over \sqrt{\Omega_m (1+z')^3 + \Omega_\Lambda}}\, ,
\end{equation}
where $c$ is the speed of light. The physical (proper) impact
parameter, $\rho$, is then given by:
\begin{equation}
  \rho = \frac{\Delta\theta}{1+z} D_{\rm C} \, ,
\end{equation}
where $\Delta\theta$ is the angular separation in radians between the
QSO and the galaxy.

To convert the observed total flux $F$ (\ergscmm) of an emission line
from a galaxy at redshift $z$ to a luminosity $L$~(\ergs) we use the
relationship:
\begin{equation}
  L = 4 \pi D_{\rm L}(z)^2 \, F \, ,
\end{equation}
where $D_{\rm L}(z) = (1+z)D_{\rm C}(z)$ is the luminosity distance.

All column density measurements have units of absorbers per cm$^2$
where they are not stated explicitly, and all logarithms are to the
base 10.

The paper is structured as follows: Section 2 describes the galaxy and
QSO absorption line samples. Section 3 presents our analysis of \HI\
absorption associations across the three QSO sightlines. Section 4
presents our analysis of galaxy-absorber associations across the three
sightlines. Section 5 describes the simulated galaxies and mock
sightlines with which we compare our observations. Sections 6 and 7
discuss and summarise our results.

\section{Data}

\subsection{Galaxy data}

The galaxy redshifts in the Q0107 field are taken from two
samples. The first sample is comprised of spectra taken with the
\textsl{Canada-France-Hawaii Telescope} (\textsl{CFHT}) multi-object
spectrograph (MOS).  In their Sections 2.3, 2.4 and 2.5,
\citet{Morris06} describe the R imaging used to select galaxy
candidates, the subsequent MOS observations and reduction steps
performed to extract the 1-d spectra. There are 32 galaxies with
redshifts in this sample.  The second sample of redshifts were
compiled from spectra taken using the COSMIC spectrograph on the
200-inch \textsl{Hale Telescope} by Weymann et al. (private
communication). M. Rauch kindly provided us with this sample,
consisting of 28 further galaxies with measured redshifts. The
positions, redshifts, apparent R magnitudes and estimates for a
minimum and maximum absolute B magnitude for all of these galaxies
were presented in \citet{Morris06}. We reproduce these parameters in
Table~\ref{tab:gal}. Note that we do not have errors for the redshifts
provided by M. Rauch. For these galaxies we assume a redshift error
equal to the median redshift error for the \textsl{CFHT} galaxy
sample. The typical $1\sigma$ redshift error for a galaxy is 0.0007,
corresponding to a velocity error of 130--180~\kms, depending on the
redshift.

For the \textsl{CFHT} spectra sample, galaxy candidates were selected
from imaging taken on the same night as the MOS observations. The
imaging was reduced and SExtractor \citep{Bertin96} was used to create
object catalogues. Galaxy candidates were selected based on morphology
and brightness.

Figure~\ref{fig:image} shows the imaging used to select galaxy
candidates and the positions of galaxies with known redshifts.  The
proper \textsl{CFHT} baffles were absent during the imaging, causing
the diffuse arcs of scattered light in the background. This scattered
light makes it very difficult to assign a completeness limit to the
imaging, as it affects some parts of the image more severely than
others. SExtractor was used to create an object catalogue from the
imaging. A histogram of number counts for SExtractor-identified
objects roughly follows a power law up to magnitude of 21.5-22 before
dropping. Thus we can say that the completeness drops significantly
past R~$=21.5$. The redshift sample completeness drops sharply past
R~$=20.5$ (see Figure~\ref{fig:numcount}).  We note that the galaxy
sample is not intended to be complete to a magnitude limit within some
radius of the QSOs, and the distribution is roughly centred around
LBQS~0107-025A, with very few redshifts north of LBQS~0107-0232.

The fraction of all possible galaxy candidates targeted for
spectroscopy varied from $\sim 50\%$ for bright targets, to $<10\%$
for the faintest spectroscopic targets ($R\approx21$).

\subsection{QSOs and absorption line data}

\subsubsection*{QSOs}

This QSO triplet was discovered in the Large Bright Quasar Survey
\citep{Foltz87}. The positions and redshifts of LBQS~0107-025A (QSO
A), LBQS~0107-025B (QSO B), LBQS~0107-0232 (QSO C) are given in
Table~\ref{tab:qso}. Their positions are also shown in
Figure~\ref{fig:image}. The angular separations between the three QSOs
are 1.29 arcmin (A--B), 1.92 arcmin (B--C) and 2.93 arcmin (A--C). One
arcmin corresponds to a transverse separation of 190~kpc, 355~kpc and
455~kpc at redshifts of 0.2, 0.5 and 0.9 with our assumed cosmology.

The \lya\ forest in each sightline of the two brighter QSOs,
LBQS~0107-025A and LBQS~0107-025B, has been studied extensively
\citep{Dinshaw95, Dinshaw97} using UV \textsl{HST} spectra. These
analyses provided some of the first evidence for large scale
($\sim400$~kpc) coherence in the \lya\ forest. The analysis by
\citet{Young01} added \textsl{HST} spectra for the third nearby line
of sight towards the QSO LBQS~0107-0232.

\subsubsection*{Absorption lines}

\citet{Petry06} used all the available UV \textsl{HST} Faint Object
Spectrograph (FOS) and Goddard High Resolution Spectrograph (GHRS)
spectra to create a catalogue of absorption lines seen in each QSO's
spectrum. LBQS~0107-0232 has observations from a single FOS wavelength
setting covering the wavelength range 1572--2311~\AA. LBQS~0107-025A
and LBQS~0107-025B were both observed with three wavelength settings
-- one using GHRS (1212--1498~\AA) and two using FOS (1572--2311~\AA\
and 2222--3277~\AA). The resolution of the FOS spectra is 1.39~\AA\
and 1.97~\AA\ (FWHM) for the lower and upper wavelength ranges. The
GHRS spectra have a resolution of 0.8~\AA. The final combined S/N per
resolution element is $\sim 30$ for the FOS spectra and $\sim 8$ for
the GHRS spectra. Table.~\ref{tab:qobs} summarises the details of the
UV QSO observations used in our analysis.

\citet{Petry06} aimed to measure velocity coincidences between
absorption across the three sightlines, and so they investigated and
corrected several causes of velocity shifts between the different FOS
spectra. We use their line wavelengths and fitted equivalent widths in
our analysis. We verified that their line-detection algorithm was
working correctly (and did not depend on, say, the method of combining
the spectra or continuum fitting) by performing an independent
combination of the archival FOS and GHRS spectra of the three QSOs,
applying the wavelength shifts given by Petry et al. to each
exposure. We confirmed that the lines they flagged as detections also
appeared in our combined spectra.

The line positions and equivalent widths from \citet{Petry06} along
with our line identifications are listed in Tables~\ref{tab:a140} to
\ref{tab:b270}.  The equivalent widths and FWHM values in these tables
are for the Gaussian line fitted to the feature.  The detection
significance is the equivalent width of a feature divided by the error
in the equivalent width ($\sigma_{\rm det}$ -- see Petry et al. for a
more detailed description). We note that Petry et al. did not adjust
the wavelength scales of the GHRS spectra.

Petry et al. did not search for metal lines inside the \lya\ forest,
other than rest wavelength galactic absorption. They also did not
search for higher order Lyman series lines, rather showing that the
number of lines in the \lyb\ region was consistent with the expected
$dN/dz$ for \lya\ alone. Therefore, we made independent line
identifications, guided by the Petry et al. and \citet{Dinshaw97} line
identifications.  This was done by assuming every line blueward of the
QSO \lya\ emission was a \lya\ transition, and searching for
corresponding strong metal (\CIV, \SiIV, \OVI, \SiIII) and higher
order \HI\ Lyman series transitions within a narrow range around the
expected redshift position.  This process resulted in several new
identifications compared to the Petry et al.~line lists, generally
higher-order Lyman series lines in the forest region. We also
identified a number of strong low-ionisation metal lines associated
with the \lya\ line at $z=0.557$ towards QSO C. Although higher order
Lyman transitions are not available for this system, it is likely a
sub-DLA with \NHI~$>10^{19}$~\cmm.  The associated strong \SiII,
\SiIII, \FeII, \CII\ and \OI\ lines suggest it may be similar to the
super-solar sub-DLAs in \cite{Prochaska06b}. These new identifications
mean that several of the lines used in the analyses by \citet{Young01}
and \citet{Petry06} were incorrectly assumed to be \lya. We discuss
what implications this has for the \citet{Petry06} results in
Section~\ref{sec:lya-triple-coinc}. The line identifications are
listed in Tables~\ref{tab:a140} to \ref{tab:b270}. In some cases
absorption features appeared at the expected position of a metal line
or higher order Lyman series line, but appeared to be blended with
unrelated absorption. This is indicated in the tables by a '?' next to
the line identification.

We checked the wavelength scales of the GHRS spectra by comparing the
expected wavelengths of galactic lines (\CII\ 1334, \SiII\ 1260,
\SiIV\ 1393, \OI\ 1302) and higher-order Lyman series lines with
absorption appearing in the FOS spectra with the observed line
positions.

For QSO B the \lyb, \lyg, Ly-4 and Ly-5 transitions associated with
the Lyman limit system at $z=0.3993$, and the galactic lines \SiII,
\OI, \CII\ and \SiIV\ are seen in the GHRS spectra. We measure the
difference between the expected position (corresponding to $z=0$ for
galactic lines, and $z=0.3993$ for the Lyman series lines) and the
observed central wavelength for a Gaussian profile fitted to each
line. We then estimate a correction to the GHRS wavelength scale by
taking the average of offsets for all these lines, excluding \SiII,
\lyg\ and \SiIV\ as they have either low S/N, poorly determined
parameters, or are blended. We find a $-0.6$~\AA\ offset from the
expected line positions, thus we add 0.6~\AA\ ($\sim 1$ resolution
element) to the wavelengths given by Petry et al. We list our adjusted
wavelengths for lines in the QSO B GHRS spectrum in
Table~\ref{tab:a140}.

For the QSO A GHRS spectrum the S/N is poorer and there are no clear,
unblended higher order Lyman series lines that show \lya\ in the FOS
spectra. Lines near the positions of Galactic \SiII, \OI, \CII\ and
\SiIV\ are present. However, \SiII\ is blended with unrelated
absorption, and the remaining three lines do not give consistent
shifts from the expected positions (\OI: $-0.72$~\AA, \CII: 0.0~\AA\
and \SiIV: $-0.18$~\AA).  As there is no clear offset, we decide not
to apply a wavelength shift to the QSO A GHRS spectrum.
Table~\ref{tab:b140} lists the (unshifted) wavelengths for lines in
the QSO A GHRS spectrum. We note that a 0.5~\AA\ shift corresponds to
a velocity shift of $\sim 120$~\kms\ at 1250~\AA, thus even if some
error in the GHRS wavelength scale remains, we are confident it is
less than $\sim200$~\kms.

The typical $1\sigma$ statistical error on the wavelength of an
absorption line is ~0.2~\AA, corresponding to a velocity error of
30--50~\kms. Petry et al. estimate the FOS absolute wavelength
calibration errors at $\sim0.5$ pixels, or $\sim32$~\kms. As we show
above, any errors in the GHRS wavelength calibration are $< 200$~\kms,
and probably much lower than this.  We expect the absorption line
positions to have an error of $\sim 80$~\kms. This uncertainty is
smaller than the typical galaxy redshift $1\sigma$ error of $\sim
180$~\kms.

To construct the sample of \lya\ lines that we compare to galaxies, we
select all lines that have been identified as \lya\ and all
unidentified lines. Finally, we remove any lines that are within
3500~\kms\ of the QSO redshift. Lines close to the QSO emission
redshift may be clustered around, ejected from, or ionised by
radiation from the background QSO. \citet{wild_narrow_2008} measure an
increase in CIV systems close to background QSOs due to both
clustering and ejected aborbers.  Based on their results, a 3500~\kms\
cut should be ample to remove any clustered absorbers. It is also much
larger than the scales on which we expect to see a significant
reduction in the line density due to the proximity effect
\citep[e.g.][]{Scott02}.  Absorbers are known to have ejected
velocities of more than 10000~\kms\ in broad absorption lines quasars
\citep[e.g.][]{hamann_broad_1998}.  However, most ejected \CIV\
absorbers have velocities $<3000$~\kms -- Wild et al. find fewer than
40\%\ of \CIV\ systems at 3000~\kms\ blueward of the QSO redshift are
ejected.  We believe 3500~\kms\ is a reasonable compromise between
excluding \lya\ absorbers that arise in ejected systems without
removing an excessive number of intervening IGM absorbers.

Our \lya\ sample may not be the true \lya\ sample for two
reasons. Firstly, some unidentified lines may be due to transitions
other than \lya.  Secondly, some lines that have been identified with
non-\lya\ transitions may be blended with \lya\ lines. In the first
case our \lya\ list may contain additional spurious \lya\ absorbers,
and in the second case we may have missed real \lya\ absorbers.  We
believe the number of spurious \lya\ lines is very low, given the low
density of strong metal-bearing forest lines at these redshifts, and
that we have searched thoroughly for galactic and extragalactic metal
lines and higher-order Lyman lines in the forest. It is more likely
that our list of \lya\ lines is conservative due to the second case;
we have missed some \lya\ lines blended with other transitions.  If we
have missed lines, the correlation strengths we measure across
sight-lines and with galaxies will be smaller than the true
strengths. We may also underestimate the number of associated
galaxy-absorber groups (see Section 4.4) over our redshift range due
to missing some \lya\ absorbers.

Optical spectra of QSOs A and B, taken with the Multiple Mirror
Telescope Spectrograph, were presented in \citet{Dinshaw97}. These
cover wavelengths from 3250--5650~\AA\ at a resolution of 1~\AA\
(FWHM) and a S/N of $\sim10$, and they rule out any \MgII\ absorption
with rest equivalent width $> 0.3$~\AA\ from absorbers in the redshift
range $0.34 < z < 1.02$.

Wedge plots showing the \lya\ lines a galaxy positions used in our
analysis are shown in Figure~\ref{fig:pie}, and \lya\ rest equivalent
widths, redshifts and the galaxy distribution are shown in
Figure~\ref{fig:z}.

\begin{table*}
\begin{center}
{\large \textsc{\textsl{HST} Observations}}\\
\vspace{0.2cm}
\begin{tabular}{lcccccccc}
\hline
QSO Name & Proposal & Instrument & Grating & Exposure & Wavelength & S/N per & Resolution & Dispersion \\
& IDs &  &  & Time (min) & Range (\AA) & pixel & FWHM (\AA) & (\AA/pixel) \\
\hline
Q0107$-$025A (A)& 5172, 6260& GHRS& G140L& 549& 1212--1498 & 5 &  0.8& 0.143\\
                & 5320, 6592& FOS & G190H& 448& 1572--2311 & 28& 1.39& 0.36\\
                & 6100      & FOS & G270H& 146& 2222--3277 & 22& 1.97& 0.51\\
Q0107$-$025B (B)& 5172, 6260& GHRS& G140L& 337& 1212--1498 & 7 &  0.8& 0.143\\
                & 5320      & FOS & G190H& 108& 1572--2311 & 13& 1.39& 0.36\\
                & 6100      & FOS & G270H& 107& 2222--3277 & 34& 1.97& 0.51\\
Q0107$-$0232 (C)& 6100, 6592& FOS & G190H& 548& 1572--2311 & 16& 1.39& 0.36\\
\hline
\end{tabular}
\caption{\label{tab:qobs} For the UV observations of each quasar we
  show the proposal ID numbers, the instruments and gratings used, the
  total exposure time, wavelengths covered, a representative S/N per
  pixel for a combined spectrum, and the resolution and
  dispersion. For more detailed information see Petry et al. (2006).}
\end{center}
\end{table*}

\section{Re-analysis of \HI\ \lya\ coincidences across three
  sightlines}

\label{sec:lya-triple-coinc}
In this section we re-analyse absorber-absorber associations across
the three sightlines using our new line identifications. We also
describe a new way of classifying absorber coincidences across
multiple QSO sightlines.

\subsection{Re-analysis with new line identifications}

Our identification of higher order Lyman series lines and metal lines
in the forest of each QSO resulted in the identification of several
lines assumed by \cite{Petry06} to be \lya\ that are in fact due to
different transitions. Here we see what effect this has on their
analysis, and discuss their method for selecting triple \lya\
coincidences across the three sightlines.

\citet{Petry06} identified absorber pairs and triplets across the
three sightlines using nearest neighbour (NN) matching. When
identifying absorber pairs, the NN match for a given absorber is the
nearest absorber in velocity space in an adjacent sightline. Each
absorber in a given sightline has two nearest neighbours; one for each
remaining sightline. One might object to this matching method because
it uses only the velocity separations, and not the angular separations
when finding the nearest neighbour.  Indeed, we could find the nearest
absorber using the 3-d distance between each absorber pair instead of
a velocity difference. For a flat universe, the 3-d distance can be
found from the hypotenuese of the right angled triangle formed by the
comoving separation in the redshift direction (assuming the velocity
difference is due to some combination of the Hubble flow and peculiar
motion) and the transverse separation between two absorbers. However,
the transverse separations between the absorbers towards our QSOs
(always $< 0.5$ physical Mpc for our redshift range) are very small
compared to the redshift separations (75~\kms\ is $\sim 1$~Mpc at
$z=0.15$, assuming pure Hubble flow). Therefore it is reasonable to
use the velocity difference as a proxy for the 3-d separation. In
addition, it is a simpler way of matching absorbers than finding the
closest absorber in 3-d space, and has the advantage that it requires
no assumptions about peculiar velocities.

Petry et al. also introduced a variant of nearest neighbour matching:
symmetric matching. A symmetric \lya\ pair is defined as a pair of
absorbers that are nearest neighbours of each other. Unlike NN
matches, where a single absorber can be a member of many NN pairs, a
given absorber can be a member of only one symmetric pair. They use
symmetric matching to identify absorber triplets in a way that makes
no assumptions about the velocity offset between the members of a
triple.  A symmetric triplet is defined as a group of three absorbers,
one in each sightline, that are all NN of each other. They detect 12
such triplets, and find that this number is significant at the
$>99.99$\% confidence level compared to random Monte Carlo absorber
simulations.

Using our new identifications we find three of these 12 symmetric
triplets contain lines that are not due to \HI\ \lya.  We note that
two of these three have large velocity separations between the
absorbers that make up the triples -- 1062~\kms\ and 814~\kms\ --
consistent with them containing a spurious line.  A fourth triplet is
removed as it contains a line with detection significance of
2.16$\sigma$, less than our cutoff detection significance of
3$\sigma$. We are left with eight triplets, shown in
Table~\ref{tab:tripmem}. Using our own Monte-Carlo simulations (see
Appendix~\ref{sec:meth-gener-rand}) we find at least this number of
symmetric triplets in 163 out of 5000 random sets, giving a
significance level of 96.7\%, somewhat lower than the significance
reported by Petry et al. (see the top panel of
Figure~\ref{fig:symtrip}). Note that even though four of the triplets
have been removed, the significance level remains at the 97\% level as
there are fewer \lya\ lines in total, increasing the significance of
any matches. Thus the significance of the Petry et al result is
reduced, but not entirely removed by the new line identifications.

\subsubsection*{Advantages and disadvantages of symmetric triple
  matching}

While following the Petry et al symmetric matching procedure, we
discovered two potential problems with this matching
algorithm. Firstly, a small but significant proportion of the
symmetric triplets we find in the random absorber sets have
unphysically large velocity separations (larger than 4000~kms, or
$\gtrsim50$~Mpc assuming no peculiar velocity contribution). Secondly,
there are configurations of absorber groups that span the three
sightlines and have plausibly small velocity separations, but will
not be detected by the symmetric matching method. An example of such a
configuration is shown in Figure~\ref{fig:diagsym}. Finally we found
that if there are large gaps in the spectral coverage, the symmetric
matching method will identify many spurious triplets that are
comprised of absorbers matched across the gaps. We can minimise the
number of such spurious triplets by restricting our analysis to ranges
where we have complete wavelength coverage across the three
sightlines.  However, it is worth noting that this matching method is
best suited to spectra with a similar, continuous wavelength coverage
across each sightline, with slowly-varying detection limits.

To address the problem of triplets with spuriously large velocity
separations we can add a velocity cut criterion to the symmetric
matching algorithm, and only accept triplets with a maximum velocity
difference between any two triplet members below some cutoff
value. Petry et al. applied a 400~\kms\ velocity cutoff when comparing
equivalent widths of members of symmetric triplets in the real
absorbers to those in the random sets. The bottom panel of
Figure~\ref{fig:symtrip} shows the result of applying a cut of $\Delta
v < 1000$~\kms. No symmetric triples in the real absorbers are removed
with this cut but many random symmetric triples are removed, so the
significance level increases to $>99$\%. We could vary the cutoff
velocity difference to maximise the significance level of the observed
triplets, but this \textit{a posteriori} choice of a velocity cut is
unsatisfying. In addition, applying a velocity cut undermines one of
the appealing characteristics of the symmetric matching method: that
it makes no assumption about the velocity difference between triplet
members. \\
\\
Thus we conclude that using the symmetric matching method there is
still very strong evidence ($>99$\% confidence) for absorbing
structures across the three sightlines with typical velocity
separations smaller than $1000$~\kms.  However, the problems with the
symmetric matching described above led us to explore a different
method of identifying absorber triplets, which we describe below.

\subsection{A new method for identifying absorber coincidences in
  multiple sightlines}

In this section we explore a different algorithm for detecting triple
line coincidences.  It uses only a velocity cut to identify groups of
absorbers, without any nearest-neighbour matching.

The algorithm consists of the following steps: first a maximum
velocity separation, $\Delta v_{\rm max}$, is specified. Next we step
through each \lya\ absorber redshift in each sightline, and search
for absorbers in all three sightlines that have an absolute velocity
separation less than $\Delta v_{\rm max}$ from this redshift. All
absorbers that satisfy this condition are placed into a provisional
absorber `group'.  This process creates one `group' for each absorber
that contains one or more absorbers across one, two or three
sightlines, all with velocity separations less than twice $\Delta
v_{\rm max}$. Finally, to avoid counting the same structure multiple
times, we remove any group whose members are all members of another
single group (however, it is still possible for a single absorber to
be the member of more than one group).

This process locates triples classified using symmetric matching that
have a maximum velocity separation between their members that is less
than $\Delta v_{\rm max}$. It also finds configurations that symmetric
matching will miss (such as the example in Figure~\ref{fig:diagsym}),
and naturally selects pairs of absorbers across each set of two
sightlines.

We measure the number of absorber groups containing at least one
absorber in each sightline for both the real and random absorbers for
three different $\Delta v_{\rm max}$ values: 200, 500 and
1000~\kms. Table~\ref{tab:tripmem} shows all the lines that are
members of triple groups founds with $\Delta v_{\rm max} = 200$~\kms\
and $\Delta v_{\rm max} = 500$~\kms. The velocity cut method finds all
but one of the triples found using the symmetric matching method for
$\Delta v_{\rm max} = 500$~\kms. Although it is possible for multiple
absorbers in a single sightline to be part of a single triple (this
was a motivation for using the new algorithm), all of the triples
found below a $\Delta v_{\rm max}$ of 500 \kms\ have only a single
absorber in each sightline.

The number of groups found across all three sightlines in the real
absorbers compared to the number found in random absorber sets are
shown in Figure~\ref{fig:grouptrip}. There is a marginally significant
(94.3\%) excess of real absorber triples compared to the random sets
for $\Delta v_{\rm max} = 200$~\kms, with the significance decreasing
for $\Delta v_{\rm max} = 500$~\kms\ (89\%) and $\Delta v_{\rm max} =
1000$~\kms\ (81\%). This is a less significant result than that for
triplets identified with symmetric matching. If real physical
structures create configurations such as in Figure~\ref{fig:diagsym},
then symmetric matching will miss these structures, and this lower
significance level is a more accurate estimate of the probability that
there are physical structures spanning the three sightlines.

It is also possible that the configurations missed using symmetric
triple matching are not found in real observations -- perhaps there
are instrumental or physical reasons why the pairs of absorbers
comprising triple coincidences are unlikely to be found with small
separations, making configurations such as that shown in
Figure~\ref{fig:diagsym} unlikely. One possibility is that compared to
the random absorbers, real absorbers are less likely to be found with
small separations in the same sightline due to line blending -- for
example, two lines might be fitted as a single broad line if their
separation is smaller than the FWHM resolution of the FOS spectra
(1.29~\AA, or $\sim230$~\kms\ at 1700~\AA\ for the G230M grating). To
check the minimum wavelength difference at which lines can be
separated within a single sightline, we measured the wavelength
difference between adjacent lines within each sightline over the
G230M wavelength range.  The minimum separation between any two
adjacent lines (not only \lya\ transitions) is 1.8~\AA, or 320~\kms\
at 1700~\AA. The minimum separation we allow between lines in the same
sightline in our random absorber catalogues is 2.0~\AA\ (see
Appendix~\ref{sec:meth-gener-rand}), thus we do not expect real
absorbers to have fewer closely separated lines within a given
sightline than the random absorbers due to this effect.  Another
possibility is that absorption lines caused by structures spanning the
three sightlines are `isolated', and less likely to show multiple
absorption components over such small velocity ranges within each
sightline. However, there is no evidence in the literature for
reduced power in the \lya\ auto-correlation function on these scales.

In summary, our new method for selecting coincident absorption across
the three sightlines finds evidence at the $94$\% confidence level for
absorbing structures that span the three sightlines.  This is a lower
significance than the symmetric triple matching method ($97 - 99$\%).

\subsection{The rest equivalent width of symmetric triplet members}

The methods used above to identify absorber coincidences do not take
into account the equivalent width of lines comprising triple absorber
coincidences. \citet{Petry06} find that members of symmetric triplets
in the real absorbers have larger equivalent widths (averaged over
members of each triplet) compared to triplets found in the random
absorber sets. We find that this relationship still holds with our new
group identifications. We also measure the rest equivalent width of
lines that are members of symmetric triplets and compare this to the
rest EW of lines that are not members. Figure~\ref{fig:ewtrip} shows
the results: the number of lines are small, and a Kolmogorov-Smirnov
(KS) test gives a $68$\% probability that both data sets are drawn
from the same distribution, thus there is no evidence that members of
triplets have a significantly different rest EW to non-members.  We
repeat the same test using lines inside and outside the triple groups
identified using our matching method.  Figure~\ref{fig:ewtrip2} shows
the result.  Again the number of lines is very small, and a KS-test
cannot rule out the two data sets coming from a single distribution.

\section{Galaxy-absorber associations}

We have seen that identifying structures that span the three QSO
sightlines using \HI\ absorption alone is difficult due to the small
numbers of absorber coincidences and the relatively weak correlation
of \lya\ absorbers. This is a further motivation for adding the galaxy
positions to our analysis -- adding more statistical power to search
for large scale structures.  In the following sections we give a
qualitative description of the observed groups of galaxies and
absorbers, look at the associations between metal line systems and
galaxies, describe the statistical tests we use to measure the extent
of any absorber-galaxy association, and present the results of these
tests.

\subsection{Initial impressions}

Of our 60 galaxies with redshifts, 56 are at redshifts with
overlapping \lya\ coverage in at least two QSO sightlines. Of these
56, 22 have impact parameters $\rho < 500$~kpc from their nearest
sightline, and 5 have $\rho < 200$~kpc. The distribution of galaxy
$\rho$ from all of the three sightlines where there is \lya\ coverage
is shown in Figure~\ref{fig:rho}.

Before we describe our statistical tests for galaxy-absorber
associations, we look more closely at the wavelengths corresponding to
\lya\ absorption in the QSO spectra close to galaxy redshifts.  The
series of plots in Figure~\ref{fig:galspec} shows $\sim6000$~\kms\
spectral regions around one or more galaxies that overlap with spectra
in two or more sightlines. In general, all the galaxies shown in a
single plot are within a 1000~\kms\ velocity range, with the exception
of the $z=0.19$ group, which consists of three pairs of galaxies that
are spread across 4000~\kms. The left panels show the galaxy
distribution relative to the QSO sightlines in right ascension and
declination.  The transverse distances given are physical distances
relative to the B sightline at the redshift of each galaxy. The size
of the galaxy symbols are proportional to the galaxy luminosity
(though note the uncertainties in the absolute B magnitudes in
Table~\ref{tab:gal}), and colour shows the relative velocity
offset. The velocity zero point is at the mean redshift of \lya\ lines
overlapping with the galaxies, or the mean redshift of the galaxies if
no line overlaps. The open circle in the top left corner shows the
size of an L* galaxy for comparison. We found the Schechter function
parameter M*$(z)$ in the $b$ band by interpolating in redshift across
the DEEP2 values given in Table~4 of \citet{Faber07} and assuming
M*$=-20$ at $z=0$.  Galaxies are marked by their numbers from
Table~\ref{tab:gal}. If there is any \HI\ \lya\ absorption along a
sightline within 1000 \kms\ of the zero velocity position it is shown
by an inverted triangle. The area of the triangle is proportional to
the summed \lya\ rest equivalent width. If no \lya\ absorption is
detected within this range, the position of the sightline is shown by
a plus sign.

The right panels show portions of spectra corresponding to \HI\ \lya\
overlapping in velocity space with the galaxies in the left panel. Our
combined spectra and fitted continua are shown, with the lines
identified by Petry et al. shown as vertical tick marks. Solid tick
marks are lines in our \lya\ list and dotted tick marks are lines that
have been attributed to a transition other than \lya. Lines are marked
by their numbers from Tables~\ref{tab:a140} -- \ref{tab:c190}. Petry
et al. used a slightly different method to combine the spectra and fit
the continuum, thus there may be small differences in the apparent
significance of lines in these plots compared to Tables~\ref{tab:a140}
-- \ref{tab:c190}. The nearby galaxy positions are shown as small
triangles at zero flux in each spectrum. Filled triangles are galaxies
that are also shown in the corresponding left panel. In
Appendix~\ref{sec:descr-galaxy-absorb} we comment on each of the
Figure~\ref{fig:galspec} plots individually.

There are several candidates for physical associations between
absorbers and nearby galaxies. There are two large groups of galaxies,
one at $\langle z\rangle=0.202$ and another over the range
$z=0.233-0.238$. Both of these groups are near sightlines A and B,
where the GHRS spectra cover the expected \lya\ positions. Both groups
show strong (rest EW $> 0.3$\AA) \lya\ absorption in one sightline,
but no strong absorption in the neighbouring sightline.

One faint ($\sim0.03$~L*) galaxy at $z=0.053$ is within 30 kpc of
sightline A and 100 kpc of sightline B.  \lya\ absorption is seen in
both sight lines at the galaxy's redshift. Unfortunately we do not
have access to a spectrum of this galaxy, and so cannot examine its
properties in detail. Another brighter galaxy ($\sim 1.2$~L*) at
$z=0.227$ is $170$~kpc from sightlines A and B.  Very strong \lya\
absorption (rest EW $>1$~\AA) is seen in both sightlines within
200~\kms\ of the galaxy redshift. We do have a spectrum of this galaxy
and we use it to constrain the galaxy's star formation history in
Section~\ref{sec:bright-star-forming}.

There is also an example of a galaxy close to a sightline that does
not show associated strong \lya\ absorption: the galaxy at $z=0.197$
is only 110~kpc away from sightline B, but there is no \lya\
absorption with rest EW $> 0.2$\AA\ within 400~\kms\ of the galaxy
redshift. We note that there is absorption about 600~\kms\ from the
galaxy redshift that has been identified as \lyg\ from a
higher-redshift absorber that may be masking \lya\ absorption. If
there is no \lya\ absorber associated with this galaxy, this means
that a simple model of each galaxy surrounded by a spherical halo of
HI gas that always gives rise to absorption in a nearby sightline
cannot explain the data without some modification (such as a variable
covering factor).

\subsection{Galaxies and metal line systems}
\label{sec:galaxies-metal-line}

There are six QSO absorption systems that show associated metal
lines. Five of these were identified in Petry et al.; the sixth is the
new sub-DLA we have identified towards QSO C. The lines comprising
each system are listed in Table~\ref{tab:metals}. One of the systems
is a previously known `grey' Lyman limit system\footnote{A Lyman-limit
  system where significant flux remains bluewards of 912~\AA\
  (rest).}, the other is the new sub-DLA system we discovered. At the
moderate FOS and GHRS resolutions, column densities and velocity
widths of lines cannot be easily measured, so we have no detailed
information about the ionisation state or metallicity of the
systems. However, we can look at their distribution relative to the
galaxies and across sightlines.

It is interesting that every metal-line system in sightlines A and B
has a \HI\ \lya\ line within $300$~\kms\ in the nearby B or A
sightline. Given the minimum rest equivalent width of an absorber
containing a metal line ($\sim 0.45$~\AA) and the number of absorbers
above this rest EW in each sightline that have \lya\ with 300~\kms\ in
the nearby sightline, we calculate the probability of this occurring
by chance is less than $3.5$\%. This is consistent with these metal
line systems being associated with \HI\ gas structures spanning the
size of the sightline separations ($300-400$~kpc). However, absorbers
with strong metal systems are not expected to be part of a single
coherent structure so large -- indeed, sub-DLAs are thought to be much
smaller structures with typical sizes of a few tens of kpc. An
alternate explanation is that such absorbers trace the halos of
galaxies that are themselves embedded in a $\sim0.5$~Mpc structure.

\begin{figure*}
\begin{center}
\includegraphics[width=1.0\textwidth]{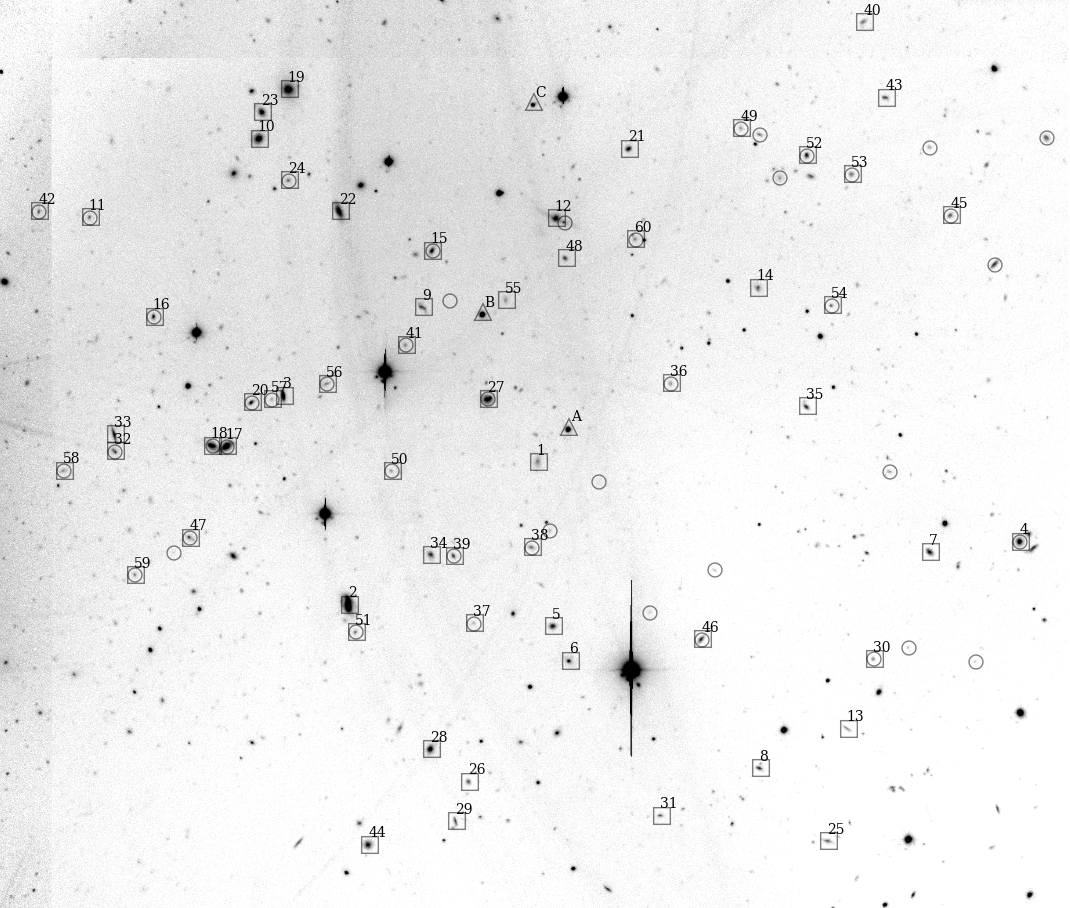}
\caption{\label{fig:image} The \textsl{CFHT} imaging used to select
  galaxy candidates for the MOS observations. North is up and East is
  to the left.  The field is around $10 \times 8$ arcmin. Triangles
  show the QSOs, squares show galaxies with redshifts, and circles
  show objects targeted with the \textsl{CFHT} MOS. The numbers next
  to each square identify the galaxies in Table
  \ref{tab:gal}. Redshifts for objects with a square but no circle
  were provided to the authors by R. Weymann and M. Rauch.  The proper
  \textsl{CFHT} baffles were absent during the imaging, causing the
  diffuse arcs of scattered light in the background.}
\end{center}
\end{figure*}

\begin{figure}
\begin{center}
\includegraphics[width=0.49\textwidth]{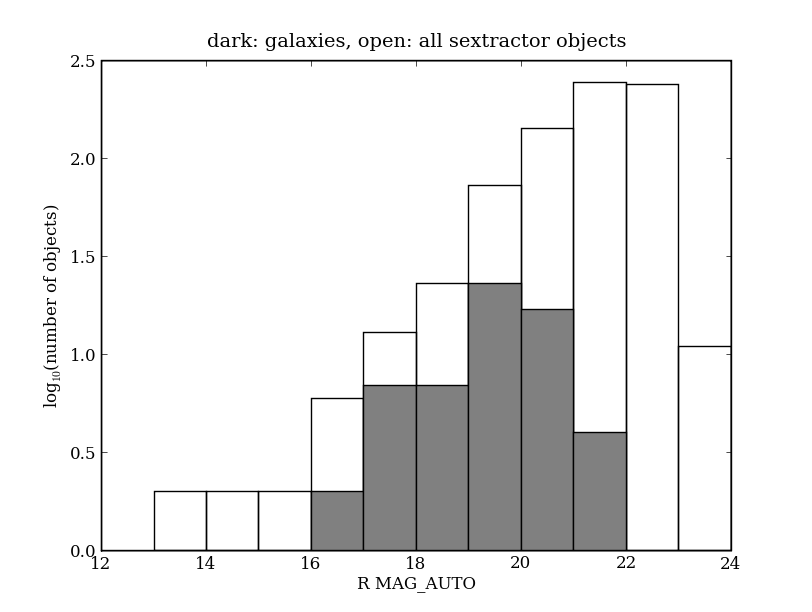}
\caption{\label{fig:numcount} Number counts for the galaxies and
  SExtractor-selected objects. The grey histogram shows the number
  counts for galaxies with redshifts, the open histogram for all
  objects. The object catalogue list appears complete to R~$=21.5$.
  The galaxy completeness drops significantly at R~$>20.5$.}
\end{center}
\end{figure}

\begin{figure*}
\begin{center}
\includegraphics[width=1.0\textwidth]{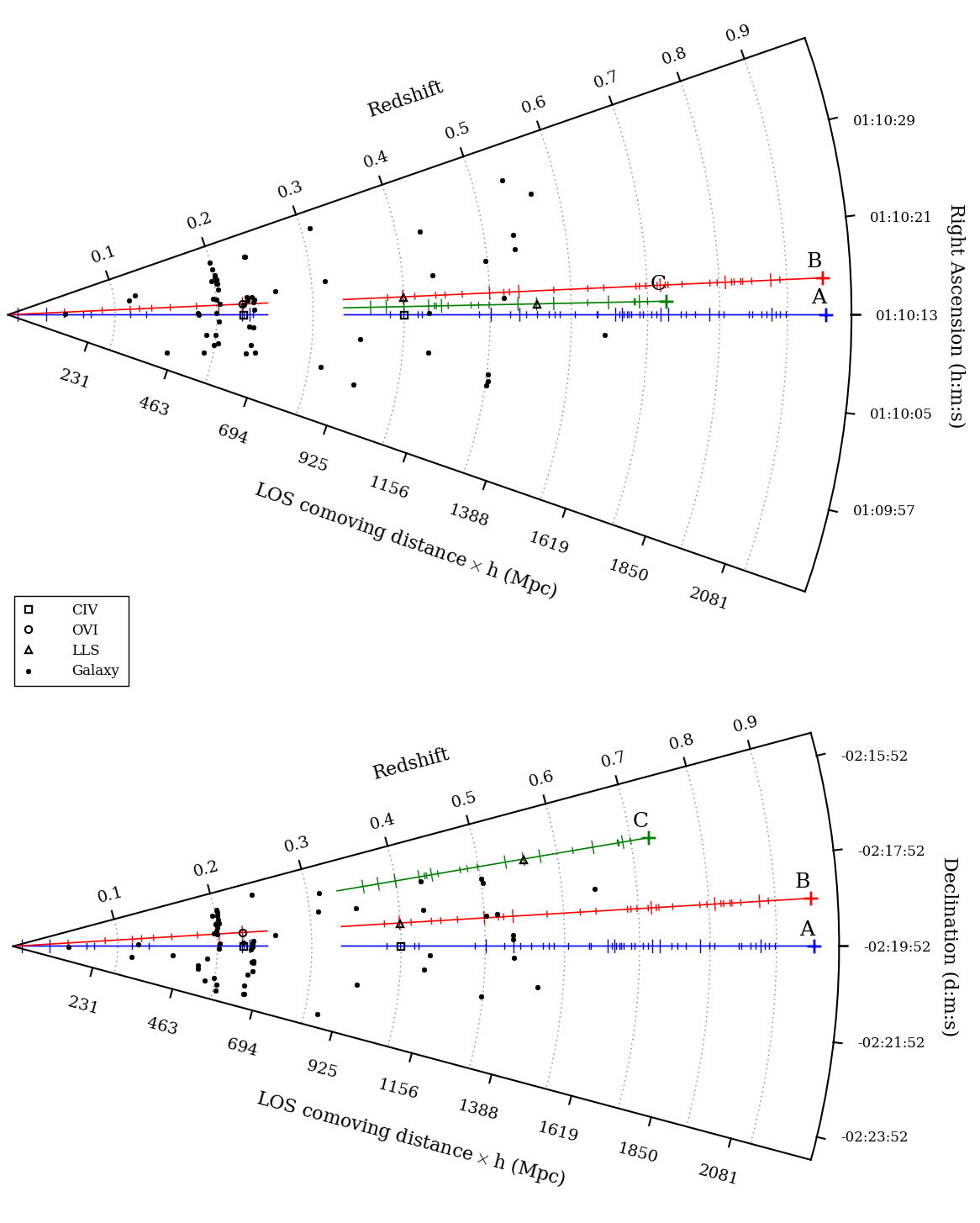}
\caption{\label{fig:pie} The distribution of galaxies and absorption
  lines in the Q0107 field. Angular separations have been exaggerated
  by 200 times to clearly separate the galaxy and absorber
  symbols. Galaxies are shown by black dots and the three QSOs are
  shown as plus signs.  Absorbers along the QSO sightlines are shown
  by strokes perpendicular to the sightline: narrow strokes represent
  \HI\ absorbers with rest EW $< 1$ \AA, wider strokes represent rest
  EW $\ge 1$ \AA. The positions of \OVI, \CIV, the Lyman-limit system
  in sightline B and probable sub-DLA in sightline C are also
  shown (see the Figure legend).}
\end{center}
\end{figure*}

\begin{figure*}
\begin{center}
\includegraphics[width=1.0\textwidth]{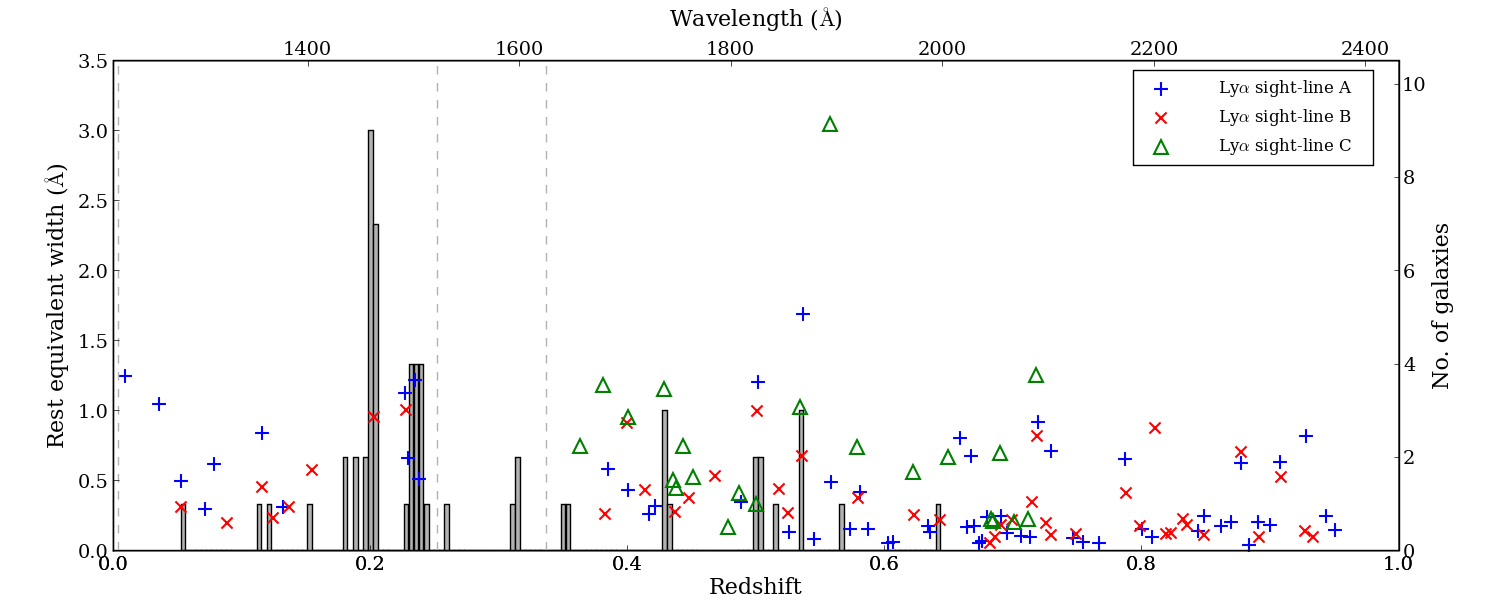}
\caption{\label{fig:z} The redshift distribution of the galaxies and
  \lya\ absorbers. The grey histogram shows our sample of galaxy
  redshifts. Plus signs, crosses and triangles show the \lya\ absorber
  redshifts and their rest equivalent widths. The dashed lines show
  the cutoff wavelengths for the GHRS spectra and FOS spectra.  QSO C
  was not observed with the GHRS, so there are no detected absorbers
  at redshifts $\lesssim0.3$ in this sightline.}
\end{center}
\end{figure*}

\begin{figure}
\begin{center}
\includegraphics[width=0.49\textwidth]{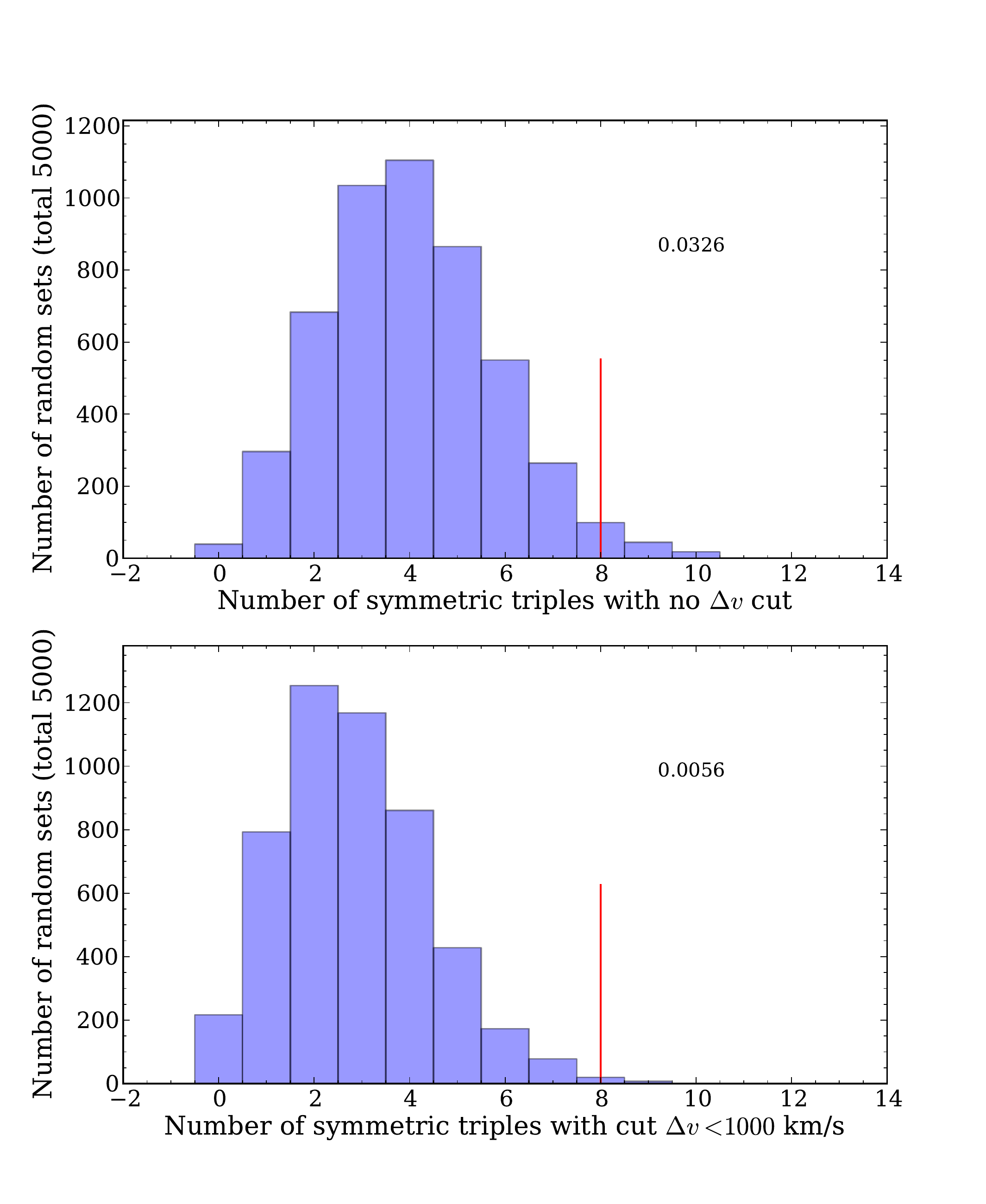}
\caption{\label{fig:symtrip} \textit{Top}: The number of symmetric
  triples across the three sightlines for a thousand sets of random
  absorbers and the real absorbers with our new line identifications
  (vertical line). This figure is directly comparable to Petry et
  al.'s Figure~11. The probability of matching or exceeding the number
  of triples in the real absorbers by chance is shown in the top
  right-hand corner; 3.3\%.  \textit{Bottom}: Similar to the top
  panel, but showing only symmetric triples containing pairs with
  velocity separations less than 1000 km/s. The significance has
  increased compared to the top panel.}
\end{center}
\end{figure}

\begin{figure}
\begin{center}
\includegraphics[width=0.49\textwidth]{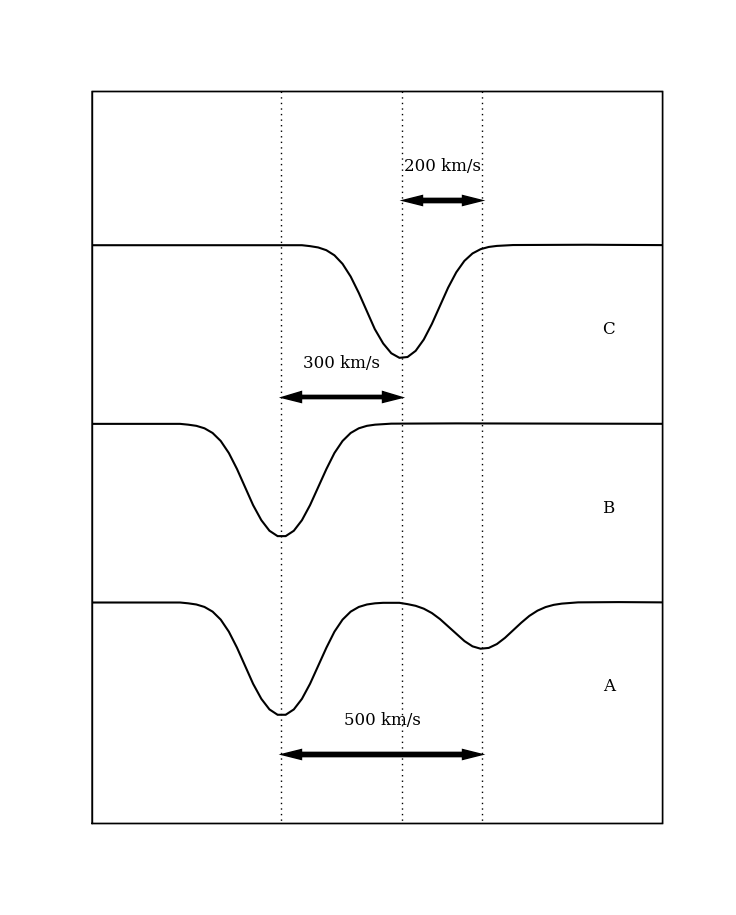}
\caption{\label{fig:diagsym} An example of a physically plausible
  group of absorbers spanning three sightlines that is not selected
  using the symmetric triple matching algorithm from \citet{Petry06},
  as there is no combination of three lines that are all nearest
  neighbours of each other. This system is selected by our group
  selection algorithm.  (Note that lines with $\approx 500$~\kms\
  separations are clearly separated in a single sightline, even in
  the moderate resolution FOS spectra).}
\end{center}
\end{figure}

\begin{figure}
\begin{center}
\includegraphics[width=0.49\textwidth]{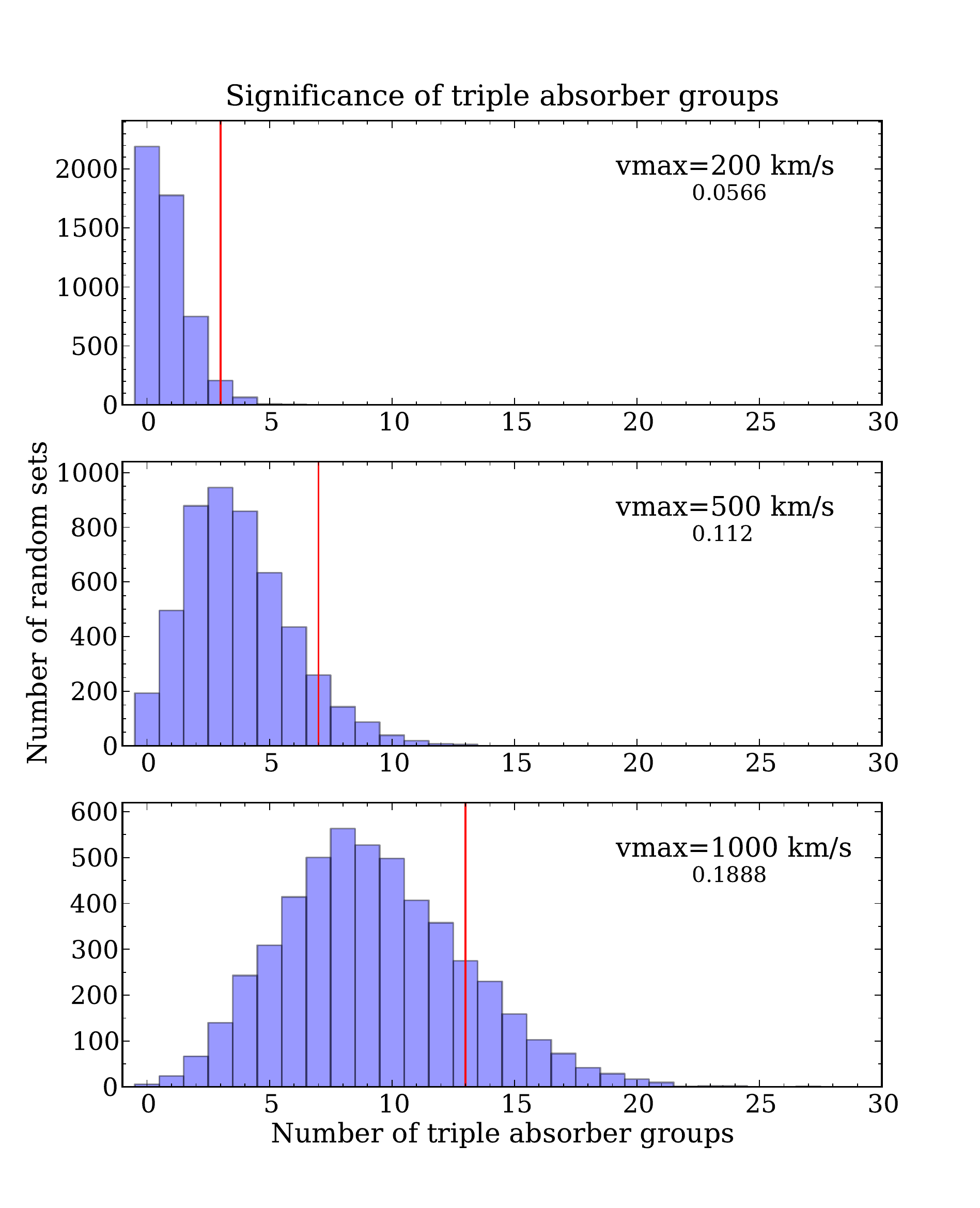}
\caption{\label{fig:grouptrip} The number of triple absorber groups in
  the real absorbers compared to the number found in random absorber
  sets for three maximum velocity differences between the absorber
  group members. The histograms show triple groups in the random sets,
  and the vertical line shows the triple groups in the real
  absorbers. The probability of matching or exceeding the number of
  triples in the real absorbers based on the distribution of the
  random absorbers is shown on each panel. There is marginal evidence
  for an excess of triple absorber groups identified using vmax~$=200$~\kms\
  (significance of 94.7\%).}
\end{center}
\end{figure}

\begin{figure}
\begin{center}
\includegraphics[width=0.49\textwidth]{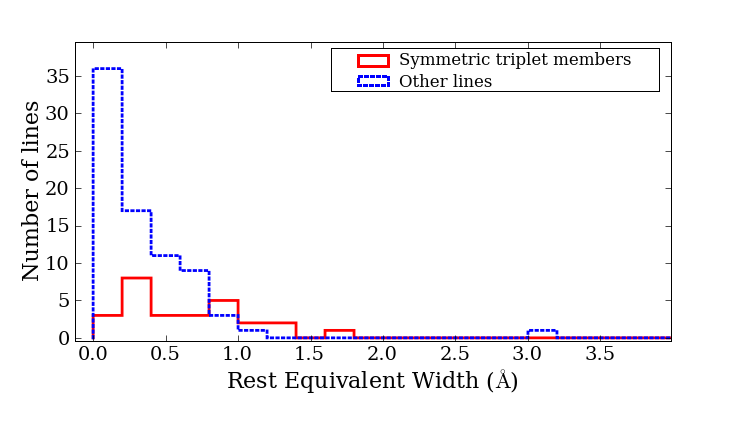}
\caption{\label{fig:ewtrip} The distribution of rest equivalent widths
  for members of symmetric absorber triplets (solid histogram) and
  lines not in triplets (dotted histogram). There is a visual
  impression that absorbers in symmetric triples have a higher
  equivalent width than the absorbers, but the number of lines is
  small, and a K-S test does not suggest the distributions are
  significantly different.}
\end{center}
\end{figure}

\begin{figure}
\begin{center}
\includegraphics[width=0.49\textwidth]{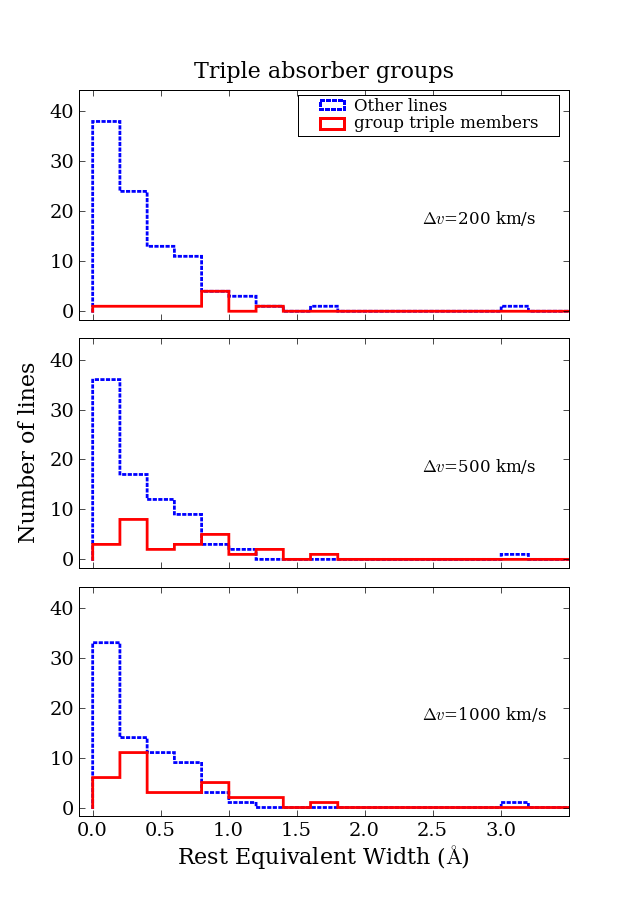}
\caption{\label{fig:ewtrip2} The distribution of rest equivalent widths
  for members of absorber group triples (solid histogram) and lines
  not inside group triples (dotted histogram) for three velocity
  cutoffs. Again there is a visual impression that triple group
  members are more likely to have higher rest EWs, particularly for
  the $\delta v = 200$~\kms, but a K-S test does not suggest the
  distributions differ significantly.}
\end{center}
\end{figure}

\begin{figure}
\begin{center}
\includegraphics[width=0.49\textwidth]{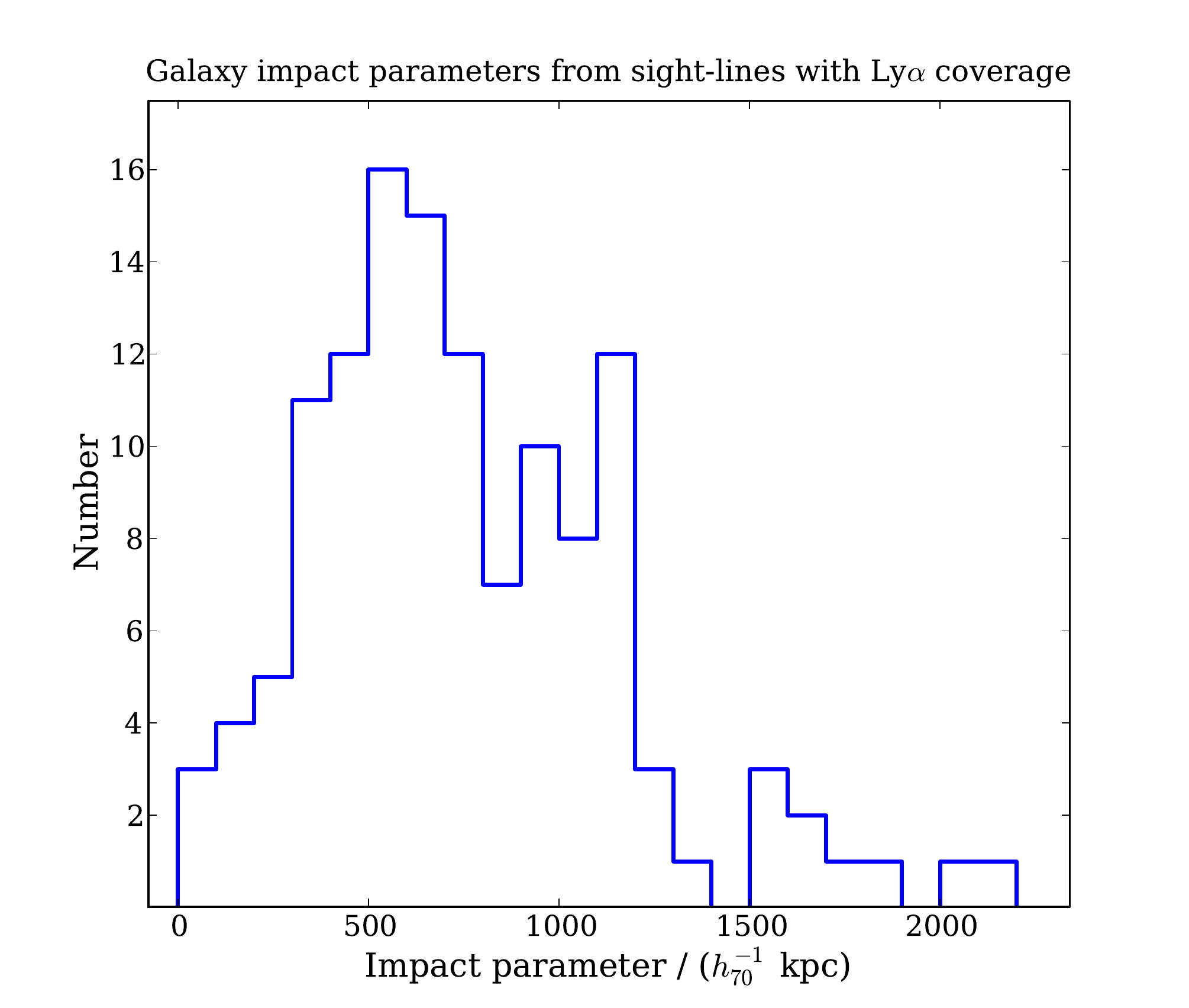}
\caption{\label{fig:rho} The distribution of impact parameters of
  galaxies from all three sightlines where there is coverage of the
  \lya\ line corresponding to the galaxy redshift.}
\end{center}
\end{figure}

No galaxies are seen close to the LLS in sightline B or the sub-DLA in
sightline C. High \NHI\ absorbers are sometimes found close to faint
dwarfs \citep{Rao03, Stocke04}.  However, such galaxies are too faint
to have been targeted in the available spectroscopy samples.  There
are several faint galaxy candidates in the imaging shown in
Figure~\ref{fig:image} close to both sight lines. Typical impact
parameters of galaxies that have been found close to LLS and DLAs
range from $<10$~kpc \citep{Gharanfoli07} to 90~kpc
\citep{Jenkins05}. This corresponds to a separation from the QSO
sightline of $~1-10$~arcsec. It is also possible that an associated
galaxy might be blended with the QSO in our ground-based imaging.

\subsubsection*{A bright galaxy with nearby metal absorption at
  $z=0.227$}
\label{sec:bright-star-forming}

This a single bright ($1.2$~L*) galaxy at $z=0.2272$, 200~kpc from
sightlines A and B. There are two strong (rest EW~$\sim 1$~\AA), broad
\lya\ absorbers (FWHM $2.26\pm0.86$ and $3.44 \pm 1.60$~\AA, or
$\sim450$ and $\sim690$~\kms) in each sight line within 200~\kms\ of
the galaxy position. The large rest EW and velocity width suggest the
\HI\ absorption is made up of several velocity components. Both the
absorbers in sightlines A and B show associated metal absorption:
\OVI\ and \CII\ is seen in sightline B (though the S/N is low at the
wavelengths of these lines and \OVI\ appears blended with unrelated
\lya ); \CIV\ and possibly \CII\ is seen in sightline A. \Ha\ and
[\OIII] emission in the galaxy's spectrum
(Figure~\ref{fig:closegalspec}) show it is undergoing some
star-formation, and \Hb\ absorption shows there is a population of
young and intermediate age stars.

We quantify the star-formation history of the galaxy by measuring the
\Ha, [\NII], [\OIII] and [\OII] emission line intensities, the
continuum drop over the 4000~\AA\ break quantified by the D(4000)
parameter, and the \Hd\ absorption line rest equivalent width (\Hd$_A$
index).

\citet{Kauffmann03} show that D(4000) and \Hd$_A$ can be used to
discriminate between two galaxy star formation scenarios.  The first
scenario models the star formation rate (SFR) in a galaxy as
decreasing exponentially from some initial formation time. The second
has the same decreasing SFR along with randomly-occurring bursts of
star formation.

The \Hd$_A$ index (as used by Kauffmann et al.) is defined in
\citet{Worthey97} as the rest equivalent width of the \Hd\ line from
$4083.50-4122.25$\AA.  The continuum over the line is defined using
the adjacent regions $4041.60-4079.75$~\AA\ and $4128.50-4161.00$~\AA.
Two continuum reference points are calculated using the mean flux and
mean wavelength in each of these regions. The continuum over the \Hd\
line is then given by a straight line joining these two reference
points.

D(4000) was first described in \citet{Bruzual83}. It is the ratio of
average flux densities $F_{\nu}$ above and below the 4000~\AA\ break.
We measure average flux densities using rest-frame wavelength ranges
$3850-3950$~\AA\ and $4000-4100$~\AA\ \citep{Balogh99}.

We find the $z=0.2272$ galaxy has \Hd$_A = 2.0 \pm 0.5$~\AA, and
D(4000)$=1.28 \pm 0.03$.  These are formal errors from the $1\sigma$
flux errors, and the true errors are likely to be somewhat larger due
to emission or absorption in the continuum
regions. Figure~\ref{fig:kauf} shows the values for this galaxy
overlayed on a reproduction of Kauffmann et al.'s Figure~6, we find
that this galaxy falls near a region where 95\% of the model galaxies
have not experienced bursts of star formation in the last 2 Gyr.

The \Ha\ emission line allows us to make a rough estimate of the star
formation rate. Nebular lines originate in gas surrounding
recently-formed stars, and thus their intensity gives a direct measure
of the star-formation rate of the galaxy (subject to many potential
pitfalls outlined in e.g. \citealt{Kewley02}).  We do not correct for
absorption from dust.  There is no measureable \Hb\ emission, and thus
we cannot use the difference between the expected and observed ratio
of \Ha\ to \Hb\ intensities to correct for any dust reddenning. Since
there is Balmer absorption visible at \Hb\ and \Hg\ it will also be
present at \Ha, and so our line intensity measurements will be smaller
than the true intensity.

Mindful of these limitations, we used \Ha\ line flux to measure a star
formation rate.  We modelled the blended \Ha\ and nearby [\NII] lines
with Gaussians, and use the deblended Gaussian to measure the \Ha\
line intensity. We fitted the continuum level around the emission
lines using a median filter, measured the summed flux over the \Ha\
line, and then converted to a luminosity using the relations given in
Section~\ref{sec:introduction}.  Using the conversion relation from
\citet{Kennicutt98}, SFR~(\msunyr)$ = 7.9 \times 10^{42}
L(\Ha)$~(\ergs), we found a star formation rate of 0.45 solar masses
per year. This is relatively small for a late-type galaxy, but normal
for an early-type.

Using the relations in \citet{Kewley02b} (their Figures~7 and 8) we
can estimate the oxygen abundance, (O/H), in the galaxy's nebular
regions using ratios of the line intensities of [\OIII], [\NII] and
\Ha.  Keeping in mind that we have not corrected for reddening or
Balmer absorption, we measure the ratios [\NII]~$\lambda6583$ / \Ha~$=
0.9$ and [\NII]~$\lambda6583$ / [\OIII]~$\lambda5007$~$=2.4$.  The
[\NII]/\Ha\ is surprisingly large, which is indirect evidence that
there is significant Balmer absorption at \Ha, and that the star
formation rate is higher than 0.45~\msunyr. If we assume that the true
\Ha\ intensity is five times larger than the measured intensity we
find a range $8.5 <$~log(O/H)$+12 < 9.4$. For [\NII]/[\OIII] we find a
range $8.2 <$~log(O/H)$+12 < 9.2$.  Thus it is likely that the
galaxy's (O/H) is larger than half solar, taking the solar oxygen
abundance log(O/H)$+12 = 8.66$ from \citet{Asplund04}. These values
are also consistent with the galaxy having undergone the bulk of its
star formation more than $1-2$~Gyr ago.

The imaging is too poor quality to differentiate whether the galaxy is
an elliptical or spiral based on its surface brightness profile.  A
fainter, stellar object is also seen very close to the galaxy, and
this object fell inside the slit targeting the galaxy. We are unsure
of its redshift, but the R band imaging suggests it does not
contribute a significant amount of flux to the galaxy spectrum.

We could imagine the following scenario for this galaxy, consistent
with the derived values above: a significant burst of star formation
occurred more than 2 Gyr ago, producing metal-enriched gas that was
ejected by winds. It is possible that the metals observed in the two
nearby sightlines were created by these winds: an average wind speed
of 100~\kms, typical of superwinds that can be associated with strong
bursts of star formation, is equivalent to 100~kpc per Gyr. This is
only one scenario; an equally plausible explanation for the enriched
gas would be that it is more closely associated with galaxies our
survey has missed.  Interestingly both absorbers show absorption by
high and low ionisation species metals, suggesting a multiphase
environment.

\subsection{Statistical tests for absorber-galaxy associations}
\label{sec:stat-tests-absorb}

There are two ways to approach a joint analysis of the galaxies and
absorbers. The first asks the question: given the galaxy distribution,
how likely are we to see the observed distribution of absorbers?  The
second asks: given the absorber distribution, how likely are we to see
the observed galaxy distribution?  We are currently obtaining a deeper
sample of galaxy redshifts over a larger area in this field using the
VIMOS and DEIMOS spectrographs, which will allow us to better address
the second question in a future paper. However, since our knowledge of
the selection function for our absorption lines is much better than
that of our current galaxy sample, in this paper we focus on the first
question.

We use two statistical matching methods to identify associations
between galaxies and absorbers. The first uses a nearest-neighbour
matching algorithm to link galaxies and absorbers, the second uses a
velocity cut and identifies absorbers across all three sightlines
that are associated with a galaxy. Both tests compare the number of
galaxies with nearby absorbers using the real absorber distribution,
and an ensemble of 5000 random absorber distributions. The method we
use to generate random absorbers is described in
Appendix~\ref{sec:meth-gener-rand}.

For the first matching method we identify the nearest \lya\ absorber
in each sightline to each galaxy in velocity space. We refer to this
as a nearest neighbour absorber (NNA). We do this both for the
galaxies and real absorbers, and the galaxies with each set of random
absorbers. We compare the distributions of velocity differences
between galaxies and their NNAs in the real and random absorbers.

The second matching method identifies galaxy-absorber `groups' in a
way similar to that we used to identify absorber groups in
Section~\ref{sec:lya-triple-coinc}.  We step through each galaxy
redshift and assign every \lya\ line in any sightline that is within
some velocity range, $\Delta v$, of that galaxy as an absorber
associated with that galaxy. Thus we have one `group' for each galaxy,
consisting of the galaxy and zero or more absorbers in zero to three
sightlines with which it is associated. Note that absorbers can be
`associated' with more than one galaxy. We do this for both the real
absorbers and random absorbers and compare the group properties
between them.

We attempted to use nearest-neighbour matching methods to link a
galaxy with absorbers in more than one sightline, but found that the
above `group' finding method was simpler to explain and implement, and
should be more clearly linked to physical structures.

\subsection{Results}

Using the statistical tests described above, we compare the
associations between galaxies and absorbers for the real absorbers and
random absorber sets.

First we use nearest-neighbour matching, looking at the smallest
distance of a galaxy from an absorber in velocity space in each sight
line. Figure~\ref{fig:NNag} shows the results -- there is a clear
excess ($>99$\% significance) of pairs at galaxy-absorber velocity
separations $<200$~\kms\ compared to the random absorbers in all three
sight lines. The signal is slightly weaker for sightline C, likely
due to the smaller wavelength overlap of the C spectrum with the
galaxy distribution. The excess at $\Delta v < 200$~\kms\ is
consistent with the excess of galaxy-absorber pairs in the larger
sample of (single) QSO sightlines and galaxies in \citet[][see their
Figure~25]{Morris06}.  This is expected, as a large fraction of the
data used in this paper is included in their analysis.

Next we look at the statistics of galaxy-absorber groups for a variety
of velocity cutoffs.  We make comparisons of three quantities between
the observed and random sets: the number of galaxies with at least one
associated absorber (in any sightline), the number of galaxies with
associated absorbers in at least two sightlines, and the number of
galaxies with associated absorbers in all three sightlines. We refer
to these three types of groups as single-LOS galaxy groups (here LOS
is an abbreviation for line-of-sight), double-LOS galaxy groups and
triple-LOS galaxy groups respectively. We measure the number of such
groups using velocity cutoffs for absorber-galaxy associations (as
described in the previous section) of 200, 500, 750 and 1000~\kms. The
results are shown in Figures~\ref{fig:gabs200} to
\ref{fig:gabs1000}. There is a significant (generally $>99\%$) excess
of galaxies with associated absorbers in one, two and three
sightlines compared to the random galaxy-absorber groups. Note that
while 56 galaxies have overlapping spectra covering \HI\ \lya\ in two
sightlines, only 16 galaxies have three sightlines overlapping in
\lya. Thus statistics are poorer for the triple LOS-galaxy groups.

\section{Comparison with simulations}
\label{sec:comp-with-simul}

In this section we compare the observations with the distribution of
gas and galaxies within cosmological hydrodynamic simulations; for
this task we use the \textit{Galaxies-Intergalactic Medium Interaction
  Calculation} (\gimic; \citealt{Crain_gimic_astroph_09}). \gimic\ is
designed to circumvent the large computational expense of simulating
large cosmological volumes ($L \gtrsim 100$~\hMpc) at high resolution
($m_{\rm gas} \lesssim 10^7$~\hMsun) to $z=0$.

Using `zoomed' initial conditions \citep{Frenk_et_al_96,
  Power_et_al_03, Navarro_et_al_04_short}, \gimic\ traces with full
gas dynamics the evolution of five roughly spherical regions drawn
from the Millennium simulation \citep{Springel_et_al_05_short}. In
order to trace a wide range of large-scale environments, the regions
were chosen such that their overdensities deviate by (-2, -1, 0,
+1, +2)$\sigma$ from the cosmic mean, where $\sigma$ is the rms mass
fluctuation, on a scale of $18$\hMpc, at $z=1.5$. The $+2\sigma$
region was additionally constrained by the requirement to be centred
on a rich galaxy cluster halo. In practice this ensures that the
simulations include rare cosmological features, since the $-2\sigma$
region is also approximately centred on a sparse void. 

Each region has an approximate comoving radius of $18$~\hMpc, except
the $+2\sigma$ region which was necessarily enlarged to a radius of
$25$~\hMpc\ in order to accommodate the rich cluster. The remainder of
the $500^3$~(\hMpc)$^3$ Millennium simulation volume is modelled with
collisionless particles at much lower resolution to provide the
correct tidal forces.

The simulations were evolved with a version of the \gadget\ code,
that includes:

\begin{enumerate}

\item a recipe for star formation designed to enforce a local
  Kennicutt-Schmidt law \citep{Schaye_and_Dalla_Vecchia_08};

\item stellar evolution and the associated delayed release of 11
  chemical elements (Wiersma et al. 2009b);

\item the contribution of metals to the cooling of gas, computed
  element-by-element, in the presence of an imposed UV-background
   (Wiersma et al. 2009a);

\item galactic winds that pollute the IGM with metals and can quench
  star formation in low-mass haloes
  \citep{Dalla_Vecchia_and_Schaye_08}.

\end{enumerate}

It does not, however, model the evolution of black holes or feedback
effects associated with them.

We consider here the intermediate-resolution \gimic\ simulations,
since they form a complete set of (-2, -1, 0, +1, +2)$\sigma$ regions for
all redshifts. Crain et al. demonstrate that the global star
formation rate density of these simulations is numerically converged for
$z\lesssim 6$, as are the specific star formation rates of galaxies
residing in dark matter haloes with circular velocities $v_{\rm c}
\gtrsim 100$~\kms. The simulations are therefore well-suited for
comparison with our observations.

A detailed analysis of the low redshift IGM-galaxy relationship in the
\gimic\ simulation will be presented in a future paper.  In
this paper we focus on identifying triple absorber coincidences and
triple absorber/galaxy groups in the simulations. We compare their
properties to our current sample of absorbers and galaxies and test
whether the number of galaxy-multiple absorber groups we see per unit
redshift in the simulations is consistent with the number per unit
redshift seen in the real data.

To do this we must generate mock spectra through the simulation that
mimic the properties of the observed spectra, and identify galaxies in
the simulation that match the properties of those in our \textsl{CFHT}
galaxy sample.

\subsection{Simulated Galaxy Properties}

As described in Crain et al., a `galaxy' in the simulations is defined
as the stellar component of self-bound substructures. We identify
these substructures using a version of the \subfind\ algorithm
\citep{Springel01} modified to consider baryonic particles (i.e. gas
and stars) as well as the dark matter when identifying substructures
within haloes identified by the friends-of-friends (FoF) algorithm
\citep{Dolag_et_al_08}. This definition is unambiguous and allows more
than one galaxy to be associated with any FoF halo.

To compare the galaxies in the simulation to our observed galaxies, we
estimate stellar masses for our observed galaxies from their rest
frame absolute B magnitude estimates. Using the mass-to-light ratios
in Figure 14 of \citet{Kauffmann03}\footnote{Kauffman's mass-to-light
  ratios were inferred from models generated using the stellar initial
  mass function from \citet{Kroupa01}.}, we find that for galaxies
overlapping the redshift range where we have spectral coverage of
\lya\ across all three sightlines ($0.35 < z < 0.7$), we are
sensitive to a minimum stellar mass of $\sim 6.0\times 10^{10}$
\msun. We adopt a cut of $6.0\times 10^{10}$ \msun\ for the total
stellar mass of galaxies in the simulations to associate with
absorbers. The stellar mass estimates obtained using rest frame B are
uncertain by around a factor of five, but as we shall see, the triple
absorber-galaxy statistic does not depend strongly on the galaxy mass
cut we adopt.

To account for the fact that our observations sample only a fraction
of all the galaxies above a given stellar mass over $0.35 < z < 0.7$,
we randomly select a fraction of the simulated galaxies above the
minimum stellar mass. Selecting $20$\% of the available galaxies gives
a number of galaxies per unit comoving Mpc$^3$ in the simulations
similar to the observed number density in the range $z=0.35-0.7$. This
fraction is consistent with our estimated completeness level for the
observed galaxy sample above the luminosity corresponding to our
stellar mass cut, and over this redshift range.

We do not add a redshift error to the simulated galaxies. The velocity
difference we use to find groups is 1000~\kms, so we do not expect the
addition of a 180~\kms\ error (comparable to error on the observed
galaxy positions) to significantly affect the number of groups found
in the simulation.

\subsection{Triple sightlines in the simulation}

To compare the simulations to the observations, we must compare a long
narrow observed region that evolves continuously with redshift, with
the small \gimic\ regions, each of which is at a single redshift
corresponding to that of the snapshot. The redshift path probed by the
observations where we have coverage of the three sightlines is 0.37.
Compare this path to the simulated regions: one \gimic\ region is
roughly $40h^{-1}$ comoving Mpc across, equivalent to a $\Delta z
\approx 0.013$, a velocity range of $\sim 2000$ \kms\ or wavelength
range of 11 \AA\ at $z=0.5$.

We generate sightlines through the simulation with the same angular
geometry and separations as the real sightlines. We choose snapshots
at three redshifts (0.25, 0.5 and 1.0) spanning the redshift range of
our observations. To probe a redshift path length at least as long as
the observed path length and to maximise the number of groups we can
find, we must generate as many sets of three sightlines through the
simulations as possible. However, if we are to treat these sets of
sightlines as independent, we must ensure that two sightlines do not
sample the same large-scale structure in a given simulation
snapshot. With these goals in mind, we use the following strategy for
generating sightlines through a snapshot: we generate many sets of
three sightlines, all parallel to one of the simulation axes -- for
example, the Z axis.  We generate triple sightlines with random X, Y
positions, ensuring that they are at least $5h^{-1}$ comoving Mpc from
the edge of the region at the point where they pass through the
region's Z centre.  We also ensure that every set of sightlines is
separated by a minimum transverse distance from neighbouring
sets. This is done by ensuring the centre of the triplet (defined by
the mean X and Y position of the triplet) is separated by at least
twice the largest distance between two sightlines within a single
triplet, from any neighbouring set of sightlines. We generate as many
sets of sightlines possible along the Z-axis with random X-Y
positions that satisfy these conditions. We then repeat this process,
generating two more sets of sightlines parallel to the Y and Z axes.

Since the simulations cover a constant volume in comoving space, we
can sample many more independent sightlines at $z=0.25$ than $z=1$. At
$z=0.25$ we generate 100 sets of triples per axis for a total of 900
sightlines. At $z=0.5$ we generate 35 triples per axis (315
sightlines) and at $z=1$, 15 triples per axis (135 sightlines).

Sample sets of sightlines generated through the $+2\sigma$ region are
shown in Figure~\ref{fig:sim_nhi_trip}. Even though we try to minimise
sampling the same structures multiple times, inevitably this will
occur to some degree, and so the Poisson errors on the number of
groups detected in our simulation are probably an underestimate of the
true error.

\subsection{Simulated absorber properties}
\label{sec:simul-absorb-prop}

Once we generate sightlines, we must calculate the absorption
properties of the gas along each sightline. To do this we use the
program \textsc{specwizard} written by Joop Schaye, Craig M. Booth and
Tom Theuns. \textsc{specwizard} finds the contributions from gas to the
optical depth along a sightline though the simulation using the
method described in \citet{Theuns_sph_sim_98}. Tables generated with
\textsc{cloudy} \citep{Ferland97}, assuming an ionising background
from \citep{HaardtMadau96}, determine the fraction of hydrogen in the
form of \HI.

\textsc{specwizard} calculates the optical depth, $\tau$, as a
function of velocity along the sightline. This optical depth can be
converted to the transmission, $T$, using $T=e^{-\tau}$. We convolved
the transmission with an instrumental spread function and added noise
to simulate the normalised \HI\ flux measured in observed spectra. We
use a resolution FWHM of $6.6$~\kms, S/N$=50$ per pixel and pixel size
of 3~\kms, thus the mock spectra have a higher S/N and resolution that
the \textsl{HST} FOS spectra we use in the analysis.  The higher
resolution allows us to fit Voigt profiles to the absorption and
measure the column densities and velocity widths of lines, so that we
can see if there is any change in these properties inside and outside
galaxy-absorber groups.  We describe below how we select a subset of
absorption features that would have been detected in a FOS-quality
spectrum from our higher quality spectra. A wavelength scale was
generated for the spectra assuming the \HI\ absorption is produced
over a small redshift range centred at the redshift of the snapshot.
The continuum level of each spectrum was assumed to be equal to the
largest flux value in the spectrum and calculated after convolution
with the instrumental profile, but before noise is added. This mimics
a continuum fitting process, which would choose the point of highest
flux over a region of a few tens of \AA\ (comparable to the length of
our simulated spectra) and consider that to be the continuum level. At
these low redshifts there is not a large amount of absorption in the
forest, and the difference between the inferred continuum level and
the true level is a few percent or less.

We fit Voigt profiles to these spectra in an automated manner using
\textsc{vpfit}\footnote{\url{http://www.ast.cam.ac.uk/~rfc/vpfit.html}}. The
fitting process generates an initial guess comprising several
absorption lines, then adjusts the model parameters to minimise
$\chi^2$. After minimising $\chi^2$, if the $\chi^2$ per degree of
freedom is greater than 1.1, another absorption component is added at
the point of largest deviation between the model and the data, and
$\chi^2$ is re-minimised using the new model. Absorption components
can also be removed if both their column density and $b$-parameter
drop below threshold values. This process is repeated until a $\chi^2$
per degree of freedom $\le 1.1$ is reached, or \textsc{vpfit} iterates
over more than thirty add/remove cycles. A large fraction of the
fitted models were visually inspected and were found to fit the data
adequately.

To compare to the linelists generated using the FOS spectra, we select
only systems with a column density larger than $10^{13.5}$ \cmm; this
roughly corresponds to the $3 \sigma$ detection limit of the FOS
spectra across the three sightlines (assuming the absorption falls on
the linear part of the curve of growth).  This cutoff value has a more
important effect than the galaxy mass cutoff on the number of groups
found, as we discuss below. 

\gimic\ was designed to simulate the transport of metals into the IGM,
and a comparison between the properties of simulated metal lines and
observations will be made in a future paper.  However, since we
believe we have removed the majority of the metal lines from our
observed line list, we have not included metal transitions in the
simulated spectra.  If our observed line lists contain significant
numbers of features that are blended with metal lines, the simulations
would appear to have fewer lines or lines with smaller equivalent
widths than the observations.

The fraction of H in the form of \HI\ (and so the amount of \HI\
absorption) is strongly dependent on the ionising background for the
highly ionised IGM. The background at low redshifts is not well known,
so we check the simulated spectra are consistent with observations by
measuring $dN/dz$ for \HI\ absorbers with $13.2<$~log(\NHI)~$<14.0$.
\citet{Lehner07} give $dN/dz$ at $z<1$ found by fitting Voigt profiles
to STIS spectra over the same column density range, with a resolution
comparable to that used in our simulated spectra. We find $dN/dz$
using all our mock spectra, combining the estimates for each simulated
region using volume weights of $1/12$, $1/6$, $1/2$, $1/6$, $1/12$ for
the (-2, -1, 0, +1, +2)$\sigma$ regions
\citep[see][]{Crain_gimic_astroph_09}. The mock spectra under-predict
the line density in this range at $z\sim 0$ by a factor of $\sim2$. We
corrected for this by multiplying the optical depth used to generate
the mock spectra by 1.2 \citep[see e.g.][for a discussion on the
validity of this process at low redshifts]{Dave99}, then generating
new spectra and fitting them with \textsc{vpfit}.

From the fitted models we selected lines that satisfied the following
criteria: $b$ parameter $< 100$~\kms, errors in $b < 40$\%, and error
in log(\NHI(\cmm)) $< 0.3$.  We exclude lines with large $b$
parameters because in real data such features can be introduced (or
divided out) in the continuum fitting process, and so are generally
excluded from line analyses. Requiring a minimum error in the $b$
parameter and log(\NHI) values ensures that any poorly constrained
lines, that are usually heavily blended with stronger components, are
not included in our line lists. Finally, we excluded lines that occur
close to the edge of the high-resolution region in the
simulation. Recall that each \gimic\ simulation consists of a smaller,
high-resolution region where dark matter, baryons and gas physics are
simulated, surrounded by a larger, low resolution region where only
dark matter is simulated. At the division between the pure dark matter
and high resolution regions, gas physics will not be simulated
correctly. Excluding lines close to the high resolution region's edge
ensures that lines affected by this incorrect gas physics are not
included in our simulated line list. This final requirement restricted
us to a $\sim20 \times 20 \times 20$~$h^{-1}$ comoving Mpc cube in the
centre of each simulation.

\subsection{Simulated galaxy-absorber groups}
\label{sec:simul-galaxy-absorb}

Using the galaxies and absorbers from the simulation, selected as
described in the previous sections, we identify galaxy-absorber groups
in the same way as we do for the real observations. We convert the
size of our $8 \times 10$ arcmin field to an area in comoving Mpc at
each snapshot redshift (0.25, 0.5 and 1.0), and allow any galaxy
within this area, centred on each simulated set of triple sightlines,
to be a group member.

To check our velocity cutoffs for associating galaxies and absorbers,
we measure the number of galaxy-absorber pairs as a function of the
velocity offsets between galaxies and absorber. The number of pairs
drops significantly at separations larger than 500 \kms\ in all
simulations except for the $+2\sigma$ region, where there are still a
significant number of pairs out to 1000 \kms. Thus the simulations
suggest a maximum velocity cut of 1000 \kms\ is appropriate for
identifying associated galaxy-absorber pairs.

We show the number of triple LOS-galaxy groups per unit redshift at
the three snapshot redshifts in Figure~\ref{fig:dgdz}. The values for
each region are shown, and the combined value using volume weights
given in Section~\ref{sec:simul-absorb-prop} is also shown. The
uncertainty on the value for each region is taken to be the Poisson
error in the number of groups detected in sightlines through that
simulated region. These are combined in quadrature with the
appropriate weights to give the uncertainty in the combined value.

There are very few groups found in the low density volumes, and the
vast majority of the groups are detected in the $+2\sigma$ region,
even when volume weighting is taken into account. This is potentially
a cause for concern -- the $+2\sigma$ region was arbitrarily selected
to contain a galaxy cluster at $z=0$, and so may not be truly
representative of similar density regions without such a
cluster. However, as long as the influence of the cluster is
restricted to a relatively small percentage of the total volume
analysed, it should not have a large effect on the cluster properties.

The frequency of these groups decreases with decreasing redshift. This
is most likely due to the reduction in the volume fraction of gas with
\NHI\ $> 13.5$ with decreasing redshift, visible in
Figure~\ref{fig:sim_nhi}. Since the group finding algorithm requires
at least three absorbers but only one galaxy per group, we expect it
is much more sensitive to the density of absorption lines than the
galaxy density. Changing the mass cuts and column density cuts used to
select galaxies and absorbers confirms this expectation. Changing the
mass cut has very little effect on the frequency of groups, but
changing the column density cut from $10^{13.5}$ \cmm\ to $10^{13.7}$
\cmm\ changes the frequency by a factor of five.

Using our initial column density cutoff of $10^{13.5}$ \cmm\ and
completeness of 20\% resulted in too few groups being found at $z=0.5$
in the simulations compared to the observations, by a factor of
$\sim10$. However, if we increased the assumed completeness to 40\%
and decreased the column density cutoff to $10^{13.2}$ \cmm, we found
the number of simulated and observed groups was consistent (these
limits are used to generate Figure~\ref{fig:dgdz}). These values are
at the extreme ranges of the expected galaxy completeness and column
density sensitivities, but are not unreasonable.

One possible effect that could cause the simulations to have lower
density of lines, and thus a lower frequency of groups, than the
observed values is due to line blending.  Close groups of weak
features in the mock spectra that fall below our \NHI\ cutoff may
still form a feature with an inferred \NHI\ above the cutoff after
they are convolved with the instrumental spread function of the FOS
spectrum. In this case we would underestimate the line density in the
simulations compared to the observations. To check the magnitude of
this effect, we created a set of mock spectra at a similar resolution
and S/N to the FOS spectra. We detected features and fit Gaussian
profiles to them in a similar manner to that used to detect lines in
the real FOS spectra. Finally we selected groups of galaxies and
triple absorbers using these new mock spectra. We found that the
number density of groups found using the FOS-resolution mock spectra
were consistent within the 1$\sigma$ errors with the number densities
calculated above.  Therefore we conclude that line blending does not
have a significant effect on our estimate of the frequency of groups.

We are currently obtaining more galaxy redshifts in this field, and
have been awarded time on COS to observe the three QSOs at high
resolution; it will be interesting to see if this tension between
observations and simulations persists with the new, larger data sets.

\subsection{Properties of absorbers and galaxies inside and 
outside groups}

We examined the properties of galaxies and absorbers inside the triple
LOS-galaxy groups compared to galaxies outside the groups. For each
combination of redshift and region density, we looked at the $b$
parameter and column density of absorbers inside groups compared to
those outside groups, and the stellar and total halo masses of
galaxies inside and outside groups. There are too few groups found in
the $-1\sigma$ and $-2\sigma$ density regions to make a useful
comparison, however we find no evidence for any difference in these
properties inside and outside groups in the $0\sigma$, $+1\sigma$ and
$+2\sigma$ regions. This suggests the algorithm for selecting triple
LOS-galaxy groups does not preferentially select a single type of
absorber or galaxy environment.

Figure~\ref{fig:sim_nhi_trip} shows the position of our
randomly-placed triple sightlines in the $+2\sigma$ simulation.  The
sightlines shown by darker dots contain triple LOS-galaxy groups. This
is a projection of the density along the 30~Mpc simulated region, so
we must take care interpreting apparent 2-d filaments as true 3-d
structures.  Nevertheless, at $z=1$, these groups appear to fall both
inside and outside filamentary structures. At $z=0.5$ and $z=0.25$, as
the comoving separation of the triplets and the characteristic column
density of filamentary structures drops, the groups tend to be found
in knots and filaments. We see multiple galaxies over a small redshift
range in the observed triple LOS-galaxy groups($z=0.501$ and
$z=0.535$), which is consistent with them arising such structures.

\section{Discussion}

The simulations in Figure~\ref{fig:sim_nhi} suggest two things: (1)
the filaments visible over this 2-d projection maintain their geometry
in comoving space through redshift. Since the comoving separation
between the sightlines changes with redshift, the triple samples large
scale structure on different characteristic scales at different
redshifts. (2) The column density that traces the filamentary
structures falls with decreasing redshift.  At $z=1$, filamentary
structures are traced by gas with log\NHI\ of 14.3, where as at $z=0$
it is traced by gas with log\NHI\ of about 13.5, and higher column
densities now traced more collapsed, spherical structures. This
implies that triple absorbers identified above a single minimum
detectable equivalent width (which can be converted to a column
density) will trace different kinds of structures at different
redshifts -- filaments and even `void' regions at higher redshifts,
and groups of halos at low redshift. Thus triples will tend to
correspond to different kinds environments at different redshifts,
depending on the combination of comoving separation and the column
density sensitivity.

The excess of groups with absorbers in one or more sightlines compared
to a distribution of random absorbers tells us the same thing as the
nearest-neighbour test -- that \lya\ absorbers are correlated with
galaxies on these scales. The significance of this excess appears to
decrease as we accept absorbers with larger velocity separations from
a galaxy into a group. This suggests that single absorbers are only
correlated with any single galaxy on scales of $\lesssim
500$~\kms. This this is consistent with the results of
\citet{Morris06}.

As we move to groups with absorbers in two or three sightlines, the
number of groups becomes smaller. Thus the statistics are more
uncertain and our interpretation is more tentative. However, we see a
clear excess of galaxies with nearby absorbers in two and three
sightlines at all velocity differences compared to a random absorber
distribution. Interestingly, the excess of triple LOS-galaxy groups is
very significant ($>99$\%) for velocity separations of $750$ and
$1000$~\kms. This suggests that the observed triple LOS-galaxy groups
trace Mpc-scale structures that appear in both \HI\ gas and
galaxies. We emphasize that while a $1000$~\kms\ separation within a
single real structure appears unrealistically large if we interpret it
as entirely due to the Hubble flow (a Mpc corresponds to $\sim
70$~\kms\ at $z=0.5$), if some of the absorbers arise in winds of the
magnitude described in the introduction, then such large velocity
differences can still be present in a single structure less than a
couple of Mpc across.

The galaxies and absorbers that make up the seven triple LOS-galaxy
groups identified in the real absorbers are shown in the
Figure~\ref{fig:galspec} $\langle z\rangle = 0.501$ and $\langle
z\rangle = 0.535$ panels. For both structures, there is no obvious
link between any one of the galaxies and its associated absorber, in
the sense that none of the absorbers are close enough to the galaxies
to be caused by a `halo' of gas associated with that galaxy. A more
appealing explanation is that they are caused by a group of galaxies
at a knot or filament of the cosmic web and its associated
absorption. This absorption may be due to large diffuse IGM gas that
spans the three sightlines, or more compact gas clouds closely
associated with individual galaxies in the structure that have not
been targeted in our spectroscopy.

In future papers we intend to present GMOS, VIMOS and DEIMOS data
providing a much larger sample of galaxy redshifts in this field with
a better-known selection function, along with \textsl{HST} Cosmic
Origins Spectrograph (COS) spectra of the three QSOs. The COS spectra
will resolve \HI\ \lya\ forest lines, detect \OVI\ and other metal
line absorbers, and allow us to measure the temperature, column
density and ionisation state of the IGM.

\section{Summary}

We have analysed the distribution of galaxies and the intergalactic
medium using three closely spaced sightlines towards background
\zem~$\sim1$ QSOs.

1. We identify metal line systems in two sightlines nearby a single
bright ($\sim$~L*) galaxy. The galaxy has an impact parameter of
$\sim170$~kpc from both sightlines and the absorbers are within
200~\kms\ of the galaxy redshift. Using the galaxy's spectrum we
estimate the star formation rate and metallicity from nebular emission
lines and find a SFR $>0.45$~\msunyr\ and (O/H) metallicity
$\gtrsim0.5$ solar.

2. We have identified a new probable sub-damped \lya\ system at
$z=0.557$ that shows many strong associated low-ionisation metal
lines. In previous analyses that showed evidence for triple \lya\
absorber coincidences across the three sightlines, several of these
lines were assumed to be \lya. We show that in a re-analysis using the
corrected line identifications evidence for triple coincidences
remains, though at a reduced significance.  Using symmetric triplet
matching, there is an excess of real symmetric triples compared to
random sets of absorber at the $97$\% confidence level, or $99$\% if a
maximum velocity difference between triplet members is imposed.  This
is still a significant result, but less so than previous results which
showed a significance $>99.99$\%.

3. We show that the symmetric matching method used to detect triple
coincidences across three sightlines in Petry et al can fail to detect
configurations of absorbers that are consistent with a physical
structure spanning the three sightlines. We present a new matching
method that does detect such configurations.  The significance of
detected triple coincidences is reduced using this new method (94\% vs
97\% confidence level using symmetric matching). We speculate that
symmetric matching may be more effective at differentiating between
triple coincidences caused by large absorbing structures and those
caused by random alignments.

4. For the six \lya\ absorbers in QSO sightlines A and B that also
show metal absorption, there is a \lya\ absorber at a similar redshift
in the adjacent sightline. The probability of this occurring by chance
is $\sim 3.5$\%. This is consistent with these metal systems being
associated with gaseous structures seen in \HI\ that span the two
sightlines (300-400~kpc). Most absorbers with strong metal systems are
not expected to arise in such large structures, but to be closely
associated with a galaxy halo. We suggest that these metal absorbers
may trace the halos of galaxies that are themselves embedded in a
$\sim0.5$~Mpc absorbing \HI\ structure.

5. We measured the velocity separations between galaxies and their
nearest-neighbour absorber in velocity space for all three
sightlines. For each sightline we find a significant excess ($>99$\%)
of pairs with velocity separations $< 200$~\kms\ compared to a random
distribution of absorbers.

6. We describe a new method for measuring associations between
galaxies and absorbers across multiple close sightlines. We compare
the number of galaxies with associated absorbers in one or more, two
or more, and three sightlines found in the real sample to the number
found in random sets of absorbers. In each case there is an excess of
groups with real absorbers compared to the random absorber sets. We
use this method to identify two apparent large-scale ($\sim1-2$~Mpc)
structures at $z=0.501$ and $z=0.535$ that show \HI\ absorption across
all three sightlines and are each associated with several galaxies.

7. We generate mock spectra using simulations, and examine the
properties of galaxy-absorber groups that are similar to those found
in the real observations. The number density of such groups in the
simulations is consistent with the observations, taking into account
the small sample size and the poorly-defined galaxy selection
function. We test the galaxy mass (stellar and total halo mass),
absorber $b$ parameter and column density for galaxies and absorbers
inside and outside triple LOS-galaxy groups seen in the
simulations. We find no significant difference in these properties
inside and outside such groups. The simulations suggest that groups
consisting of a galaxy and \HI\ absorption in three nearby sightlines
trace the filamentary large-scale structure at redshifts $z<0.5$.

Finally, we revisit the questions we posed in the introduction: our
sample confirms that \HI\ absorbers with a rest EW limit of $\sim
0.24$~\AA\ are more likely to be found close to galaxies, consistent
with the results of other surveys \citep{Lanzetta95, Penton02,
  Morris06}. We have found a galaxy that appears linked to metal
absorption in two nearby sightlines. It is possible that the gas was
enriched and ejected by the galaxy more than 2 Gyr ago.  We also find
evidence that galaxies are associated with absorption spanning the
three sightlines.

The other questions -- constraining the geometry of the absorbing gas
and whether gas is preferentially associated with groups of galaxies
-- can be better addressed with improved spectra and a larger sample
of galaxies, and so we defer their discussion to future papers.

\section{Acknowledgements}

We thank the referee for their helpful comments and a careful reading
of the paper.  NC thanks the Science and Technology Facilities Council
for the rolling grant that funds his position. We thank Romeel Dav\'e,
Ben Oppenheimer, Joop Schaye, Adrian Jenkins, Sabrina Pakzad and Greg
Davies for useful conversations and correspondence.

Simulations were carried out using the HPCx facility at the Edinburgh
Parallel Computing Centre (EPCC) as part of the EC's DEISA `Extreme
Computing Initiative', and with the Cosmology Machine at the Institute
for Computational Cosmology at Durham University.

Many programs used for this analysis were written in
Python\footnote{\url{http://www.python.org}} using the Numpy and Scipy
packages\footnote{\url{http://www.scipy.org}}. Figures were produced
using
Matplotlib\footnote{\url{http://www.matplotlib.sourceforge.net}}. The
programs used are available from NC on request, or can be downloaded
from an online code
repository\footnote{\url{http://bitbucket.org/nhmc/pyserpens}}.

Our analysis used observations made with the NASA/ESA Hubble Space
Telescope, obtained from the Hubble Legacy Archive, which is a
collaboration between the Space Telescope Science Institute
(STScI/NASA), the Space Telescope European Coordinating Facility
(ST-ECF/ESA) and the Canadian Astronomy Data Centre (CADC/NRC/CSA).

\bibliographystyle{mn2e}

\begin{thebibliography}{}

\bibitem[\protect\citeauthoryear{{Adelberger}, {Shapley}, {Steidel}, {Pettini},
  {Erb} \& {Reddy}}{{Adelberger} et~al.}{2005}]{Adelberger05}
{Adelberger} K.~L.,  {Shapley} A.~E.,  {Steidel} C.~C.,  {Pettini} M.,  {Erb}
  D.~K.,    {Reddy} N.~A.,  2005, \apj, 629, 636

\bibitem[\protect\citeauthoryear{{Asplund}, {Grevesse}, {Sauval}, {Allende
  Prieto} \& {Kiselman}}{{Asplund} et~al.}{2004}]{Asplund04}
{Asplund} M.,  {Grevesse} N.,  {Sauval} A.~J.,  {Allende Prieto} C.,
  {Kiselman} D.,  2004, \aap, 417, 751

\bibitem[\protect\citeauthoryear{{Bahcall}, {Jannuzi}, {Schneider}, {Hartig},
  {Bohlin} \& {Junkkarinen}}{{Bahcall} et~al.}{1991}]{Bahcall91}
{Bahcall} J.~N.,  {Jannuzi} B.~T.,  {Schneider} D.~P.,  {Hartig} G.~F.,
  {Bohlin} R.,    {Junkkarinen} V.,  1991, \apjl, 377, L5+

\bibitem[\protect\citeauthoryear{Bahcall \& Spitzer}{Bahcall \&
  Spitzer}{1969}]{bahcall_absorption_1969}
Bahcall J.~N.,  Spitzer L.,  1969, Astrophysical Journal, 156, L63

\bibitem[\protect\citeauthoryear{{Balogh}, {Morris}, {Yee}, {Carlberg} \&
  {Ellingson}}{{Balogh} et~al.}{1999}]{Balogh99}
{Balogh} M.~L.,  {Morris} S.~L.,  {Yee} H.~K.~C.,  {Carlberg} R.~G.,
  {Ellingson} E.,  1999, \apj, 527, 54

\bibitem[\protect\citeauthoryear{{Bechtold}, {Crotts}, {Duncan} \&
  {Fang}}{{Bechtold} et~al.}{1994}]{Bechtold94a}
{Bechtold} J.,  {Crotts} A.~P.~S.,  {Duncan} R.~C.,    {Fang} Y.,  1994, \apjl,
  437, L83

\bibitem[\protect\citeauthoryear{{Bergeron} \& {Boisse}}{{Bergeron} \&
  {Boisse}}{1991}]{Bergeron91}
{Bergeron} J.,  {Boisse} P.,  1991, \aap, 243, 344

\bibitem[\protect\citeauthoryear{{Bertin} \& {Arnouts}}{{Bertin} \&
  {Arnouts}}{1996}]{Bertin96}
{Bertin} E.,  {Arnouts} S.,  1996, \aaps, 117, 393

\bibitem[\protect\citeauthoryear{Boksenberg \& Sargent}{Boksenberg \&
  Sargent}{1978}]{boksenberg_existence_1978}
Boksenberg A.,  Sargent W. L.~W.,  1978, Astrophysical Journal, 220, 42

\bibitem[\protect\citeauthoryear{{Bruzual A.}}{{Bruzual A.}}{1983}]{Bruzual83}
{Bruzual A.} G.,  1983, \apj, 273, 105

\bibitem[\protect\citeauthoryear{{Chen}, {Kennicutt} Jr. \& {Rauch}}{{Chen}
  et~al.}{2005}]{Chen05}
{Chen} H.-W.,  {Kennicutt} Jr. R.~C.,    {Rauch} M.,  2005, \apj, 620, 703

\bibitem[\protect\citeauthoryear{{Chen}, {Lanzetta} \& {Webb}}{{Chen}
  et~al.}{2001}]{Chen01}
{Chen} H.-W.,  {Lanzetta} K.~M.,    {Webb} J.~K.,  2001, \apj, 556, 158

\bibitem[\protect\citeauthoryear{{Chen}, {Lanzetta}, {Webb} \&
  {Barcons}}{{Chen} et~al.}{1998}]{Chen98}
{Chen} H.-W.,  {Lanzetta} K.~M.,  {Webb} J.~K.,    {Barcons} X.,  1998, \apj,
  498, 77

\bibitem[\protect\citeauthoryear{{Chen}, {Lanzetta}, {Webb} \&
  {Barcons}}{{Chen} et~al.}{2001}]{Chen01b}
{Chen} H.-W.,  {Lanzetta} K.~M.,  {Webb} J.~K.,    {Barcons} X.,  2001, \apj,
  559, 654

\bibitem[\protect\citeauthoryear{{Chen}, {Prochaska}, {Weiner}, {Mulchaey} \&
  {Williger}}{{Chen} et~al.}{2005}]{Chen05corr}
{Chen} H.-W.,  {Prochaska} J.~X.,  {Weiner} B.~J.,  {Mulchaey} J.~S.,
  {Williger} G.~M.,  2005, \apjl, 629, L25

\bibitem[\protect\citeauthoryear{{Churchill}, {Mellon}, {Charlton}, {Jannuzi},
  {Kirhakos}, {Steidel} \& {Schneider}}{{Churchill}
  et~al.}{2000}]{Churchill00b}
{Churchill} C.~W.,  {Mellon} R.~R.,  {Charlton} J.~C.,  {Jannuzi} B.~T.,
  {Kirhakos} S.,  {Steidel} C.~C.,    {Schneider} D.~P.,  2000, \apj, 543, 577

\bibitem[\protect\citeauthoryear{{Colberg}, {Krughoff} \& {Connolly}}{{Colberg}
  et~al.}{2005}]{Colberg05}
{Colberg} J.~M.,  {Krughoff} K.~S.,    {Connolly} A.~J.,  2005, \mnras, 359,
  272

\bibitem[\protect\citeauthoryear{{Crain}, {Theuns}, {Dalla Vecchia}, {Eke},
  {Frenk}, {Jenkins}, {Kay}, {Peacock}, {Pearce}, {Schaye}, {Springel},
  {Thomas}, {White} \& {Wiersma}}{{Crain}
  et~al.}{2009}]{Crain_gimic_astroph_09}
{Crain} R.~A.,  {Theuns} T.,  {Dalla Vecchia} C.,  {Eke} V.~R.,  {Frenk} C.~S.,
   {Jenkins} A.,  {Kay} S.~T.,  {Peacock} J.~A.,  {Pearce} F.~R.,  {Schaye} J.,
   {Springel} V.,  {Thomas} P.~A.,  {White} S.~D.~M.,    {Wiersma} R.~P.~C.,
  2009, ArXiv e-prints

\bibitem[\protect\citeauthoryear{{Dalla Vecchia} \& {Schaye}}{{Dalla Vecchia}
  \& {Schaye}}{2008}]{Dalla_Vecchia_and_Schaye_08}
{Dalla Vecchia} C.,  {Schaye} J.,  2008, \mnras, 387, 1431

\bibitem[\protect\citeauthoryear{{Dav{\'e}}, {Hernquist}, {Katz} \&
  {Weinberg}}{{Dav{\'e}} et~al.}{1999}]{Dave99}
{Dav{\'e}} R.,  {Hernquist} L.,  {Katz} N.,    {Weinberg} D.~H.,  1999, \apj,
  511, 521

\bibitem[\protect\citeauthoryear{{Dinshaw}, {Foltz}, {Impey}, {Weymann} \&
  {Morris}}{{Dinshaw} et~al.}{1995}]{Dinshaw95}
{Dinshaw} N.,  {Foltz} C.~B.,  {Impey} C.~D.,  {Weymann} R.~J.,    {Morris}
  S.~L.,  1995, \nat, 373, 223

\bibitem[\protect\citeauthoryear{{Dinshaw}, {Weymann}, {Impey}, {Foltz},
  {Morris} \& {Ake}}{{Dinshaw} et~al.}{1997}]{Dinshaw97}
{Dinshaw} N.,  {Weymann} R.~J.,  {Impey} C.~D.,  {Foltz} C.~B.,  {Morris}
  S.~L.,    {Ake} T.,  1997, \apj, 491, 45

\bibitem[\protect\citeauthoryear{{Dolag}, {Borgani}, {Murante} \&
  {Springel}}{{Dolag} et~al.}{2008}]{Dolag_et_al_08}
{Dolag} K.,  {Borgani} S.,  {Murante} G.,    {Springel} V.,  2008, ArXiv
  e-prints, 808

\bibitem[\protect\citeauthoryear{{Eke}, {Baugh}, {Cole}, {Frenk}, {Norberg},
  {Peacock}, {Baldry}, {Bland-Hawthorn}, {Bridges}, {Cannon}, {Colless},
  {Collins}, {Couch}, {Dalton}, {de Propris}, {Driver}, {Efstathiou}, {Ellis},
  {Glazebrook}, {Jackson}, {Lahav}}{{Eke} et~al.}{2004}]{Eke04}
{Eke} V.~R.,  {Baugh} C.~M.,  {Cole} S.,  {Frenk} C.~S.,  {Norberg} P.,
  {Peacock} J.~A.,  {Baldry} I.~K.,  {Bland-Hawthorn} J.,  {Bridges} T.,
  {Cannon} R.,  {Colless} M.,  {Collins} C.,  {Couch} W.,  {Dalton} G.,  {de
  Propris} R.,  {Driver} S.~P.,  {Efstathiou} G.,  {Ellis} R.~S.,  {Glazebrook}
  K.,  {Jackson} C.,  {Lahav} O.,  {Lewis} I.,  {Lumsden} S.,  {Maddox} S.,
  {Madgwick} D.,  {Peterson} B.~A.,  {Sutherland} W.,    {Taylor} K.,  2004,
  \mnras, 348, 866

\bibitem[\protect\citeauthoryear{{Faber}, {Willmer}, {Wolf}, {Koo}, {Weiner},
  {Newman}, {Im}, {Coil}, {Conroy}, {Cooper}, {Davis}, {Finkbeiner}, {Gerke},
  {Gebhardt}, {Groth}, {Guhathakurta}, {Harker}, {Kaiser}, {Kassin},
  {Kleinheinrich}, {Konidaris}}{{Faber} et~al.}{2007}]{Faber07}
{Faber} S.~M.,  {Willmer} C.~N.~A.,  {Wolf} C.,  {Koo} D.~C.,  {Weiner} B.~J.,
  {Newman} J.~A.,  {Im} M.,  {Coil} A.~L.,  {Conroy} C.,  {Cooper} M.~C.,
  {Davis} M.,  {Finkbeiner} D.~P.,  {Gerke} B.~F.,  {Gebhardt} K.,  {Groth}
  E.~J.,  {Guhathakurta} P.,  {Harker} J.,  {Kaiser} N.,  {Kassin} S.,
  {Kleinheinrich} M.,  {Konidaris} N.~P.,  {Kron} R.~G.,  {Lin} L.,  {Luppino}
  G.,  {Madgwick} D.~S.,  {Meisenheimer} K.,  {Noeske} K.~G.,  {Phillips}
  A.~C.,  {Sarajedini} V.~L.,  {Schiavon} R.~P.,  {Simard} L.,  {Szalay} A.~S.,
   {Vogt} N.~P.,    {Yan} R.,  2007, \apj, 665, 265


\bibitem[\protect\citeauthoryear{{Fadda}, {Girardi}, {Giuricin}, {Mardirossian}
  \& {Mezzetti}}{{Fadda} et~al.}{1996}]{Fadda96}
{Fadda} D.,  {Girardi} M.,  {Giuricin} G.,  {Mardirossian} F.,    {Mezzetti}
  M.,  1996, \apj, 473, 670

\bibitem[\protect\citeauthoryear{{Ferland}}{{Ferland}}{1997}]{Ferland97}
{Ferland} G.,  1997, Hazy, a Brief Introduction to Cloudy.
University of Kentucky, Department of Physics and Astronomy Internal Report

\bibitem[\protect\citeauthoryear{{Foltz}, {Chaffee} Jr., {Hewett}, {MacAlpine},
  {Turnshek}, {Weymann} \& {Anderson}}{{Foltz} et~al.}{1987}]{Foltz87}
{Foltz} C.~B.,  {Chaffee} Jr. F.~H.,  {Hewett} P.~C.,  {MacAlpine} G.~M.,
  {Turnshek} D.~A.,  {Weymann} R.~J.,    {Anderson} S.~F.,  1987, \aj, 94, 1423

\bibitem[\protect\citeauthoryear{{Franx}, {Illingworth} \& {de Zeeuw}}{{Franx}
  et~al.}{1991}]{Franx91}
{Franx} M.,  {Illingworth} G.,    {de Zeeuw} T.,  1991, \apj, 383, 112

\bibitem[\protect\citeauthoryear{{Frenk}, {Evrard}, {White} \&
  {Summers}}{{Frenk} et~al.}{1996}]{Frenk_et_al_96}
{Frenk} C.~S.,  {Evrard} A.~E.,  {White} S.~D.~M.,    {Summers} F.~J.,  1996,
  \apj, 472, 460

\bibitem[\protect\citeauthoryear{{Gharanfoli}, {Kulkarni}, {Chun} \&
  {Takamiya}}{{Gharanfoli} et~al.}{2007}]{Gharanfoli07}
{Gharanfoli} S.,  {Kulkarni} V.~P.,  {Chun} M.~R.,    {Takamiya} M.,  2007,
  \aj, 133, 130

\bibitem[\protect\citeauthoryear{{Haardt} \& {Madau}}{{Haardt} \&
  {Madau}}{1996}]{HaardtMadau96}
{Haardt} F.,  {Madau} P.,  1996, \apj, 461, 20

\bibitem[\protect\citeauthoryear{Hamann}{Hamann}{1998}]{hamann_broad_1998}
Hamann F.,  1998, Astrophysical Journal, 500, 798

\bibitem[\protect\citeauthoryear{{Heckman}, {Armus} \& {Miley}}{{Heckman}
  et~al.}{1990}]{Heckman90}
{Heckman} T.~M.,  {Armus} L.,    {Miley} G.~K.,  1990, \apjs, 74, 833

\bibitem[\protect\citeauthoryear{{Jenkins}, {Bowen}, {Tripp} \&
  {Sembach}}{{Jenkins} et~al.}{2005}]{Jenkins05}
{Jenkins} E.~B.,  {Bowen} D.~V.,  {Tripp} T.~M.,    {Sembach} K.~R.,  2005,
  \apj, 623, 767

\bibitem[\protect\citeauthoryear{{Kacprzak}, {Churchill}, {Steidel} \&
  {Murphy}}{{Kacprzak} et~al.}{2008}]{Kacprzak08}
{Kacprzak} G.~G.,  {Churchill} C.~W.,  {Steidel} C.~C.,    {Murphy} M.~T.,
  2008, \aj, 135, 922

\bibitem[\protect\citeauthoryear{{Kauffmann}, {Heckman}, {White},
  {Charlot}, {Tremonti}, {Brinchmann}, {Bruzual}, {Peng}, {Seibert},
  {Bernardi}, {Blanton}, {Brinkmann}, {Castander}, {Cs{\'a}bai},
  {Fukugita}, {Ivezic}, {Munn}, {Nichol}, {Padmanabhan},
  {Thakar}}{{Kauffmann} et~al.}{2003}]{Kauffmann03} {Kauffmann} G.,
  {Heckman} T.~M., {White} S.~D.~M., {Charlot} S., {Tremonti} C.,
  {Brinchmann} J., {Bruzual} G., {Peng} E.~W., {Seibert} M.,
  {Bernardi} M., {Blanton} M., {Brinkmann} J., {Castander} F.,
  {Cs{\'a}bai} I., {Fukugita} M., {Ivezic} Z., {Munn} J.~A., {Nichol}
  R.~C., {Padmanabhan} N., {Thakar} A.~R., {Weinberg} D.~H., {York}
  D., 2003, \mnras, 341, 33


\bibitem[\protect\citeauthoryear{{Kennicutt}
  Jr.}{{Kennicutt}}{1998}]{Kennicutt98}
{Kennicutt} Jr. R.~C.,  1998, \araa, 36, 189

\bibitem[\protect\citeauthoryear{{Kewley} \& {Dopita}}{{Kewley} \&
  {Dopita}}{2002}]{Kewley02b}
{Kewley} L.~J.,  {Dopita} M.~A.,  2002, \apjs, 142, 35

\bibitem[\protect\citeauthoryear{{Kewley}, {Geller}, {Jansen} \&
  {Dopita}}{{Kewley} et~al.}{2002}]{Kewley02}
{Kewley} L.~J.,  {Geller} M.~J.,  {Jansen} R.~A.,    {Dopita} M.~A.,  2002,
  \aj, 124, 3135

\bibitem[\protect\citeauthoryear{{Kroupa}}{{Kroupa}}{2001}]{Kroupa01}
{Kroupa} P.,  2001, \mnras, 322, 231

\bibitem[\protect\citeauthoryear{{Lanzetta}, {Bowen}, {Tytler} \&
  {Webb}}{{Lanzetta} et~al.}{1995}]{Lanzetta95}
{Lanzetta} K.~M.,  {Bowen} D.~V.,  {Tytler} D.,    {Webb} J.~K.,  1995, \apj,
  442, 538

\bibitem[\protect\citeauthoryear{{Le Brun}, {Bergeron} \& {Boisse}}{{Le Brun}
  et~al.}{1996}]{LeBrun96}
{Le Brun} V.,  {Bergeron} J.,    {Boisse} P.,  1996, \aap, 306, 691

\bibitem[\protect\citeauthoryear{{Lehner}, {Savage}, {Richter}, {Sembach},
  {Tripp} \& {Wakker}}{{Lehner} et~al.}{2007}]{Lehner07}
{Lehner} N.,  {Savage} B.~D.,  {Richter} P.,  {Sembach} K.~R.,  {Tripp} T.~M.,
    {Wakker} B.~P.,  2007, \apj, 658, 680

\bibitem[\protect\citeauthoryear{{Morris} \& {Jannuzi}}{{Morris} \&
  {Jannuzi}}{2006}]{Morris06}
{Morris} S.~L.,  {Jannuzi} B.~T.,  2006, \mnras, 367, 1261

\bibitem[\protect\citeauthoryear{{Morris}, {Weymann}, {Dressler}, {McCarthy},
  {Smith}, {Terrile}, {Giovanelli} \& {Irwin}}{{Morris}
  et~al.}{1993}]{Morris93}
{Morris} S.~L.,  {Weymann} R.~J.,  {Dressler} A.,  {McCarthy} P.~J.,  {Smith}
  B.~A.,  {Terrile} R.~J.,  {Giovanelli} R.,    {Irwin} M.,  1993, \apj, 419,
  524

\bibitem[\protect\citeauthoryear{{Morris}, {Weymann}, {Savage} \&
  {Gilliland}}{{Morris} et~al.}{1991}]{Morris91}
{Morris} S.~L.,  {Weymann} R.~J.,  {Savage} B.~D.,    {Gilliland} R.~L.,  1991,
  \apjl, 377, L21

\bibitem[\protect\citeauthoryear{{Navarro} et~al.,}{{Navarro}
  et~al.}{2004}]{Navarro_et_al_04_short}
{Navarro} J.~F.,  et~al., 2004, \mnras, 349, 1039

\bibitem[\protect\citeauthoryear{{Penton}, {Stocke} \& {Shull}}{{Penton}
  et~al.}{2002}]{Penton02}
{Penton} S.~V.,  {Stocke} J.~T.,    {Shull} J.~M.,  2002, \apj, 565, 720

\bibitem[\protect\citeauthoryear{{Petry}, {Impey}, {Fenton} \& {Foltz}}{{Petry}
  et~al.}{2006}]{Petry06}
{Petry} C.~E.,  {Impey} C.~D.,  {Fenton} J.~L.,    {Foltz} C.~B.,  2006, \aj,
  132, 2046

\bibitem[\protect\citeauthoryear{{Power}, {Navarro}, {Jenkins}, {Frenk},
  {White}, {Springel}, {Stadel} \& {Quinn}}{{Power}
  et~al.}{2003}]{Power_et_al_03}
{Power} C.,  {Navarro} J.~F.,  {Jenkins} A.,  {Frenk} C.~S.,  {White} S.~D.~M.,
   {Springel} V.,  {Stadel} J.,    {Quinn} T.,  2003, \mnras, 338, 14

\bibitem[\protect\citeauthoryear{{Prochaska}, {O'Meara}, {Herbert-Fort},
  {Burles}, {Prochter} \& {Bernstein}}{{Prochaska} et~al.}{2006}]{Prochaska06b}
{Prochaska} J.~X.,  {O'Meara} J.~M.,  {Herbert-Fort} S.,  {Burles} S.,
  {Prochter} G.~E.,    {Bernstein} R.~A.,  2006, \apjl, 648, L97

\bibitem[\protect\citeauthoryear{{Ramella}, {Geller} \& {Huchra}}{{Ramella}
  et~al.}{1989}]{Ramella89}
{Ramella} M.,  {Geller} M.~J.,    {Huchra} J.~P.,  1989, \apj, 344, 57

\bibitem[\protect\citeauthoryear{{Rao}, {Nestor}, {Turnshek}, {Lane}, {Monier}
  \& {Bergeron}}{{Rao} et~al.}{2003}]{Rao03}
{Rao} S.~M.,  {Nestor} D.~B.,  {Turnshek} D.~A.,  {Lane} W.~M.,  {Monier}
  E.~M.,    {Bergeron} J.,  2003, \apj, 595, 94

\bibitem[\protect\citeauthoryear{{Rauch} \& {Haehnelt}}{{Rauch} \&
  {Haehnelt}}{1995}]{Rauch95}
{Rauch} M.,  {Haehnelt} M.~G.,  1995, \mnras, 275, L76

\bibitem[\protect\citeauthoryear{{Schaye}}{{Schaye}}{2001}]{Schaye01}
{Schaye} J.,  2001, \apj, 559, 507

\bibitem[\protect\citeauthoryear{{Schaye} \& {Dalla Vecchia}}{{Schaye} \&
  {Dalla Vecchia}}{2008}]{Schaye_and_Dalla_Vecchia_08}
{Schaye} J.,  {Dalla Vecchia} C.,  2008, \mnras, 383, 1210

\bibitem[\protect\citeauthoryear{{Scott}, {Bechtold}, {Morita}, {Dobrzycki} \&
  {Kulkarni}}{{Scott} et~al.}{2002}]{Scott02}
{Scott} J.,  {Bechtold} J.,  {Morita} M.,  {Dobrzycki} A.,    {Kulkarni} V.~P.,
   2002, \apj, 571, 665

\bibitem[\protect\citeauthoryear{{Sheth}, {Bernardi}, {Schechter}, {Burles},
  {Eisenstein}, {Finkbeiner}, {Frieman}, {Lupton}, {Schlegel}, {Subbarao},
  {Shimasaku}, {Bahcall}, {Brinkmann} \& {Ivezi{\'c}}}{{Sheth}
  et~al.}{2003}]{Sheth03}
{Sheth} R.~K.,  {Bernardi} M.,  {Schechter} P.~L.,  {Burles} S.,  {Eisenstein}
  D.~J.,  {Finkbeiner} D.~P.,  {Frieman} J.,  {Lupton} R.~H.,  {Schlegel}
  D.~J.,  {Subbarao} M.,  {Shimasaku} K.,  {Bahcall} N.~A.,  {Brinkmann} J.,
  {Ivezi{\'c}} {\v Z}.,  2003, \apj, 594, 225

\bibitem[\protect\citeauthoryear{{Smette}, {Robertson}, {Shaver}, {Reimers},
  {Wisotzki} \& {Koehler}}{{Smette} et~al.}{1995}]{Smette95}
{Smette} A.,  {Robertson} J.~G.,  {Shaver} P.~A.,  {Reimers} D.,  {Wisotzki}
  L.,    {Koehler} T.,  1995, \aaps, 113, 199

\bibitem[\protect\citeauthoryear{{Springel} et~al.,}{{Springel}
  et~al.}{2005}]{Springel_et_al_05_short}
{Springel} V.,  et~al., 2005, \nat, 435, 629

\bibitem[\protect\citeauthoryear{{Springel}, {White}, {Tormen} \&
  {Kauffmann}}{{Springel} et~al.}{2001}]{Springel01}
{Springel} V.,  {White} S.~D.~M.,  {Tormen} G.,    {Kauffmann} G.,  2001,
  \mnras, 328, 726

\bibitem[\protect\citeauthoryear{{Steidel}, {Dickinson} \& {Persson}}{{Steidel}
  et~al.}{1994}]{Steidel94b}
{Steidel} C.~C.,  {Dickinson} M.,    {Persson} S.~E.,  1994, \apjl, 437, L75

\bibitem[\protect\citeauthoryear{{Stocke}, {Keeney}, {McLin}, {Rosenberg},
  {Weymann} \& {Giroux}}{{Stocke} et~al.}{2004}]{Stocke04}
{Stocke} J.~T.,  {Keeney} B.~A.,  {McLin} K.~M.,  {Rosenberg} J.~L.,  {Weymann}
  R.~J.,    {Giroux} M.~L.,  2004, \apj, 609, 94

\bibitem[\protect\citeauthoryear{{Stocke}, {Shull}, {Penton}, {Donahue} \&
  {Carilli}}{{Stocke} et~al.}{1995}]{Stocke95}
{Stocke} J.~T.,  {Shull} J.~M.,  {Penton} S.,  {Donahue} M.,    {Carilli} C.,
  1995, \apj, 451, 24

\bibitem[\protect\citeauthoryear{{Theuns}, {Leonard} \& {Efstathiou}}{{Theuns}
  et~al.}{1998}]{Theuns_loz_evol_98}
{Theuns} T.,  {Leonard} A.,    {Efstathiou} G.,  1998, \mnras, 297, L49

\bibitem[\protect\citeauthoryear{{Theuns}, {Leonard}, {Efstathiou}, {Pearce} \&
  {Thomas}}{{Theuns} et~al.}{1998}]{Theuns_sph_sim_98}
{Theuns} T.,  {Leonard} A.,  {Efstathiou} G.,  {Pearce} F.~R.,    {Thomas}
  P.~A.,  1998, \mnras, 301, 478

\bibitem[\protect\citeauthoryear{{Theuns}, {Viel}, {Kay}, {Schaye}, {Carswell}
  \& {Tzanavaris}}{{Theuns} et~al.}{2002}]{Theuns02}
{Theuns} T.,  {Viel} M.,  {Kay} S.,  {Schaye} J.,  {Carswell} R.~F.,
  {Tzanavaris} P.,  2002, \apjl, 578, L5

\bibitem[\protect\citeauthoryear{{Tripp}, {Lu} \& {Savage}}{{Tripp}
  et~al.}{1998}]{Tripp98}
{Tripp} T.~M.,  {Lu} L.,    {Savage} B.~D.,  1998, \apj, 508, 200

\bibitem[\protect\citeauthoryear{Wagoner}{Wagoner}{1967}]{wagoner_effects_1967}
Wagoner R.~V.,  1967, Astrophysical Journal, 149, 465

\bibitem[\protect\citeauthoryear{{Weymann}, {Jannuzi}, {Lu}, {Bahcall},
  {Bergeron}, {Boksenberg}, {Hartig}, {Kirhakos}, {Sargent}, {Savage},
  {Schneider}, {Turnshek} \& {Wolfe}}{{Weymann} et~al.}{1998}]{Weymann98}
{Weymann} R.~J.,  {Jannuzi} B.~T.,  {Lu} L.,  {Bahcall} J.~N.,  {Bergeron} J.,
  {Boksenberg} A.,  {Hartig} G.~F.,  {Kirhakos} S.,  {Sargent} W.~L.~W.,
  {Savage} B.~D.,  {Schneider} D.~P.,  {Turnshek} D.~A.,    {Wolfe} A.~M.,
  1998, \apj, 506, 1

\bibitem[\protect\citeauthoryear{{Wiersma}, {Schaye} \& {Smith}}{{Wiersma}
  et~al.}{2009}]{Wiersma_Schaye_and_Smith_09}
{Wiersma} R.~P.~C.,  {Schaye} J.,    {Smith} B.~D.,  2009a, \mnras, pp 20--+

\bibitem[\protect\citeauthoryear{{Wiersma}, {Schaye}, {Theuns}, {Dalla Vecchia}
  \& {Tornatore}}{{Wiersma} et~al.}{2009}]{Wiersma09b}
{Wiersma} R.~P.~C.,  {Schaye} J.,  {Theuns} T.,  {Dalla Vecchia} C.,
  {Tornatore} L.,  2009b, ArXiv e-prints

\bibitem[\protect\citeauthoryear{Wild, Kauffmann, White, York, Lehnert,
  Heckman, Hall, Khare, Lundgren, Schneider \& vanden Berk}{Wild
  et~al.}{2008}]{wild_narrow_2008}
Wild V.,  Kauffmann G.,  White S.,  York D.,  Lehnert M.,  Heckman T.,  Hall
  P.~B.,  Khare P.,  Lundgren B.,  Schneider D.~P.,    vanden Berk D.,  2008,
  Monthly Notices of the Royal Astronomical Society, 388, 227

\bibitem[\protect\citeauthoryear{{Wilman}, {Morris}, {Jannuzi}, {Dav{\'e}} \&
  {Shone}}{{Wilman} et~al.}{2007}]{Wilman07}
{Wilman} R.~J.,  {Morris} S.~L.,  {Jannuzi} B.~T.,  {Dav{\'e}} R.,    {Shone}
  A.~M.,  2007, \mnras, 375, 735

\bibitem[\protect\citeauthoryear{{Worthey} \& {Ottaviani}}{{Worthey} \&
  {Ottaviani}}{1997}]{Worthey97}
{Worthey} G.,  {Ottaviani} D.~L.,  1997, \apjs, 111, 377

\bibitem[\protect\citeauthoryear{{Young}, {Impey} \& {Foltz}}{{Young}
  et~al.}{2001}]{Young01}
{Young} P.~A.,  {Impey} C.~D.,    {Foltz} C.~B.,  2001, \apj, 549, 76

\end{thebibliography}

\clearpage

\begin{table*}
\begin{minipage}{140mm}
\begin{center}
{\large \sc Galaxy Properties}\\
\vspace{0.2cm}
\begin{tabular}{cccccccc}
\hline
ID & R.A.(J2000)& Decl.(J2000) &$z$&$z_{\rm err}$&R-mag&B$_{\rm max}$& B$_{\rm min}$ \\
\hline
1  & 1:10:14.2 & -2:20:10.2 & 0.0528 &      - & 19.3 & -15.7 & -16.6 \\
2  & 1:10:21.0 & -2:21:27.1 & 0.1145 &      - & 16.4 & -20.5 & -21.3 \\
3  & 1:10:23.4 & -2:19:34.8 & 0.1204 &      - & 18.1 & -18.9 & -19.7 \\
4  & 1:09:56.9 & -2:20:52.5 & 0.1550 & 0.0004 & 18.1 & -19.6 & -20.4 \\
5  & 1:10:13.7 & -2:21:38.7 & 0.1817 &      - & 19.0 & -19.1 & -19.9 \\
6  & 1:10:13.1 & -2:21:57.3 & 0.1820 &      - & 19.0 & -19.1 & -19.8 \\
7  & 1:10:00.1 & -2:20:58.2 & 0.1905 &      - & 18.8 & -19.4 & -20.1 \\
8  & 1:10:06.2 & -2:22:54.9 & 0.1905 &      - & 19.5 & -18.7 & -19.5 \\
9  & 1:10:18.4 & -2:18:47.0 & 0.1971 &      - & 19.3 & -18.9 & -19.7 \\
10 & 1:10:24.3 & -2:17:16.3 & 0.1973 &      - & 17.5 & -20.8 & -21.5 \\
11 & 1:10:30.4 & -2:17:58.9 & 0.1995 & 0.0007 & 20.6 & -17.7 & -18.5 \\
12 & 1:10:13.6 & -2:17:58.8 & 0.1996 &      - & 17.6 & -20.7 & -21.4 \\
13 & 1:10:03.1 & -2:22:33.5 & 0.1997 &      - & 20.3 & -18.1 & -18.8 \\
14 & 1:10:06.3 & -2:18:36.1 & 0.1999 &      - & 19.5 & -18.8 & -19.6 \\
15 & 1:10:18.1 & -2:18:16.5 & 0.2009 & 0.0014 & 19.1 & -19.3 & -20.0 \\
16 & 1:10:28.1 & -2:18:52.4 & 0.2009 & 0.0004 & 19.6 & -18.7 & -19.4 \\
17 & 1:10:25.4 & -2:20:02.2 & 0.2016 & 0.0007 & 17.2 & -21.1 & -21.8 \\
18 & 1:10:26.0 & -2:20:01.8 & 0.2020 & 0.0007 & 17.8 & -20.6 & -21.3 \\
19 & 1:10:23.2 & -2:16:49.8 & 0.2025 &      - & 16.8 & -21.5 & -22.2 \\
20 & 1:10:24.5 & -2:19:38.5 & 0.2027 & 0.0004 & 19.4 & -19.0 & -19.7 \\
21 & 1:10:11.0 & -2:17:21.2 & 0.2029 &      - & 19.1 & -19.3 & -20.0 \\
22 & 1:10:21.4 & -2:17:55.4 & 0.2029 &      - & 17.5 & -20.9 & -21.6 \\
23 & 1:10:24.2 & -2:17:02.0 & 0.2029 &      - & 18.3 & -20.1 & -20.8 \\
24 & 1:10:23.2 & -2:17:38.8 & 0.2032 & 0.0005 & 20.3 & -18.1 & -18.8 \\
25 & 1:10:03.8 & -2:23:34.1 & 0.2033 &      - & 19.4 & -19.0 & -19.7 \\
26 & 1:10:16.7 & -2:23:02.6 & 0.2033 &      - & 20.0 & -18.4 & -19.1 \\
27 & 1:10:16.0 & -2:19:36.4 & 0.2272 & 0.0007 & 17.7 & -20.9 & -21.6 \\
28 & 1:10:18.1 & -2:22:45.1 & 0.2318 &      - & 18.0 & -20.7 & -21.4 \\
29 & 1:10:17.2 & -2:23:23.9 & 0.2325 &      - & 19.6 & -19.1 & -19.8 \\
30 & 1:10:02.1 & -2:21:55.9 & 0.2334 & 0.0007 & 20.5 & -18.2 & -18.9 \\
31 & 1:10:09.8 & -2:23:20.6 & 0.2334 &      - & 20.4 & -18.4 & -19.0 \\
32 & 1:10:29.5 & -2:20:04.9 & 0.2349 & 0.0005 & 19.6 & -19.1 & -19.8 \\
33 & 1:10:29.5 & -2:19:55.8 & 0.2366 &      - & 18.7 & -20.1 & -20.8 \\
34 & 1:10:18.1 & -2:21:00.4 & 0.2367 &      - & 19.0 & -19.7 & -20.4 \\
35 & 1:10:04.6 & -2:19:39.9 & 0.2373 &      - & 19.4 & -19.3 & -20.0 \\
36 & 1:10:09.5 & -2:19:27.7 & 0.2380 & 0.0007 & 21.0 & -17.8 & -18.5 \\
37 & 1:10:16.5 & -2:21:37.2 & 0.2381 & 0.0007 & 21.5 & -17.2 & -17.9 \\
38 & 1:10:14.5 & -2:20:56.2 & 0.2382 & 0.0007 & 20.2 & -18.5 & -19.2 \\
39 & 1:10:17.3 & -2:21:00.9 & 0.2388 & 0.0008 & 19.9 & -18.9 & -19.6 \\
40 & 1:10:02.6 & -2:16:12.8 & 0.2421 &      - & 19.7 & -19.1 & -19.7 \\
41 & 1:10:19.0 & -2:19:07.3 & 0.2616 & 0.0007 & 20.2 & -18.8 & -19.4 \\
42 & 1:10:32.2 & -2:17:55.5 & 0.3093 & 0.0007 & 19.5 & -20.1 & -20.6 \\
43 & 1:10:01.8 & -2:16:53.4 & 0.3130 &      - & 19.7 & -19.8 & -20.3 \\
44 & 1:10:20.3 & -2:23:36.6 & 0.3146 &      - & 18.4 & -21.2 & -21.7 \\
45 & 1:09:59.4 & -2:17:56.8 & 0.3512 & 0.0007 & 19.9 & -20.0 & -20.5 \\
46 & 1:10:08.3 & -2:21:45.6 & 0.3524 & 0.0006 & 19.3 & -20.6 & -21.1 \\
47 & 1:10:26.8 & -2:20:51.3 & 0.4279 & 0.0006 & 19.9 & -20.7 & -21.0 \\
48 & 1:10:13.3 & -2:18:20.4 & 0.4283 &      - & 19.9 & -20.7 & -21.0 \\
49 & 1:10:07.0 & -2:17:10.0 & 0.4292 & 0.0007 & 21.0 & -19.5 & -19.8 \\
50 & 1:10:19.5 & -2:20:15.2 & 0.4343 & 0.0005 & 21.4 & -19.2 & -19.5 \\
51 & 1:10:20.8 & -2:21:42.1 & 0.4983 & 0.0003 & 20.3 & -20.7 & -20.9 \\
52 & 1:10:04.6 & -2:17:24.6 & 0.5014 & 0.0007 & 19.3 & -21.8 & -22.0 \\
53 & 1:10:03.0 & -2:17:34.7 & 0.5021 & 0.0007 & 20.1 & -21.1 & -21.2 \\
54 & 1:10:03.7 & -2:18:45.5 & 0.5024 & 0.0007 & 20.8 & -20.4 & -20.5 \\
55 & 1:10:15.4 & -2:18:42.8 & 0.5167 & 0.0007 & 20.6 & -20.7 & -20.8 \\
56 & 1:10:21.9 & -2:19:28.4 & 0.5349 & 0.0007 & 20.3 & -21.1 & -21.2 \\
57 & 1:10:23.8 & -2:19:36.8 & 0.5351 & 0.0007 & 21.8 & -19.6 & -19.7 \\
58 & 1:10:31.3 & -2:20:15.5 & 0.5364 & 0.0010 & 21.0 & -20.4 & -20.5 \\
59 & 1:10:28.7 & -2:21:11.5 & 0.5669 & 0.0005 & 20.9 & -20.7 & -20.8 \\
60 & 1:10:10.8 & -2:18:09.9 & 0.6438 & 0.0007 & 20.6 & -21.3 & -21.4 \\
\hline
\end{tabular}
\caption{\label{tab:gal}Properties for the galaxy sample. The columns
  show, from left to right: galaxy ID number, right ascension and
  declination (with units of hours, minutes, seconds and degrees,
  arcminutes, arcseconds), galaxy redshifts and $1\sigma$ redshift
  errors, apparent R magnitude, and maximum and minimum rest-frame
  absolute B magnitudes. For objects with no redshift error we assume
  an error of 0.0007.}
\end{center}
\end{minipage}
\end{table*}

\begin{table*}
\begin{minipage}{140mm}

\begin{center}
{\large \sc QSO Properties}\\
\vspace{0.2cm}
\begin{tabular}{cccccc}
\hline
ID & Name & R.A.(J2000) & Decl.(J2000) & $z_{\rm em}$ & R-mag  \\
\hline
A & LBQS~0107-025A & 01:10:13:14 & -02:19:52.9 & 0.960 & 18.1 \\
B & LBQS~0107-025B & 01:10:16.25 & -02:18:51.0 & 0.956 & 17.4 \\
C & LBQS~0107-0232 & 01:10:14.43 & -02:16:57.6 & 0.726 & 18.4 \\
\hline
\end{tabular}
\caption{\label{tab:qso}The QSO properties. The columns show, from
  left to right: ID letter, name, right ascension and declination
  (with units of hours, minutes, seconds and degrees, arcminutes and
  arcseconds), emission redshift and R magnitude.}
\end{center}
\end{minipage}
\end{table*}



\begin{table*}
\begin{minipage}{140mm}
\begin{center}
{\large \sc LBQS 0107-0125A GHRS G140L}\\
\vspace{0.2cm}
\begin{tabular}{ccccccc}
\hline
ID & Wavelength(\AA)& EW$_{\rm obs}$ (\AA) & FWHM & S$_\sigma$ &  Identification & $z_{\rm abs}$  \\
\hline
   1 & $1226.53\pm 0.33$ & $ 1.26\pm 0.54$ &$ 1.80\pm 0.97$ &   6.12 & -  &  -  \\ 
   2 & $1258.54\pm 0.19$ & $ 1.09\pm 0.28$ &$ 1.73\pm 0.48$ &  14.30 & -  &  - \\ 
   3 & $1260.30\pm 0.16$ & $ 0.73\pm 0.25$ &$ 1.23\pm 0.35$ &   9.61 & \SiII\ 1260/\lyb & 0.0000/0.2286  \\ 
   4 & $1265.50\pm 0.09$ & $ 0.43\pm 0.11$ &$ 0.80\pm 0.23$ &   5.79 & \lyb & 0.234   \\ 
   5 & $1279.39\pm 0.15$ & $ 0.53\pm 0.14$ &$ 1.22\pm 0.41$ &   6.94 & -  &  - \\ 
   6 & $1301.41\pm 0.13$ & $ 0.32\pm 0.10$ &$ 0.92\pm 0.34$ &   4.87 & \OI\ 1302  &  0.0000 \\ 
   7 & $1310.02\pm 1.78$ & $ 0.67\pm 1.57$ &$ 4.69$         &   9.47 & -  &  - \\ 
   8 & $1334.53\pm 0.15$ & $ 0.68\pm 0.13$ &$ 1.63\pm 0.38$ &   9.53 & \CII\ 1334 & 0.0000   \\ 
   9 & $1355.44\pm 0.18$ & $ 0.94\pm 0.20$ &$ 2.02\pm 0.56$ &  11.88 & -  & - \\ 
  10 & $1375.08\pm 0.27$ & $ 0.36\pm 0.15$ &$ 1.47\pm 0.77$ &   5.38 & -  & - \\ 
  11 & $1393.58\pm 0.12$ & $ 0.34\pm 0.11$ &$ 0.80\pm 0.32$ &   4.11 & \SiIV\ 1393 & 0.0000   \\ 
  12 & $1491.20\pm 0.30$ & $ 1.39\pm 0.42$ &$ 2.26\pm 0.86$ &   8.60 & -  &  -  \\ 
  13 & $1493.50\pm 0.15$ & $ 0.82\pm 0.25$ &$ 1.03\pm 0.37$ &   4.85 & \lya  & 0.2286   \\ 
  14 & $1500.44\pm 0.32$ & $ 1.51\pm 0.51$ &$ 2.45\pm 1.07$ &  10.37 & \lya & 0.234   \\ 
  15 & $1504.52\pm 0.09$ & $ 0.64\pm 0.16$ &$ 0.80\pm 0.24$ &   4.99 & -  &  -  \\ 
\hline
\end{tabular}
\caption{\label{tab:a140} Detected lines in the GHRS G140L combined
  spectrum.  The columns show, from left to right: line ID number,
  wavelength and $1\sigma$ error, observed equivalent width and
  $1\sigma$ error of a fitted Gaussian, FWHM and $1\sigma$ error of
  the fitted Gaussian, detection significance, line identification and
  redshift, if applicable. The wavelengths, observed equivalent widths
  and detection significances are from \citet{Petry06}.}
\end{center}
\end{minipage}
\end{table*}

\begin{table*}
\begin{minipage}{140mm}
\begin{center}
{\large \sc LBQS 0107-0125B GHRS G140L}\\
\vspace{0.2cm}
\begin{tabular}{ccccccc}
\hline
ID & Wavelength(\AA)& EW$_{\rm obs}$ (\AA) & FWHM (\AA) & S$_\sigma$ &  Identification & $z_{\rm abs}$\\
\hline
   1 & $1260.26 \pm 0.17$ & $ 0.74\pm 0.15$ & $ 1.77\pm 0.43$ &  9.11 & \SiII\ 1260& $-0.0001$ \\ 
   2 & $1266.42 \pm 0.22$ & $ 0.66\pm 0.19$ & $ 1.68\pm 0.61$ &  8.37 & \OVI\ 1031 & 0.2273 \\ 
   3 & $1274.16 \pm 0.13$ & $ 0.30\pm 0.10$ & $ 0.86\pm 0.34$ &  4.13 & \OVI\ 1037 & 0.2273 \\ 
   4 & $1280.00 \pm 0.15$ & $ 0.33\pm 0.10$ & $ 1.10\pm 0.41$ &  5.30 & -          & -  \\ 
   5 & $1301.93 \pm 0.11$ & $ 0.60\pm 0.11$ & $ 1.45\pm 0.33$ & 12.52 & \OI\ 1302  & $-0.0002$ \\ 
   6 & $1312.49 \pm 0.15$ & $ 0.26\pm 0.08$ & $ 1.07\pm 0.42$ &  5.27 & Ly-5       & 0.3993  \\
   7 & $1323.44 \pm 0.10$ & $ 0.22\pm 0.06$ & $ 0.76\pm 0.26$ &  4.49 & -          & -  \\ 
   8 & $1329.24 \pm 0.08$ & $ 0.50\pm 0.08$ & $ 1.18\pm 0.22$ &  9.96 & Ly-4       & 0.3993  \\ 
   9 & $1334.66 \pm 0.10$ & $ 0.82\pm 0.09$ & $ 1.86\pm 0.25$ & 16.80 & \CII\ 1334 & 0.0001 \\ 
  10 & $1356.06 \pm 0.12$ & $ 0.51\pm 0.10$ & $ 1.44\pm 0.34$ & 10.03 & -          & -  \\ 
  11 & $1361.26 \pm 0.19$ & $ 1.05\pm 0.17$ & $ 2.79\pm 0.56$ & 19.78 & \lyg       & 0.3993 \\ 
  12 & $1366.95 \pm 0.14$ & $ 0.27\pm 0.08$ & $ 1.09\pm 0.36$ &  5.38 & -          & -  \\ 
  13 & $1381.48 \pm 0.12$ & $ 0.36\pm 0.08$ & $ 1.16\pm 0.30$ &  7.10 & -          & -  \\ 
  14 & $1393.54 \pm 0.19$ & $ 0.38\pm 0.11$ & $ 1.43\pm 0.50$ &  7.03 & \SiIV\ 1393& $-0.0002$ \\ 
  15 & $1403.60 \pm 0.09$ & $ 0.67\pm 0.09$ & $ 1.39\pm 0.24$ & 11.37 & -          & -  \\ 
  16 & $1435.29 \pm 0.10$ & $ 0.42\pm 0.10$ & $ 0.96\pm 0.27$ &  6.11 & \lyb       & 0.3993  \\ 
  17 & $1458.91 \pm 0.07$ & $ 0.54\pm 0.11$ & $ 0.78\pm 0.19$ &  5.91 & \lyg?/\lya?& 0.5012/0.2001 \\ 
  18 & $1462.24 \pm 0.08$ & $ 1.15\pm 0.14$ & $ 1.39\pm 0.20$ & 12.85 & -          & -  \\ 
  19 & $1492.06 \pm 0.43$ & $ 1.24\pm 0.45$ & $ 3.44\pm 1.60$ & 13.90 & \lya       & 0.2273 \\ 
\hline
\end{tabular}
\caption{\label{tab:b140} Detected lines in the GHRS G140L combined
  spectrum of LBQS 0107-025B.  See Table~\ref{tab:a140} for a
  description of each column.}
\end{center}
\end{minipage}
\end{table*}

\begin{table*}
\begin{minipage}{140mm}
\begin{center}
{\large \sc LBQS 0107-0125A FOS G190H}\\
\vspace{0.2cm}
\begin{tabular}{ccccccc}
\hline
ID & Wavelength(\AA)& EW$_{\rm obs}$ (\AA) & FWHM (\AA) & S$_\sigma$ &  Identification & $z_{\rm abs}$\\
\hline
   1 & $1637.49\pm 0.25$ & $ 0.55\pm 0.18$ & $ 1.58\pm 0.62$ &  4.27 & \CII\ 1334 &0.2272   \\ 
   2 & $1671.28\pm 0.13$ & $ 0.88\pm 0.11$ & $ 2.09\pm 0.31$ & 11.30 & \AlII\ 1670/\lyg &0.0000/0.7188  \\ 
   3 & $1683.17\pm 0.09$ & $ 0.81\pm 0.09$ & $ 1.54\pm 0.20$ & 10.74 & - & -    \\ 
   4 & $1701.30\pm 0.16$ & $ 0.61\pm 0.09$ & $ 2.15\pm 0.39$ &  9.64 & \lya &0.3995  \\ 
   5 & $1721.69\pm 0.13$ & $ 0.38\pm 0.08$ & $ 1.24\pm 0.31$ &  5.41 & - & -    \\ 
   6 & $1727.48\pm 0.38$ & $ 0.46\pm 0.12$ & $ 2.87\pm 0.92$ &  7.07 & - & -   \\ 
   7 & $1738.09\pm 0.08$ & $ 0.79\pm 0.07$ & $ 1.70\pm 0.18$ & 13.84 & \lyg/Ly-7 & 0.7865/0.8766  \\ 
   8 & $1746.32\pm 0.10$ & $ 0.16\pm 0.05$ & $ 0.73\pm 0.25$ &  2.95 & Ly-6 & 0.8766  \\ 
   9 & $1759.87\pm 0.16$ & $ 0.39\pm 0.07$ & $ 1.78\pm 0.39$ &  7.40 & Ly-5 & 0.8766  \\ 
  10 & $1763.23\pm 0.09$ & $ 0.83\pm 0.09$ & $ 1.87\pm 0.23$ & 13.52 & \lyb & 0.7188  \\ 
  11 & $1773.33\pm 0.13$ & $ 0.42\pm 0.06$ & $ 1.76\pm 0.32$ &  8.79 & \lyb & 0.7286  \\ 
  12 & $1782.15\pm 0.08$ & $ 0.60\pm 0.06$ & $ 1.72\pm 0.21$ & 12.82 & Ly-4 & 0.8766  \\ 
  13 & $1808.65\pm 0.15$ & $ 0.52\pm 0.07$ & $ 2.32\pm 0.36$ & 11.46 & - & -    \\ 
  14 & $1824.96\pm 0.04$ & $ 1.82\pm 0.06$ & $ 2.42\pm 0.10$ & 40.36 & \lyg/\lya &0.8766/0.5012  \\ 
  15 & $1832.63\pm 0.11$ & $ 0.57\pm 0.06$ & $ 1.99\pm 0.26$ & 12.81 & \lyb &0.7865  \\ 
  16 & $1853.54\pm 0.22$ & $ 0.21\pm 0.06$ & $ 1.68\pm 0.55$ &  4.78 & - & -    \\ 
  17 & $1867.12\pm 0.03$ & $ 2.60\pm 0.06$ & $ 2.56\pm 0.07$ & 54.57 & - & -    \\ 
  18 & $1877.42\pm 0.14$ & $ 0.14\pm 0.04$ & $ 0.97\pm 0.34$ &  3.27 & - & -    \\ 
  19 & $1893.49\pm 0.07$ & $ 0.77\pm 0.07$ & $ 1.69\pm 0.19$ & 18.23 & - & -    \\ 
  20 & $1899.97\pm 0.41$ & $ 0.86\pm 0.20$ & $ 3.83\pm 0.97$ & 19.97 & \CIV\ 1548 & 0.2272  \\ 
  21 & $1903.04\pm 0.21$ & $ 0.44\pm 0.17$ & $ 1.98\pm 0.47$ & 10.34 & \CIV\ 1550 & 0.2272  \\ 
  22 & $1911.51\pm 0.12$ & $ 0.25\pm 0.04$ & $ 1.41\pm 0.29$ &  6.72 & - & -    \\ 
  23 & $1921.19\pm 0.11$ & $ 0.67\pm 0.06$ & $ 2.51\pm 0.27$ & 18.68 & - & -    \\ 
  24 & $1924.78\pm 0.05$ & $ 0.80\pm 0.05$ & $ 1.70\pm 0.11$ & 23.11 & \lyb & 0.8766  \\ 
  25 & $1928.64\pm 0.19$ & $ 0.25\pm 0.05$ & $ 1.89\pm 0.50$ &  7.24 & - & -  \\ 
  26 & $1947.43\pm 0.07$ & $ 0.10\pm 0.03$ & $ 0.59\pm 0.27$ &  2.75 & - & -  \\ 
  27 & $1952.16\pm 0.10$ & $ 0.11\pm 0.03$ & $ 0.68\pm 0.24$ &  3.07 & - & -  \\ 
  28 & $1955.37\pm 0.10$ & $ 0.30\pm 0.04$ & $ 1.41\pm 0.24$ &  8.01 & \lyb  & 0.907 \\ 
  29 & $1976.50\pm 0.18$ & $ 0.32\pm 0.05$ & $ 2.16\pm 0.42$ &  9.23 & \lyb? & 0.926   \\ 
  30 & $1985.65\pm 0.13$ & $ 0.29\pm 0.06$ & $ 1.38\pm 0.31$ &  8.64 & - & -  \\ 
  31 & $1987.40\pm 0.14$ & $ 0.23\pm 0.06$ & $ 1.26\pm 0.34$ &  6.92 & - & -   \\ 
  32 & $1994.20\pm 0.07$ & $ 0.47\pm 0.05$ & $ 1.49\pm 0.18$ & 14.83 & \CIV\ 1548?& 0.2881  \\ 
  33 & $1998.14\pm 0.11$ & $ 0.21\pm 0.04$ & $ 1.33\pm 0.27$ &  6.64 & \CIV\ 1550?& 0.2881  \\ 
  34 & $2007.02\pm 0.07$ & $ 0.32\pm 0.03$ & $ 1.26\pm 0.16$ & 14.42 & \lyb &0.957   \\ 
  35 & $2015.66\pm 0.07$ & $ 1.34\pm 0.05$ & $ 4.29\pm 0.18$ & 83.34 & - & -   \\ 
  36 & $2022.43\pm 0.11$ & $ 0.29\pm 0.09$ & $ 1.64\pm 0.29$ & 21.33 & - & -   \\ 
  37 & $2025.66\pm 0.17$ & $ 1.13\pm 0.22$ & $ 3.99\pm 0.86$ & 82.36 & - & -   \\ 
  38 & $2028.98\pm 0.25$ & $ 0.30\pm 0.14$ & $ 2.18\pm 0.51$ & 20.34 & - & -   \\ 
  39 & $2034.10\pm 0.35$ & $ 0.10\pm 0.04$ & $ 1.92\pm 0.94$ &  5.88 & - & -   \\ 
  40 & $2036.60\pm 0.20$ & $ 0.12\pm 0.04$ & $ 1.38\pm 0.49$ &  6.43 & - & -   \\ 
  41 & $2041.44\pm 0.04$ & $ 0.41\pm 0.03$ & $ 1.31\pm 0.11$ & 19.16 & - & -   \\ 
  42 & $2054.27\pm 0.05$ & $ 0.42\pm 0.03$ & $ 1.50\pm 0.13$ & 16.26 & - & -   \\ 
  43 & $2059.89\pm 0.10$ & $ 0.22\pm 0.03$ & $ 1.38\pm 0.23$ &  7.89 & - & -   \\ 
  44 & $2073.69\pm 0.12$ & $ 0.19\pm 0.03$ & $ 1.39\pm 0.29$ &  6.36 & - & -   \\ 
  45 & $2082.10\pm 0.11$ & $ 0.18\pm 0.03$ & $ 1.25\pm 0.26$ &  5.78 & - & -   \\ 
  46 & $2089.48\pm 0.03$ & $ 1.59\pm 0.04$ & $ 2.31\pm 0.07$ & 48.99 & \lya & 0.7188  \\ 
  47 & $2101.39\pm 0.02$ & $ 1.24\pm 0.03$ & $ 1.79\pm 0.05$ & 38.72 & \lya & 0.7286  \\ 
  48 & $2122.79\pm 0.13$ & $ 0.17\pm 0.03$ & $ 1.43\pm 0.32$ &  6.32 & - & -   \\ 
  49 & $2131.56\pm 0.19$ & $ 0.12\pm 0.03$ & $ 1.39\pm 0.47$ &  4.39 & - & -   \\ 
  50 & $2146.95\pm 0.16$ & $ 0.10\pm 0.03$ & $ 1.07\pm 0.38$ &  3.59 & - & -   \\ 
  51 & $2166.83\pm 0.15$ & $ 0.22\pm 0.04$ & $ 1.88\pm 0.37$ &  8.30 & \CIV\ 1548 & 0.3995  \\ 
  52 & $2171.74\pm 0.03$ & $ 1.18\pm 0.03$ & $ 1.94\pm 0.06$ & 44.57 & \lya & 0.7865   \\ 
  53 & $2187.61\pm 0.16$ & $ 0.29\pm 0.04$ & $ 2.25\pm 0.43$ & 11.17 & - & -  \\ 
  54 & $2196.86\pm 0.10$ & $ 0.19\pm 0.03$ & $ 1.29\pm 0.23$ &  7.52 & - & -  \\ 
  55 & $2240.78\pm 0.07$ & $ 0.27\pm 0.03$ & $ 1.41\pm 0.18$ & 10.90 & - & -  \\ 
  56 & $2246.33\pm 0.04$ & $ 0.47\pm 0.03$ & $ 1.34\pm 0.09$ & 19.01 & - & -  \\ 
  57 & $2251.63\pm 0.07$ & $ 0.23\pm 0.03$ & $ 1.27\pm 0.18$ &  9.40 & \NI\ 1200 & 0.8766   \\ 
  58 & $2262.96\pm 0.35$ & $ 0.33\pm 0.06$ & $ 3.89\pm 0.95$ & 13.79 & - & -  \\ 
  59 & $2271.83\pm 0.06$ & $ 0.39\pm 0.03$ & $ 1.52\pm 0.13$ & 15.38 & - & -  \\ 
  60 & $2281.30\pm 0.02$ & $ 1.18\pm 0.03$ & $ 1.71\pm 0.04$ & 51.35 & \lya & 0.8766  \\ 
  61 & $2288.84\pm 0.12$ & $ 0.08\pm 0.02$ & $ 0.88\pm 0.29$ &  3.54 & - & -  \\ 
  62 & $2297.18\pm 0.04$ & $ 0.39\pm 0.03$ & $ 1.32\pm 0.10$ & 16.89 & - & -  \\ 
  63 & $2308.66\pm 0.05$ & $ 0.36\pm 0.03$ & $ 1.16\pm 0.12$ & 12.49 & - & -  \\ 
\hline
\end{tabular}
\caption{\label{tab:a190} Detected lines in the FOS G190H combined
  spectrum of LBQS 0107-025A. See Table~\ref{tab:a140} for a
  description of each column.}
\end{center}
\end{minipage}
\end{table*}

\begin{table*}
\begin{minipage}{140mm}
\begin{center}
{\large \sc LBQS 0107-0125B FOS G190H}\\
\vspace{0.2cm}
\begin{tabular}{ccccccc}
\hline
ID & Wavelength(\AA)& EW$_{\rm obs}$ (\AA) & FWHM (\AA) & S$_\sigma$ &  Identification & $z_{\rm abs}$\\
\hline
   1 & $1637.82\pm 0.08$ & $ 0.51\pm 0.14$ &$ 0.55\pm 0.20$ &  2.34 & \CII\ 1334   & 0.2273 \\
   2 & $1671.19\pm 0.13$ & $ 0.82\pm 0.15$ &$ 1.55\pm 0.33$ &  6.53 & \AlII\ 1670/\lyg  & 0.0000,0.7183 \\
   3 & $1680.16\pm 0.15$ & $ 0.37\pm 0.12$ &$ 1.00\pm 0.39$ &  3.23 &- & -    \\
   4 & $1701.06\pm 0.07$ & $ 1.28\pm 0.12$ &$ 1.69\pm 0.19$ & 12.42 & \lya   & 0.3993 \\
   5 & $1717.51\pm 0.25$ & $ 0.62\pm 0.16$ &$ 1.99\pm 0.60$ &  5.57 &- & -    \\
   6 & $1738.05\pm 0.16$ & $ 0.32\pm 0.10$ &$ 1.11\pm 0.39$ &  3.57 & \lyg   & 0.7874 \\
   7 & $1746.62\pm 0.06$ & $ 0.40\pm 0.07$ &$ 0.74\pm 0.16$ &  4.49 &- & -    \\
   8 & $1759.89\pm 0.17$ & $ 0.55\pm 0.13$ &$ 1.56\pm 0.41$ &  6.77 &- & -    \\
   9 & $1762.53\pm 0.10$ & $ 1.44\pm 0.16$ &$ 2.03\pm 0.28$ & 14.08 & \lyb  & 0.7183  \\
  10 & $1784.15\pm 0.77$ & $ 0.79\pm 0.34$ &$ 4.13\pm 1.99$ & 10.67 &- & -    \\
  11 & $1823.58\pm 0.09$ & $ 1.50\pm 0.11$ &$ 2.66\pm 0.23$ & 20.90 & \lya & 0.499  \\ 
  12 & $1844.60\pm 0.07$ & $ 0.68\pm 0.07$ &$ 1.31\pm 0.17$ &  9.61 &- & -    \\
  13 & $1852.68\pm 0.12$ & $ 0.42\pm 0.08$ &$ 1.28\pm 0.29$ &  6.02 &- & -    \\
  14 & $1855.65\pm 0.10$ & $ 1.19\pm 0.11$ &$ 2.32\pm 0.25$ & 16.79 & \lyb  & 0.8094  \\
  15 & $1866.73\pm 0.10$ & $ 1.05\pm 0.11$ &$ 2.09\pm 0.25$ & 13.14 &- & -    \\
  16 & $1919.41\pm 0.34$ & $ 0.60\pm 0.12$ &$ 3.56\pm 0.83$ & 10.05 &- & -    \\
  17 & $1924.62\pm 0.08$ & $ 0.93\pm 0.08$ &$ 2.04\pm 0.21$ & 16.25 & \lyb       & 0.8763  \\
  18 & $1950.21\pm 0.21$ & $ 0.34\pm 0.08$ &$ 1.78\pm 0.51$ &  5.63 & \SiIV\ 1393 & 0.3993  \\ 
  19 & $1956.40\pm 0.17$ & $ 0.51\pm 0.10$ &$ 1.96\pm 0.45$ &  7.97 & \lyb       & 0.9065  \\
  20 & $1971.91\pm 0.15$ & $ 0.42\pm 0.08$ &$ 1.72\pm 0.38$ &  7.24 &- & -    \\
  21 & $1997.16\pm 0.34$ & $ 0.37\pm 0.10$ &$ 2.72\pm 0.93$ &  7.15 &- & -    \\
  22 & $2044.48\pm 0.15$ & $ 0.10\pm 0.03$ &$ 0.40\pm 0.63$ &  2.16 &- & -    \\
  23 & $2048.84\pm 0.23$ & $ 0.17\pm 0.06$ &$ 1.42\pm 0.55$ &  3.76 &- & -    \\
  24 & $2054.50\pm 0.12$ & $ 0.32\pm 0.05$ &$ 1.44\pm 0.29$ &  6.88 &- & -    \\
  25 & $2064.73\pm 0.38$ & $ 0.38\pm 0.11$ &$ 3.07\pm 1.11$ &  7.83 &- & -    \\
  26 & $2083.57\pm 0.10$ & $ 0.60\pm 0.06$ &$ 2.02\pm 0.25$ & 13.31 &- & -    \\
  27 & $2088.83\pm 0.03$ & $ 1.42\pm 0.05$ &$ 1.81\pm 0.08$ & 31.87 & \lya   & 0.7183 \\
  28 & $2097.15\pm 1.13$ & $ 0.35\pm 0.19$ &$ 5.76\pm 4.18$ &  8.11 &- & -    \\
  29 & $2101.69\pm 0.15$ & $ 0.20\pm 0.05$ &$ 1.23\pm 0.37$ &  4.47 &- & -    \\
  30 & $2125.45\pm 0.22$ & $ 0.22\pm 0.06$ &$ 1.66\pm 0.57$ &  4.81 &- & -    \\
  31 & $2166.32\pm 0.10$ & $ 0.30\pm 0.05$ &$ 1.22\pm 0.23$ &  6.80 & \CIV\ 1548 & 0.3993  \\
  32 & $2170.02\pm 0.16$ & $ 0.24\pm 0.06$ &$ 1.42\pm 0.39$ &  5.52 & \CIV\ 1550 & 0.3993   \\
  33 & $2172.83\pm 0.07$ & $ 0.75\pm 0.06$ &$ 1.89\pm 0.19$ & 17.19 & \lya       & 0.7874  \\
  34 & $2185.55\pm 0.17$ & $ 0.32\pm 0.06$ &$ 1.88\pm 0.42$ &  7.33 &- & -    \\
  35 & $2199.64\pm 0.03$ & $ 1.59\pm 0.05$ &$ 2.00\pm 0.08$ & 37.62 & \lya &0.8094    \\
  36 & $2210.67\pm 0.13$ & $ 0.23\pm 0.05$ &$ 1.25\pm 0.31$ &  5.26 &- & -    \\
  37 & $2215.33\pm 0.49$ & $ 0.24\pm 0.08$ &$ 3.10\pm 1.18$ &  5.76 &- & -    \\
  38 & $2226.93\pm 0.09$ & $ 0.42\pm 0.05$ &$ 1.47\pm 0.21$ &  9.99 &- & -    \\
  39 & $2230.35\pm 0.11$ & $ 0.34\pm 0.05$ &$ 1.48\pm 0.26$ &  8.17 &- & -    \\  
  40 & $2246.48\pm 0.26$ & $ 0.26\pm 0.06$ &$ 2.14\pm 0.64$ &  6.37 & - & - \\ 
  41 & $2280.94\pm 0.03$ & $ 1.08\pm 0.04$ &$ 1.71\pm 0.08$ & 29.45 & \lya & 0.8763  \\ 
  42 & $2297.60\pm 0.08$ & $ 0.25\pm 0.04$ &$ 1.09\pm 0.20$ &  6.67 & - & -  \\ 
\hline
\end{tabular}
\caption{\label{tab:b190} Detected lines in the FOS G190H combined
  spectrum of LBQS 0107-025B.  See Table~\ref{tab:a140} for a
  description of each column.}
\end{center}
\end{minipage}
\end{table*}

\begin{table*}
\begin{minipage}{140mm}
\begin{center}
{\large \sc LBQS 0107-0232 FOS G190H}\\
\vspace{0.2cm}
\begin{tabular}{ccccccc}
\hline
ID & Wavelength(\AA)& EW$_{\rm obs}$ (\AA) & FWHM (\AA) & S$_\sigma$ &  Identification & $z_{\rm abs}$\\
\hline
   1 & $1656.72\pm 0.19$ & $ 1.02\pm 0.31$& $ 1.39\pm 0.50$ &  4.24 & - & - \\ 
   2 & $1670.43\pm 0.14$ & $ 0.71\pm 0.17$& $ 1.20\pm 0.33$ &  4.65 & \AlII\ 1670 & 0.000 \\ 
   3 & $1677.88\pm 0.30$ & $ 1.64\pm 0.34$& $ 3.30\pm 0.86$ & 12.34 & - & - \\ 
   4 & $1702.02\pm 0.28$ & $ 1.34\pm 0.22$& $ 3.68\pm 0.78$ & 14.63 & \lya & 0.4001  \\ 
   5 & $1732.50\pm 0.34$ & $ 0.85\pm 0.23$& $ 2.63\pm 0.89$ &  5.99 & \lyb & 0.689     \\ 
   6 & $1736.11\pm 0.18$ & $ 1.65\pm 0.24$& $ 2.72\pm 0.48$ & 11.29 & - & - \\ 
   7 & $1744.17\pm 0.55$ & $ 0.73\pm 0.32$& $ 2.90\pm 1.57$ &  6.17 & - & - \\ 
   8 & $1746.91\pm 0.22$ & $ 0.65\pm 0.25$& $ 1.58\pm 0.48$ &  6.06 & - & - \\ 
   9 & $1754.24\pm 0.16$ & $ 1.08\pm 0.14$& $ 2.72\pm 0.43$ & 12.77 & - & - \\ 
  10 & $1761.39\pm 0.10$ & $ 0.53\pm 0.26$& $ 1.16\pm 0.35$ &  7.71 & \lyb& 0.7174  \\ 
  11 & $1763.62\pm 0.68$ & $ 0.77\pm 0.34$& $ 3.35\pm 1.70$ & 12.55 & - & - \\ 
  12 & $1781.29\pm 0.89$ & $ 0.54\pm 0.29$& $ 3.36\pm 1.50$ & 13.30 & \FeII\ 1143 & 0.5569  \\ 
  13 & $1783.14\pm 0.13$ & $ 0.32\pm 0.25$& $ 1.35\pm 0.52$ &  7.98 & \FeII\ 1144 & 0.5569  \\ 
  14 & $1796.23\pm 0.20$ & $ 0.26\pm 0.07$& $ 1.50\pm 0.49$ &  4.12 & - & - \\ 
  15 & $1807.05\pm 0.37$ & $ 0.62\pm 0.15$& $ 3.24\pm 0.95$ &  6.63 & - & - \\ 
  16 & $1822.39\pm 0.14$ & $ 0.51\pm 0.10$& $ 1.53\pm 0.36$ &  4.94 & - & - \\ 
  17 & $1853.41\pm 0.09$ & $ 0.79\pm 0.09$& $ 1.64\pm 0.22$ &  7.83 & \SiII\ 1190 & 0.5569  \\ 
  18 & $1857.74\pm 0.09$ & $ 0.72\pm 0.10$& $ 1.49\pm 0.24$ &  6.65 & \SiII\ 1193 & 0.5569  \\ 
  19 & $1864.76\pm 0.12$ & $ 1.58\pm 0.14$& $ 2.70\pm 0.28$ & 13.02 & - & - \\ 
  20 & $1878.49\pm 0.07$ & $ 1.76\pm 0.10$& $ 2.36\pm 0.17$ & 16.81 & \SiIII\ 1206 & 0.5569  \\ 
  21 & $1892.65\pm 0.07$ & $ 4.75\pm 0.16$& $ 4.28\pm 0.18$ & 55.70 & \lya& 0.5569  \\ 
  22 & $1918.61\pm 0.07$ & $ 1.17\pm 0.09$& $ 1.86\pm 0.17$ & 16.09 & - & - \\ 
  23 & $1962.31\pm 0.06$ & $ 1.33\pm 0.09$& $ 1.94\pm 0.16$ & 19.03 & \SiII\ 1260 & 0.5569  \\ 
  24 & $1971.14\pm 0.14$ & $ 0.92\pm 0.11$& $ 2.50\pm 0.37$ & 12.83 & - & - \\ 
  25 & $2003.89\pm 0.08$ & $ 1.11\pm 0.10$& $ 2.03\pm 0.21$ & 20.70 & - & - \\ 
  26 & $2027.25\pm 0.07$ & $ 0.61\pm 0.07$& $ 1.20\pm 0.16$ & 11.46 & \OI\ 1302 & 0.5569 \\ 
  27 & $2045.29\pm 0.57$ & $ 0.39\pm 0.26$& $ 1.83\pm 1.00$ &  8.95 & - & - \\ 
  28 & $2046.95\pm 0.38$ & $ 0.36\pm 0.25$& $ 1.48\pm 0.58$ &  8.51 & - & - \\ 
  29 & $2053.70\pm 0.04$ & $ 1.19\pm 0.05$& $ 1.77\pm 0.09$ & 30.91 & \lya & 0.689 \\ 
  30 & $2067.00\pm 0.16$ & $ 0.35\pm 0.06$& $ 2.12\pm 0.39$ & 12.03 & - & - \\ 
  31 & $2077.75\pm 0.06$ & $ 1.43\pm 0.10$& $ 1.89\pm 0.12$ & 66.47 & \CII\ 1334 & 0.5569  \\ 
  32 & $2079.65\pm 0.17$ & $ 0.39\pm 0.10$& $ 1.58\pm 0.31$ & 19.73 & - & - \\ 
  33 & $2087.85\pm 0.02$ & $ 2.16\pm 0.03$& $ 2.53\pm 0.04$ & 164.89& \lya & 0.7174 \\ 
  34 & $2092.54\pm 0.29$ & $ 0.05\pm 0.02$& $ 1.31\pm 0.68$ &  5.67 & - & - \\ 
  35 & $2113.12\pm 0.12$ & $ 0.20\pm 0.03$& $ 1.42\pm 0.28$ &  9.35 & - & - \\ 
  36 & $2137.71\pm 0.08$ & $ 0.18\pm 0.03$& $ 1.04\pm 0.19$ &  8.48 & - & - \\ 
  37 & $2263.61\pm 0.25$ & $ 0.27\pm 0.07$& $ 2.15\pm 0.64$ &  6.23 & - & - \\ 
\hline
\end{tabular}
\caption{\label{tab:c190} Detected lines in the FOS G190H combined
  spectrum of LBQS 0107-0232 (QSO C).  See Table~\ref{tab:a140} for a
  description of each column.}
\end{center}
\end{minipage}
\end{table*}

\begin{table*}
\begin{minipage}{140mm}
\begin{center}
{\large \sc LBQS 0107-0125A FOS G270H}\\
\vspace{0.2cm}
\begin{tabular}{ccccccc}
\hline
ID & Wavelength(\AA)& EW$_{\rm obs}$ (\AA) & FWHM (\AA) & S$_\sigma$ &  Identification & $z_{\rm abs}$\\
\hline
   1 & $2247.19\pm 0.20$ & $ 0.39\pm 0.08$& $ 1.88\pm 0.47$ &  5.31 & - &-   \\ 
   2 & $2251.98\pm 0.22$ & $ 0.36\pm 0.08$& $ 1.92\pm 0.53$ &  5.01 & \NI\ 1200 & 0.8769  \\ 
   3 & $2272.78\pm 0.21$ & $ 0.37\pm 0.08$& $ 1.95\pm 0.51$ &  5.25 & - & -  \\ 
   4 & $2281.72\pm 0.06$ & $ 1.39\pm 0.08$& $ 2.21\pm 0.15$ & 18.87 & \lya & 0.8769  \\ 
   5 & $2297.30\pm 0.17$ & $ 0.35\pm 0.07$& $ 1.79\pm 0.41$ &  5.66 & - & -   \\ 
   6 & $2308.96\pm 0.18$ & $ 0.34\pm 0.07$& $ 1.86\pm 0.43$ &  5.86 & - & -   \\ 
   7 & $2318.24\pm 0.06$ & $ 1.21\pm 0.07$& $ 2.34\pm 0.15$ & 21.69 & \lya & 0.907 \\ 
   8 & $2342.76\pm 0.06$ & $ 1.58\pm 0.06$& $ 3.45\pm 0.15$ & 39.88 & \lya & 0.926  \\ 
   9 & $2361.46\pm 0.06$ & $ 0.49\pm 0.03$& $ 1.72\pm 0.13$ & 21.48 & - & -   \\ 
  10 & $2370.48\pm 0.09$ & $ 0.30\pm 0.03$& $ 1.90\pm 0.22$ & 17.45 & - & -   \\ 
  11 & $2374.96\pm 0.10$ & $ 0.17\pm 0.02$& $ 1.38\pm 0.23$ & 11.61 & - & -   \\ 
  12 & $2379.06\pm 0.03$ & $ 0.83\pm 0.02$& $ 2.04\pm 0.07$ & 66.83 & \lya & 0.957   \\ 
  13 & $2382.87\pm 0.05$ & $ 0.50\pm 0.16$& $ 1.59\pm 0.20$ & 40.65 & \FeII\ 2382 & 0.0000  \\ 
  14 & $2384.81\pm 0.86$ & $ 0.31\pm 0.17$& $ 3.30\pm 1.49$ & 23.24 & - & -   \\
  15 & $2389.70\pm 0.06$ & $ 0.49\pm 0.03$& $ 2.11\pm 0.16$ & 30.08 & - & -   \\
  16 & $2396.46\pm 0.24$ & $ 0.13\pm 0.03$& $ 1.97\pm 0.57$ &  6.70 & \SiIV\ 1393 &0.7194 \\ 
  17 & $2411.81\pm 0.19$ & $ 0.16\pm 0.04$& $ 1.68\pm 0.45$ &  5.39 & \SiIV\ 1402 &0.7194 \\ 
  18 & $2586.76\pm 0.22$ & $ 0.66\pm 0.09$& $ 3.58\pm 0.55$ &  6.94 & \FeII\ 2586 &0.0000 \\ 
  19 & $2600.38\pm 0.09$ & $ 0.51\pm 0.06$& $ 1.73\pm 0.23$ &  5.16 & \FeII\ 2600 &0.0000 \\ 
  20 & $2796.37\pm 0.07$ & $ 0.77\pm 0.06$& $ 1.95\pm 0.18$ &  6.88 & \MgII\ 2796 &0.0000 \\ 
  21 & $2803.58\pm 0.07$ & $ 0.65\pm 0.06$& $ 1.79\pm 0.18$ &  5.76 & \MgII\ 2803 &0.0000 \\ 
\hline
\end{tabular}
\caption{\label{tab:a270} Detected lines in the FOS G270H combined
  spectrum of LBQS 0107-025A. See Table~\ref{tab:a140} for a
  description of each column.}
\end{center}
\end{minipage}
\end{table*}

\begin{table*}
\begin{minipage}{140mm}
\begin{center}
{\large \sc LBQS 0107-0125B FOS G270H}\\
\vspace{0.2cm}
\begin{tabular}{ccccccc}
\hline
ID & Wavelength(\AA)& EW$_{\rm obs}$ (\AA) & FWHM (\AA) & S$_\sigma$ &  Identification & $z_{\rm abs}$\\
\hline
   1 & $2246.72\pm 0.23$ & $ 0.22\pm 0.06$& $ 1.86\pm 0.56$ &  4.66 & - & -  \\ 
   2 & $2281.30\pm 0.05$ & $ 1.33\pm 0.06$& $ 2.38\pm 0.12$ & 28.49 &\lya & 0.8766   \\ 
   3 & $2298.37\pm 0.20$ & $ 0.19\pm 0.05$& $ 1.69\pm 0.47$ &  4.70 & - & -  \\ 
   4 & $2319.05\pm 0.04$ & $ 1.01\pm 0.04$& $ 2.05\pm 0.10$ & 26.27 &\lya  & 0.907  \\ 
   5 & $2342.12\pm 0.18$ & $ 0.28\pm 0.08$& $ 1.67\pm 0.38$ &  8.43 & - & -  \\ 
   6 & $2344.52\pm 0.22$ & $ 0.48\pm 0.09$& $ 2.54\pm 0.50$ & 15.07 &\FeII\ 2344 & 0.0000 \\ 
   7 & $2349.38\pm 0.19$ & $ 0.19\pm 0.04$& $ 1.97\pm 0.48$ &  6.02 & - & - \\ 
   8 & $2374.84\pm 0.08$ & $ 0.20\pm 0.02$& $ 1.39\pm 0.20$ &  8.39 & \FeII\ 2374&0.0000 \\ 
   9 & $2383.23\pm 0.05$ & $ 0.67\pm 0.03$& $ 2.14\pm 0.11$ & 25.09 & \FeII\ 2382&0.0000 \\ 
  10 & $2510.88\pm 0.28$ & $ 0.15\pm 0.05$& $ 1.97\pm 0.69$ &  4.11 & - & -  \\
  11 & $2587.18\pm 0.11$ & $ 0.41\pm 0.05$& $ 2.06\pm 0.27$ & 11.01 & \FeII\ 2586 &0.0000 \\ 
  12 & $2600.63\pm 0.07$ & $ 0.53\pm 0.04$& $ 1.89\pm 0.17$ & 14.69 & \FeII\ 2600 &0.0000 \\ 
  13 & $2796.64\pm 0.04$ & $ 0.91\pm 0.04$& $ 1.99\pm 0.10$ & 24.49 & \MgII\ 2796 &0.0000 \\ 
  14 & $2803.92\pm 0.05$ & $ 0.72\pm 0.04$& $ 1.78\pm 0.11$ & 19.65 & \MgII\ 2803 &0.0000 \\ 
  15 & $2853.53\pm 0.20$ & $ 0.19\pm 0.04$& $ 1.86\pm 0.49$ &  5.12 & \MgI\  2853 &0.0000 \\ 
\hline
\end{tabular}
\caption{\label{tab:b270} Detected lines in the FOS G270H combined
  spectrum of LBQS 0107-025B.  See Table~\ref{tab:a140} for a
  description of each column.}
\end{center}
\end{minipage}
\end{table*}

\begin{table*}
\begin{minipage}{180mm}
\begin{center}
{\large \sc Members of Triple Absorber Groups}\\
\vspace{0.2cm}
\begin{tabular}{ccccccccccc}
\hline
 \multicolumn{2}{|c|}{sightline A} & \multicolumn{2}{|c|}{sightline B}  & \multicolumn{2}{|c|}{sightline C} & & $\Delta v_{\rm max}$   & $\Delta v_{\rm max}$ & Largest $\Delta v$ & Triple LOS-\\ 
 ID &  REW (\AA) &  ID & REW (\AA) & ID & REW (\AA) &  Sym. & 200 &  500 & (\kms) & gal. group?\\
\hline 
 3  & $0.59 \pm 0.07$ &  3   &  $0.27 \pm 0.09$  &  3   & $1.19 \pm 0.25$ & y & n & n & 944 & n \\    
 4  & $0.44 \pm 0.06$ &  4   &  $0.91 \pm 0.09$  &  4   & $0.96 \pm 0.16$ & y & y & y & 169 & n \\ 
 14 & $1.21 \pm 0.04$ &  11  &  $1.00 \pm 0.07$  &  16  & $0.34 \pm 0.07$ & y & n & y & 422 & y \\ 
 17 & $1.69 \pm 0.04$ &  15  &  $0.68 \pm 0.07$  &  19  & $1.03 \pm 0.09$ & y & n & y & 379 & y \\ 
 23 & $0.42 \pm 0.04$ &  16  &  $0.38 \pm 0.08$  &  22  & $0.74 \pm 0.06$ & y & n & y & 402 & n \\ 
 42 & $0.25 \pm 0.02$ &  24  &  $0.19 \pm 0.03$  &  29  & $0.70 \pm 0.03$ & y & y & y & 117 & n \\ 
 45 & $0.11 \pm 0.02$ &  26  &  $0.35 \pm 0.04$  &  32  & $0.23 \pm 0.06$ & y & n & y & 565 & n \\ 
 46 & $0.93 \pm 0.02$ &  27  &  $0.83 \pm 0.03$  &  33  & $1.26 \pm 0.02$ & y & y & y & 234 & n \\ 
\hline
\end{tabular}
\caption{\label{tab:tripmem} Absorbers that are members of symmetric
  triples, or triple absorber groups found using a $\Delta v_{\rm max}
  = 200$ or $500$~\kms. The columns show the ID number from Tables 5,
  6 or 7, the rest equivalent width and $1\sigma$ error for absorbers
  in each sightline; whether the triple was found using symmetric
  matching, whether it is found using the group matching algorithm
  with $\Delta v_{\rm max} = 200$ and 500~\kms, the maximum velocity
  difference between any two triple members, and whether the absorbers
  are also members of a triple LOS-galaxy group (described in Section
  4.3).}
\end{center}
\end{minipage}
\end{table*}

\begin{table*}
\begin{minipage}{140mm}
\begin{center}
{\large \sc Metal Line systems}\\
\vspace{0.2cm}
\begin{tabular}{cccccc}
\hline
LOS & $z_{\rm abs}$ & Identification & Wavelength(\AA)& EW$_{\rm obs}$ (\AA) &   Comments  \\
\hline
A & 0.2286   & \lya         &  $1493.50\pm 0.15$ & $ 0.82\pm 0.25$ &        -         \\ 
A & 0.2272   & \CII\ 1334?  &  $1637.49\pm 0.25$ & $ 0.55\pm 0.18$ &  S/N poor        \\ 
A & 0.2272   & \CIV\ 1548   &  $1899.97\pm 0.41$ & $ 0.86\pm 0.20$ &        -         \\ 
A & 0.2272   & \CIV\ 1550   &  $1903.04\pm 0.21$ & $ 0.44\pm 0.17$ &        -         \\ 
B & 0.2273   & \OVI\ 1031?  &  $1265.82\pm 0.22$ & $ 0.66\pm 0.19$ &  S/N poor        \\ 
B & 0.2273   & \OVI\ 1037?  &  $1273.56\pm 0.13$ & $ 0.30\pm 0.10$ &  S/N poor        \\ 
B & 0.2273   & \lya         &  $1491.46\pm 0.43$ & $ 1.24\pm 0.45$ &        -         \\ 
B & 0.2273   & \CII\ 1334   &  $1637.82\pm 0.08$ & $ 0.51\pm 0.14$ &  S/N poor        \\ 
\\                                                                                 
A & 0.2881   & \CIV\ 1548?  &  $1994.20\pm 0.07$ & $ 0.47\pm 0.05$ & No coverage of \lya \\ 
A & 0.2881   & \CIV\ 1550?  &  $1998.14\pm 0.11$ & $ 0.21\pm 0.04$ & No coverage of \lya \\ 
\\                                                                                
A & 0.3995   & \lya         &  $1701.30\pm 0.16$ & $ 0.61\pm 0.09$ &         -            \\ 
A & 0.3995   & \CIV\ 1548   &  $2166.83\pm 0.15$ & $ 0.22\pm 0.04$ &\CIV\ 1550 blended with \lya  \\ 
B & 0.3993   & Ly-5         &  $1311.89\pm 0.15$ & $ 0.26\pm 0.08$ &          -           \\
B & 0.3993   & Ly-4         &  $1328.64\pm 0.08$ & $ 0.50\pm 0.08$ &          -           \\ 
B & 0.3993   & \lyg         &  $1360.66\pm 0.19$ & $ 1.05\pm 0.17$ &          -           \\ 
B & 0.3993   & \lyb         &  $1434.69\pm 0.10$ & $ 0.42\pm 0.10$ &          -           \\ 
B & 0.3993   & \lya         &  $1701.06\pm 0.07$ & $ 1.28\pm 0.12$ &Lyman limit system   \\ 
B & 0.3993   & \SiIV\ 1393  &  $1950.21\pm 0.21$ & $ 0.34\pm 0.08$ &\SiIV\ 1402 blended with \lya \\ 
B & 0.3993   & \CIV\ 1548   &  $2166.32\pm 0.10$ & $ 0.30\pm 0.05$ &          -           \\ 
B & 0.3993   & \CIV\ 1550   &  $2170.02\pm 0.16$ & $ 0.24\pm 0.06$ &          -           \\ 
C & 0.4001   & \lya         &  $1702.02\pm 0.28$ & $ 1.34\pm 0.22$ &          -           \\ 
\\                                                                                 
A & 0.5576   & \lya         &  $1893.49\pm 0.07$ & $ 0.77\pm 0.07$ &          -           \\ 
C & 0.5569   & \FeII\ 1143  &  $1781.29\pm 0.89$ & $ 0.54\pm 0.29$ &Blended with \lya?   \\ 
C & 0.5569   & \FeII\ 1144  &  $1783.14\pm 0.13$ & $ 0.32\pm 0.25$ &Blended with \lya?   \\ 
C & 0.5569   & \SiII\ 1190  &  $1853.41\pm 0.09$ & $ 0.79\pm 0.09$ &          -           \\ 
C & 0.5569   & \SiII\ 1193  &  $1857.74\pm 0.09$ & $ 0.72\pm 0.10$ &          -           \\ 
C & 0.5569   & \SiIII\ 1206 &  $1878.49\pm 0.07$ & $ 1.76\pm 0.10$ &          -           \\ 
C & 0.5569   & \lya         &  $1892.65\pm 0.07$ & $ 4.75\pm 0.16$ &Probable sub-DLA     \\ 
C & 0.5569   & \SiII\ 1260  &  $1962.31\pm 0.06$ & $ 1.33\pm 0.09$ &          -           \\ 
C & 0.5569   & \OI\ 1302    &  $2027.25\pm 0.07$ & $ 0.61\pm 0.07$ &          -           \\ 
C & 0.5569   & \CII\ 1334   &  $2077.75\pm 0.06$ & $ 1.43\pm 0.10$ &          -           \\ 
\\                                                                                
A & 0.7188   & \lyb         &  $1763.23\pm 0.09$ & $ 0.83\pm 0.09$ &          -           \\ 
A & 0.7188   & \lya         &  $2089.48\pm 0.03$ & $ 1.59\pm 0.04$ &          -           \\ 
A & 0.7194   & \SiIV\ 1393  &  $2396.46\pm 0.24$ & $ 0.13\pm 0.03$ & Outside \lya\ forest \\ 
A & 0.7194   & \SiIV\ 1402  &  $2411.81\pm 0.19$ & $ 0.16\pm 0.04$ & Outside \lya\ forest \\ 
B & 0.7183   & \lyb         &  $1762.53\pm 0.10$ & $ 1.44\pm 0.16$ &          -           \\
B & 0.7183   & \lya         &  $2088.83\pm 0.03$ & $ 1.42\pm 0.05$ &          -           \\
\\                                                                                
A & 0.8766   & \NI\ 1200    &  $2251.63\pm 0.07$ & $ 0.23\pm 0.03$ &          -           \\ 
A & 0.8766   & Ly-6         &  $1746.32\pm 0.10$ & $ 0.16\pm 0.05$ &          -           \\ 
A & 0.8766   & Ly-5         &  $1759.87\pm 0.16$ & $ 0.39\pm 0.07$ &          -           \\ 
A & 0.8766   & Ly-4         &  $1782.15\pm 0.08$ & $ 0.60\pm 0.06$ &          -           \\ 
A & 0.8766   & \lyb         &  $1924.78\pm 0.05$ & $ 0.80\pm 0.05$ &          -           \\ 
A & 0.8766   & \lya         &  $2281.30\pm 0.02$ & $ 1.18\pm 0.03$ &          -           \\ 
B & 0.8763   & \lyb         &  $1924.62\pm 0.08$ & $ 0.93\pm 0.08$ &          -           \\
B & 0.8763   & \lya         &  $2280.94\pm 0.03$ & $ 1.08\pm 0.04$ &          -           \\ 
\hline
\end{tabular}
\caption{\label{tab:metals} Absorption systems that have associated
  metal lines.  \HI\ absorbers with a similar redshift as metal line
  systems in adjacent sightlines are also shown. The columns show,
  from left to right: sightline where the line is seen; redshift;
  line identification; observed wavelength and $1\sigma$ error;
  observed equivalent width and $1\sigma$ error; and finally any
  comments about the line. Lines are grouped by
  redshift. Identifications ending in a question mark are uncertain.}
\end{center}
\end{minipage}
\end{table*}

\clearpage

\clearpage

\begin{figure*}
\begin{center}
\includegraphics[width=1.0\textwidth]{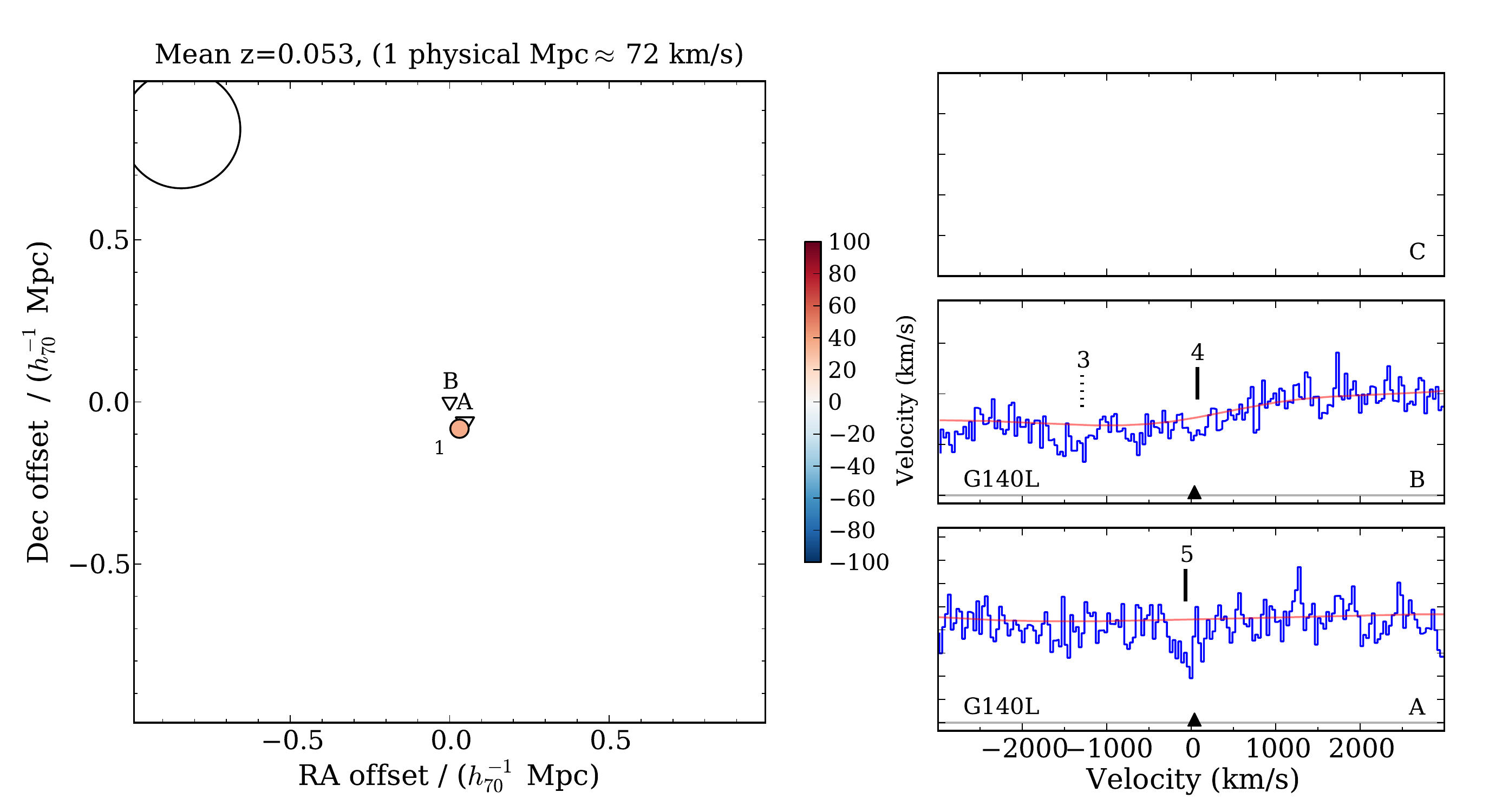}
\caption{\label{fig:galspec} Galaxies and absorbers near a given
  redshift. \textit{Left panel:} Galaxy positions relative to the
  three sightlines.  The area of the galaxy symbol is proportional to
  the galaxy's luminosity.  The open circle in the top left hand
  corner shows the size of an L* galaxy for comparison. Galaxies are
  coloured by their velocity offset, and their numbers from
  Table~\ref{tab:gal} are shown.  If there is any \HI\ \lya\
  absorption within 1000 \kms\ of the zero velocity position it is
  shown by an inverted triangle. The area of the triangle is
  proprtional to the summed \lya\ rest equivalent width. If no \lya\
  absorption is detected within this range, the position of the
  sightline is shown by a plus sign.  \textit{Right panels:} Portions
  of the QSO spectra corresponding to \HI\ \lya\ near the galaxies in
  the left panel. Our fitted continua are also shown. Flux units are
  arbitrary. The tick marks indicate line positions -- solid ticks are
  \lya\ or unidentified lines, dotted ticks are lines that have been
  identified as something other than \lya. Line numbers from
  Tables~\ref{tab:a140} to \ref{tab:b270} are shown above the tick
  marks. Triangles in the right panels show the velocity offsets for
  galaxies; filled triangles denote galaxies also shown in the left
  panel.}
\end{center}
\end{figure*}

\begin{figure*}
\begin{center}
\includegraphics[width=1.0\textwidth]{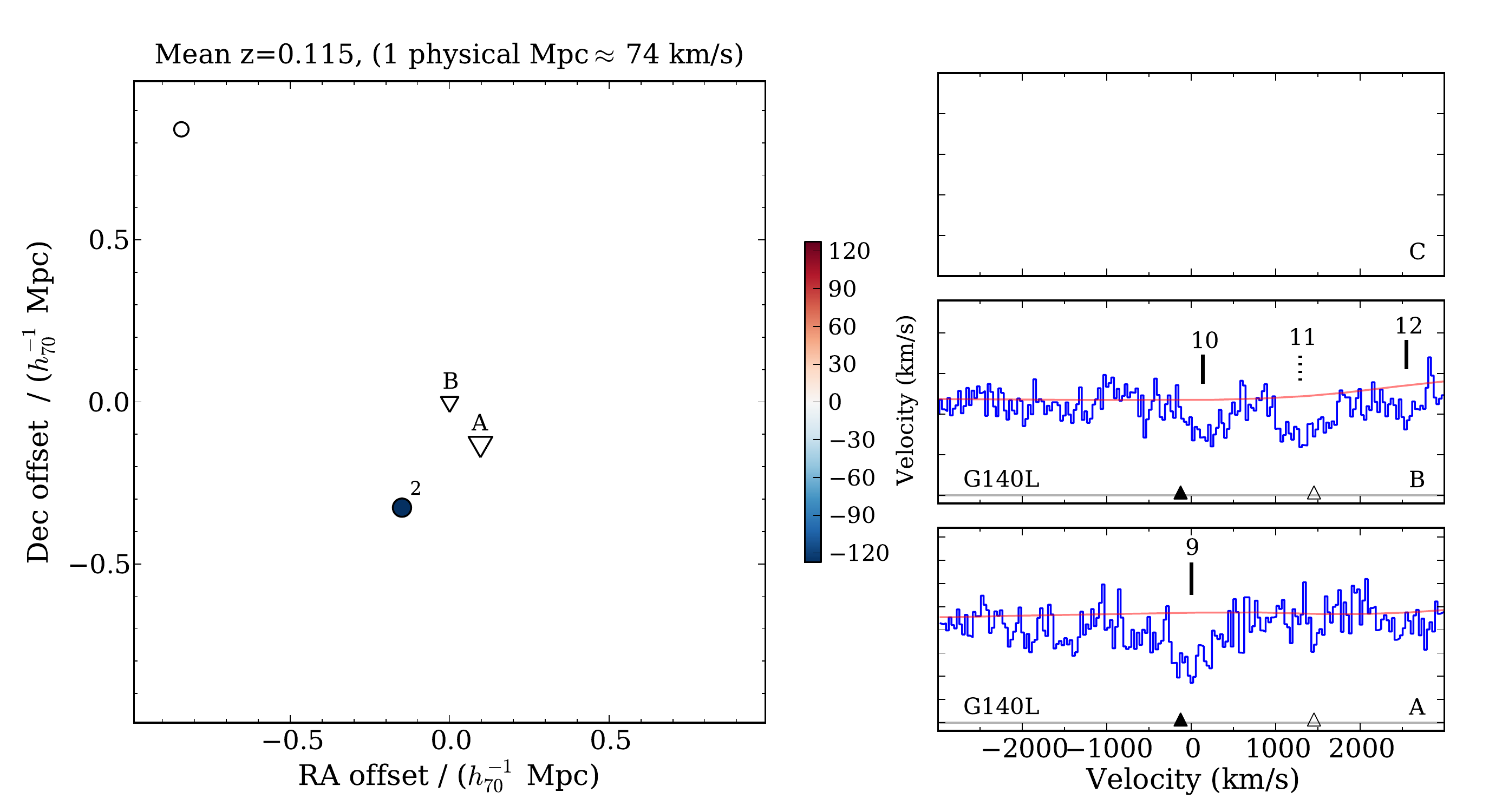}
Figure.~\ref{fig:galspec} --- continued.
\end{center}
\end{figure*}

\begin{figure*}
\begin{center}
\includegraphics[width=1.0\textwidth]{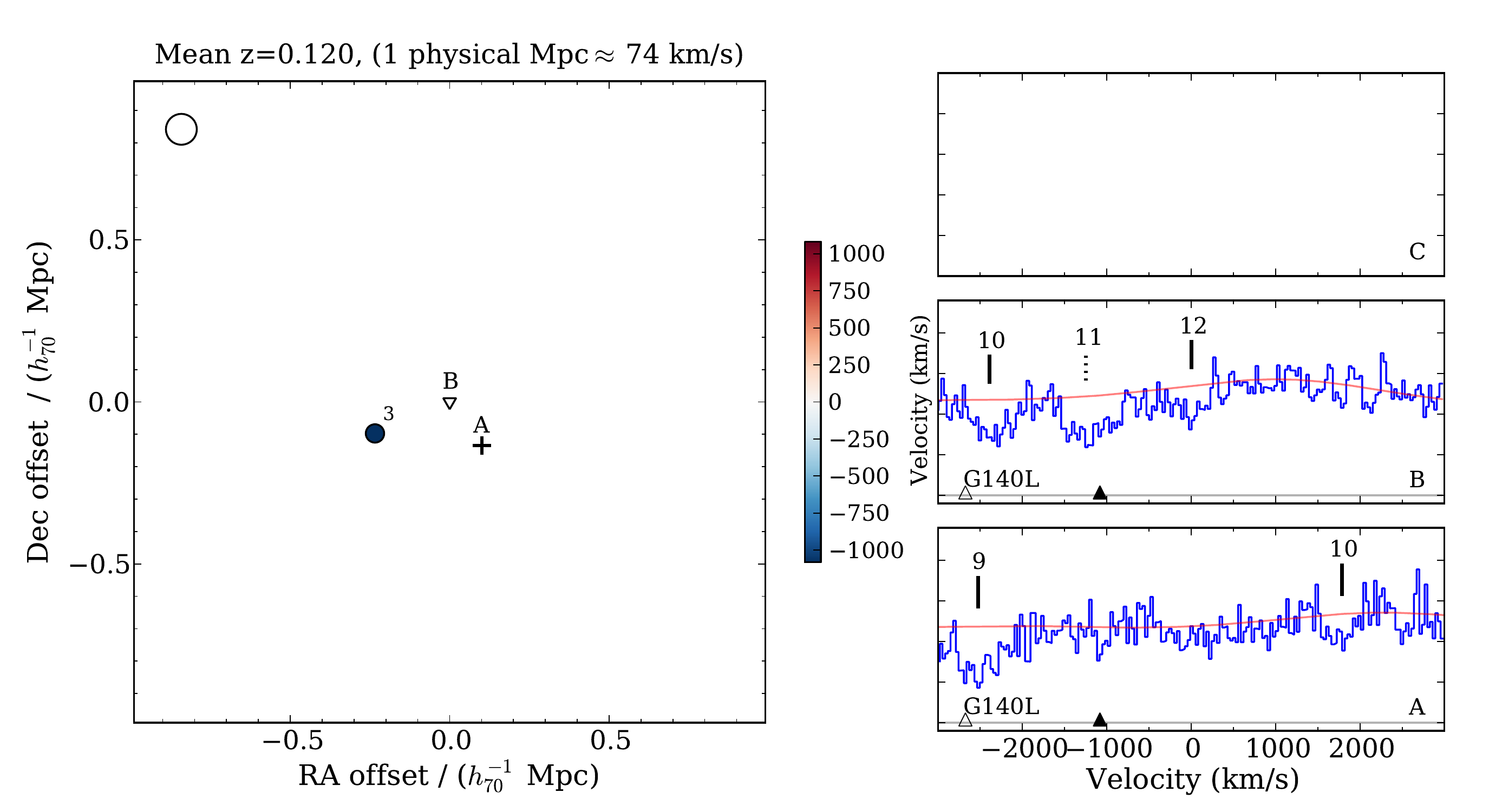}
Figure.~\ref{fig:galspec} --- continued.
\end{center}
\end{figure*}

\begin{figure*}
\begin{center}
\includegraphics[width=1.0\textwidth]{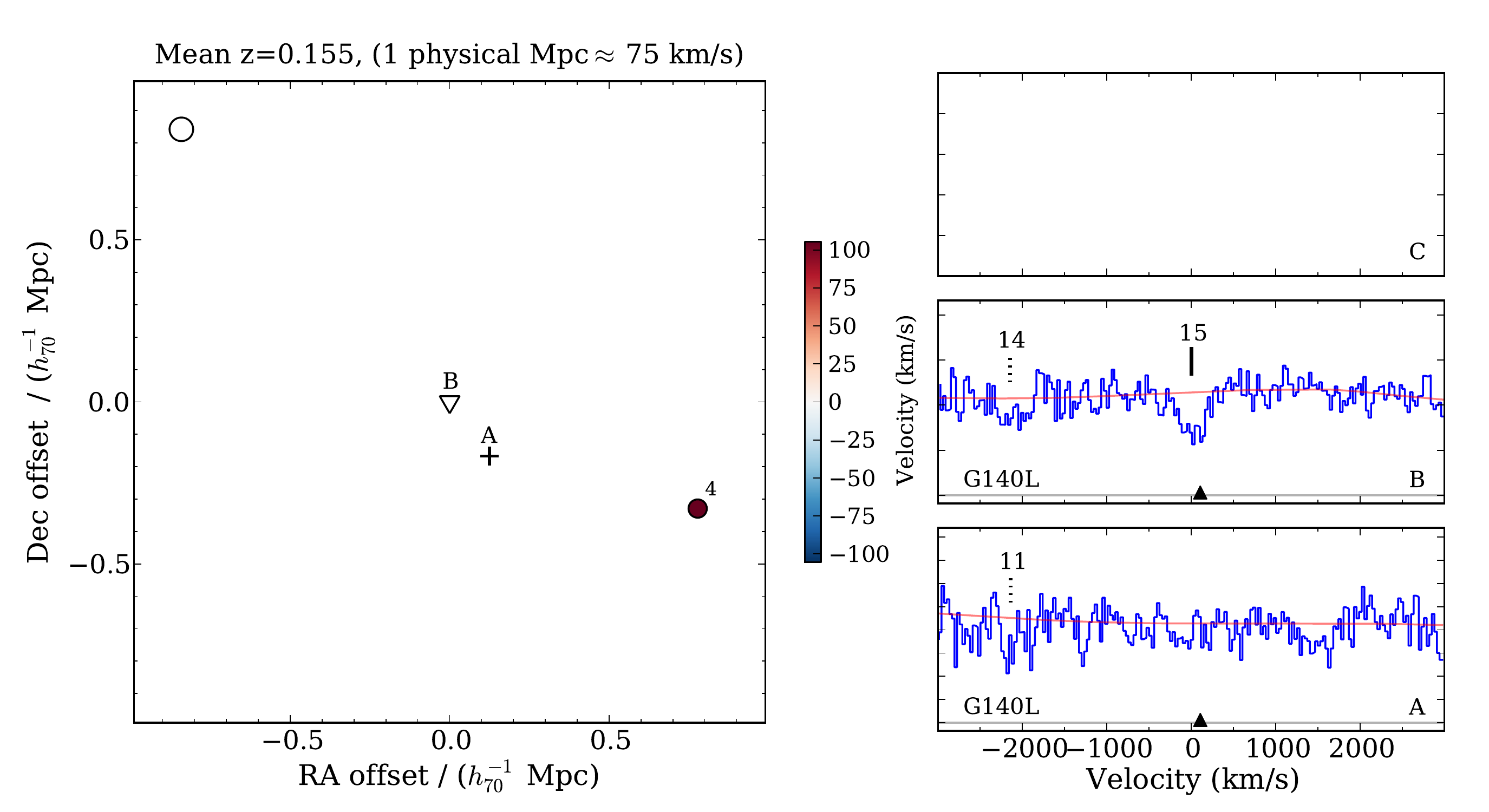}
Figure.~\ref{fig:galspec} --- continued.
\end{center}
\end{figure*}

\begin{figure*}
\begin{center}
\includegraphics[width=1.0\textwidth]{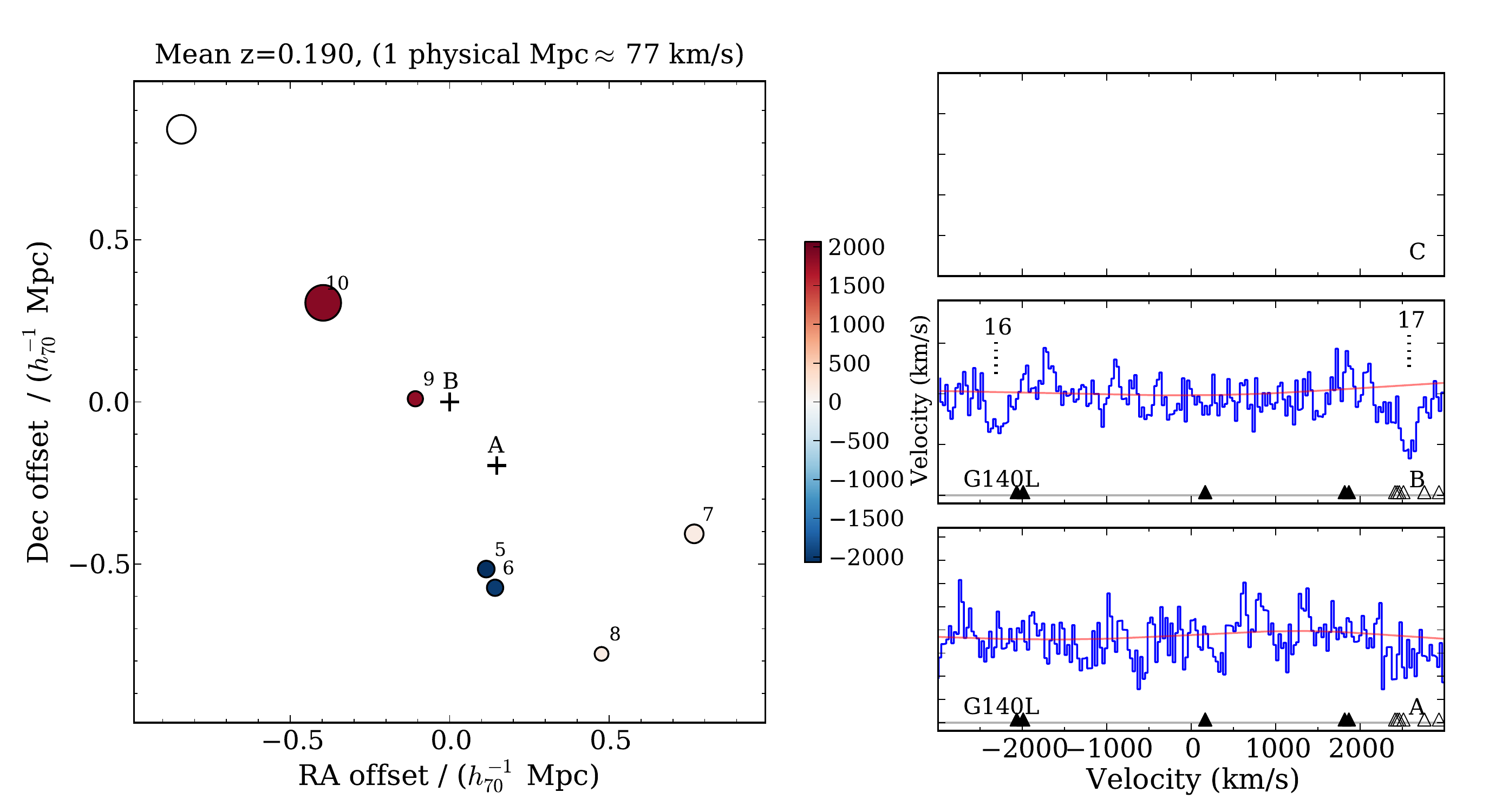}
Figure.~\ref{fig:galspec} --- continued.
\end{center}
\end{figure*}

\begin{figure*}
\begin{center}
\includegraphics[width=1.0\textwidth]{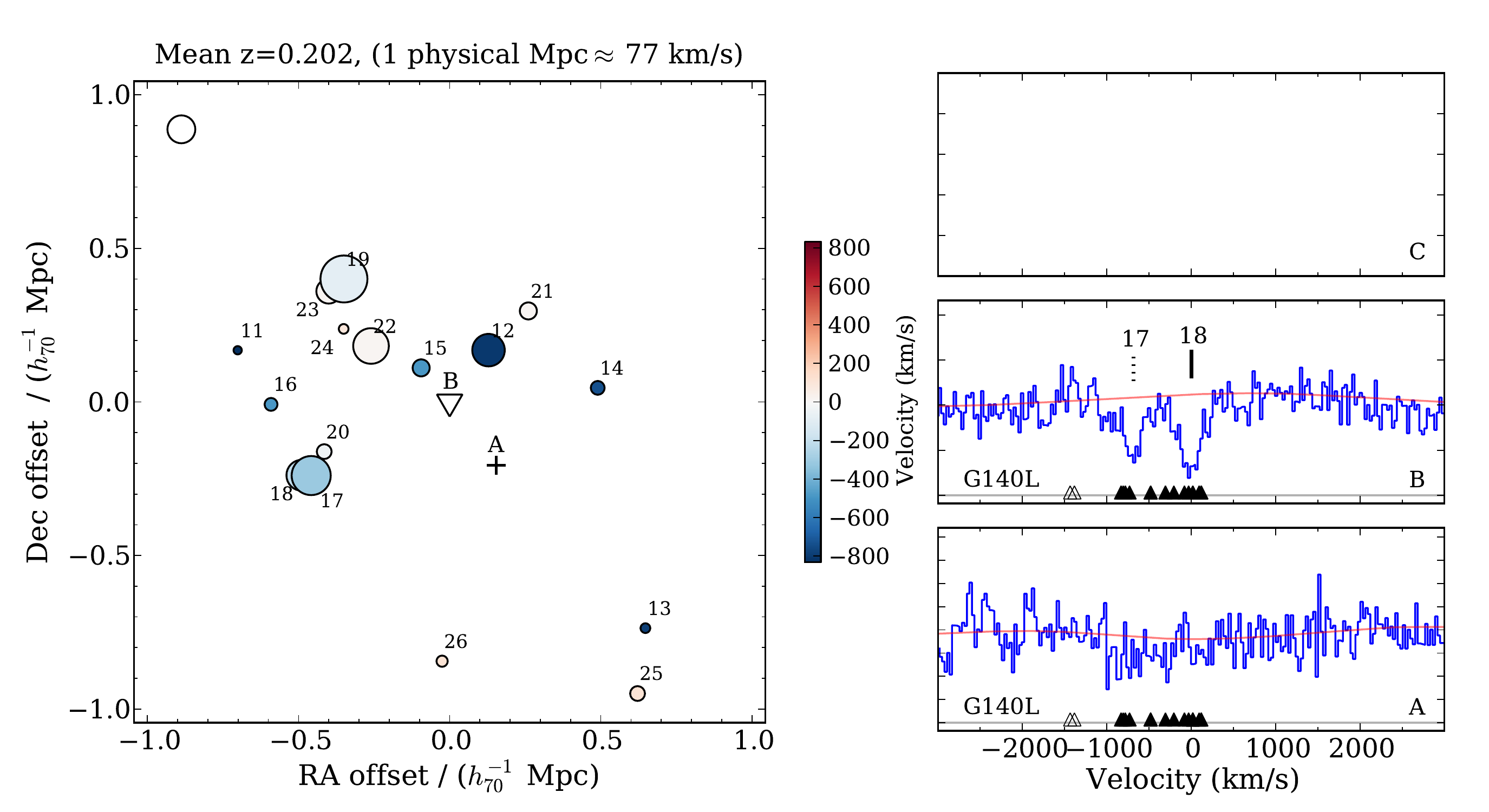}
Figure.~\ref{fig:galspec} --- continued.
\end{center}
\end{figure*}

\clearpage

\begin{figure*}
\begin{center}
\includegraphics[width=1.0\textwidth]{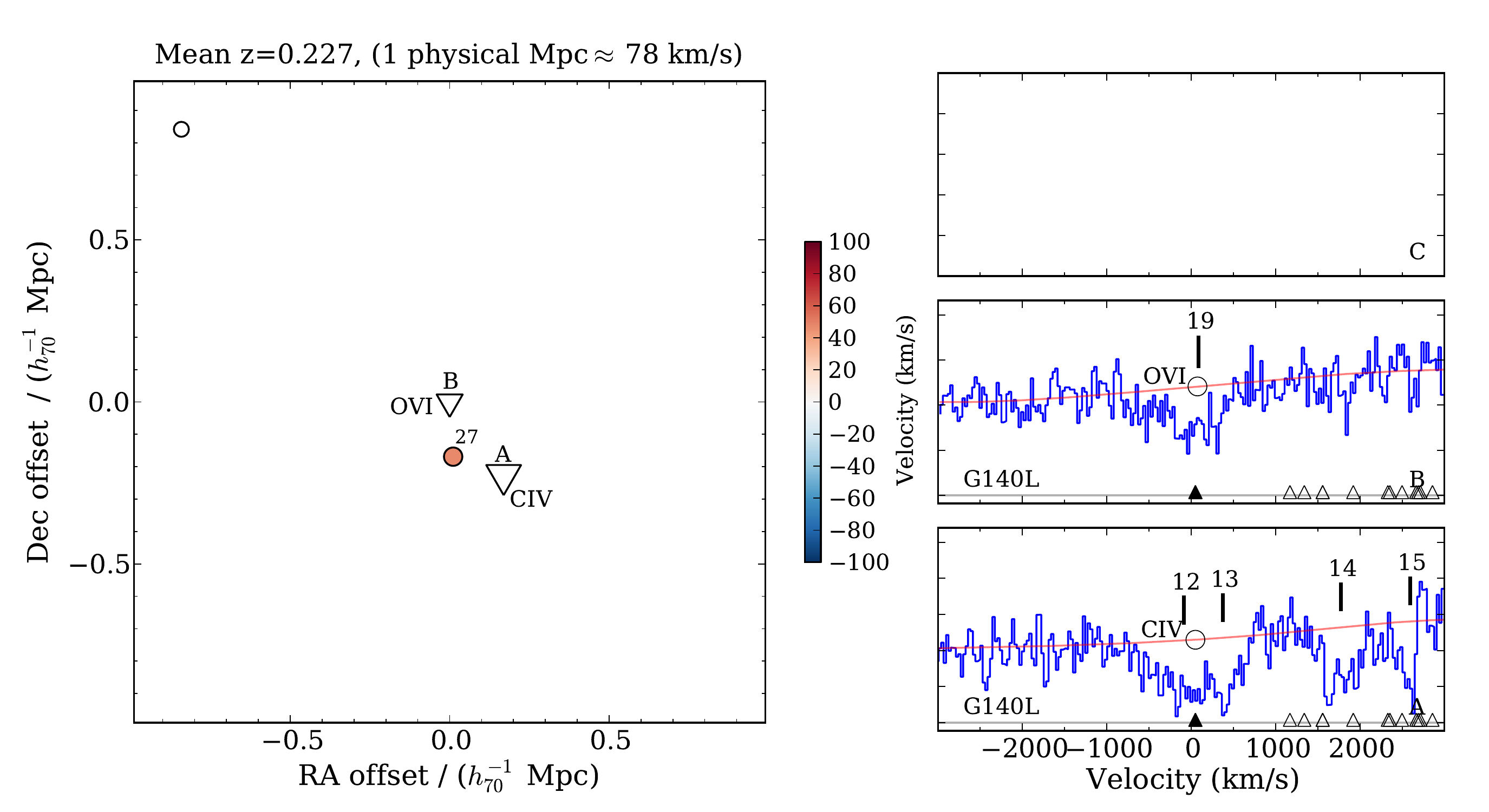}
Figure.~\ref{fig:galspec} --- continued. Circles in the right panels
show the positions of \CIV\ and an \OVI\ system in the A and B
spectra.
\end{center}
\end{figure*}

\begin{figure*}
\begin{center}
\includegraphics[width=1.0\textwidth]{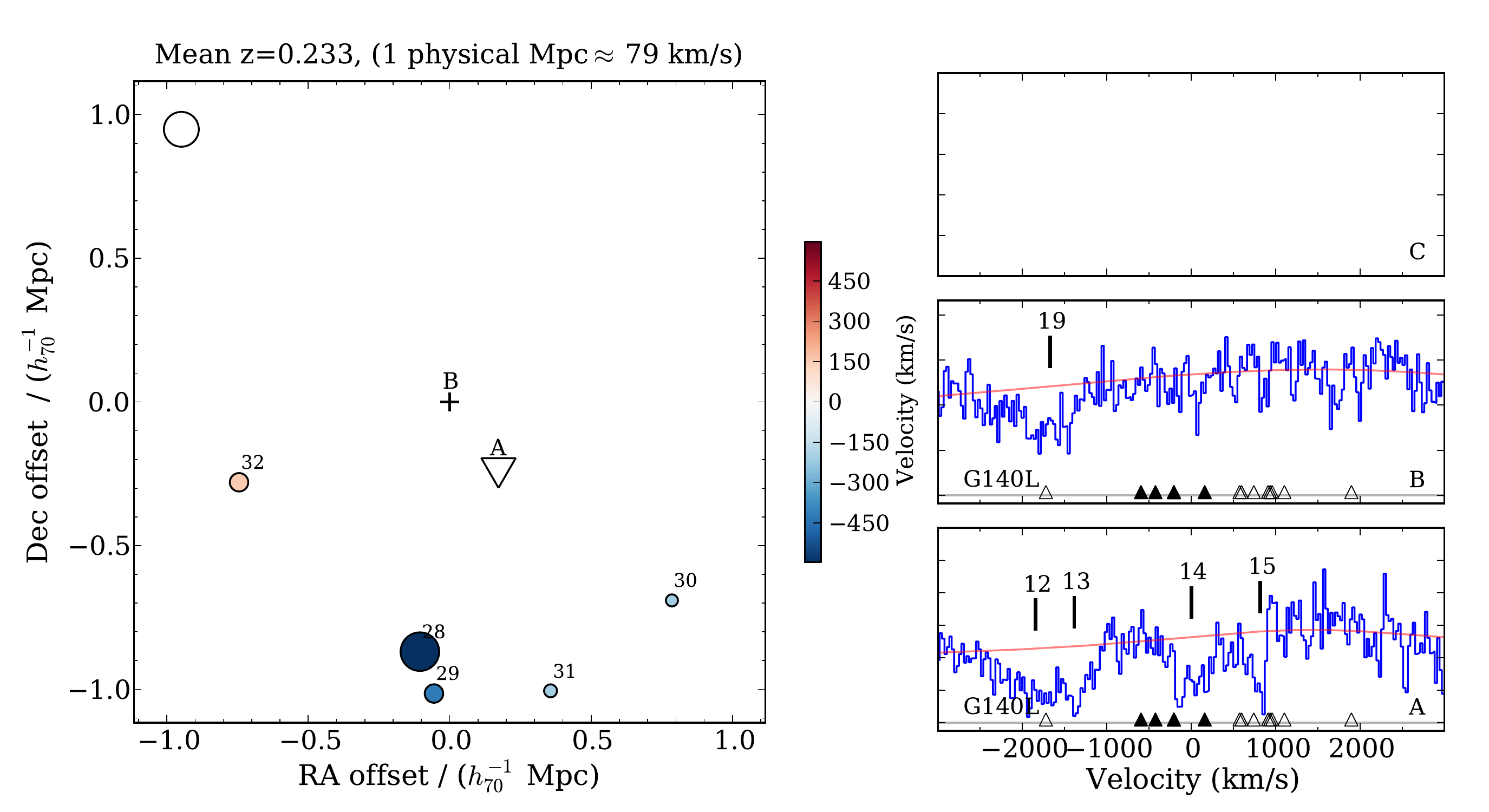}
Figure.~\ref{fig:galspec} --- continued.
\end{center}
\end{figure*}

\begin{figure*}
\begin{center}
\includegraphics[width=1.0\textwidth]{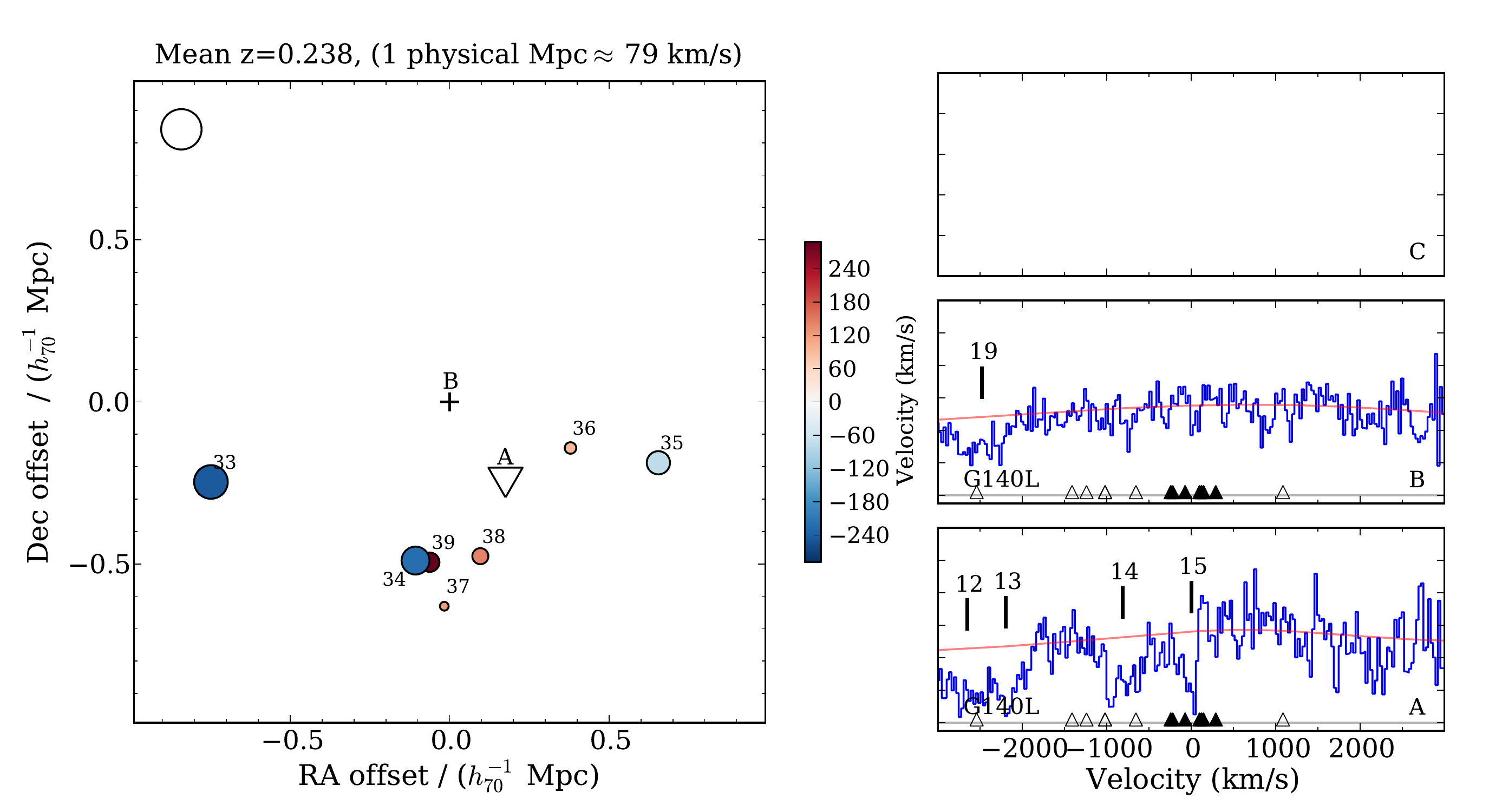}
Figure.~\ref{fig:galspec} --- continued.
\end{center}
\end{figure*}

\begin{figure*}
\begin{center}
\includegraphics[width=1.0\textwidth]{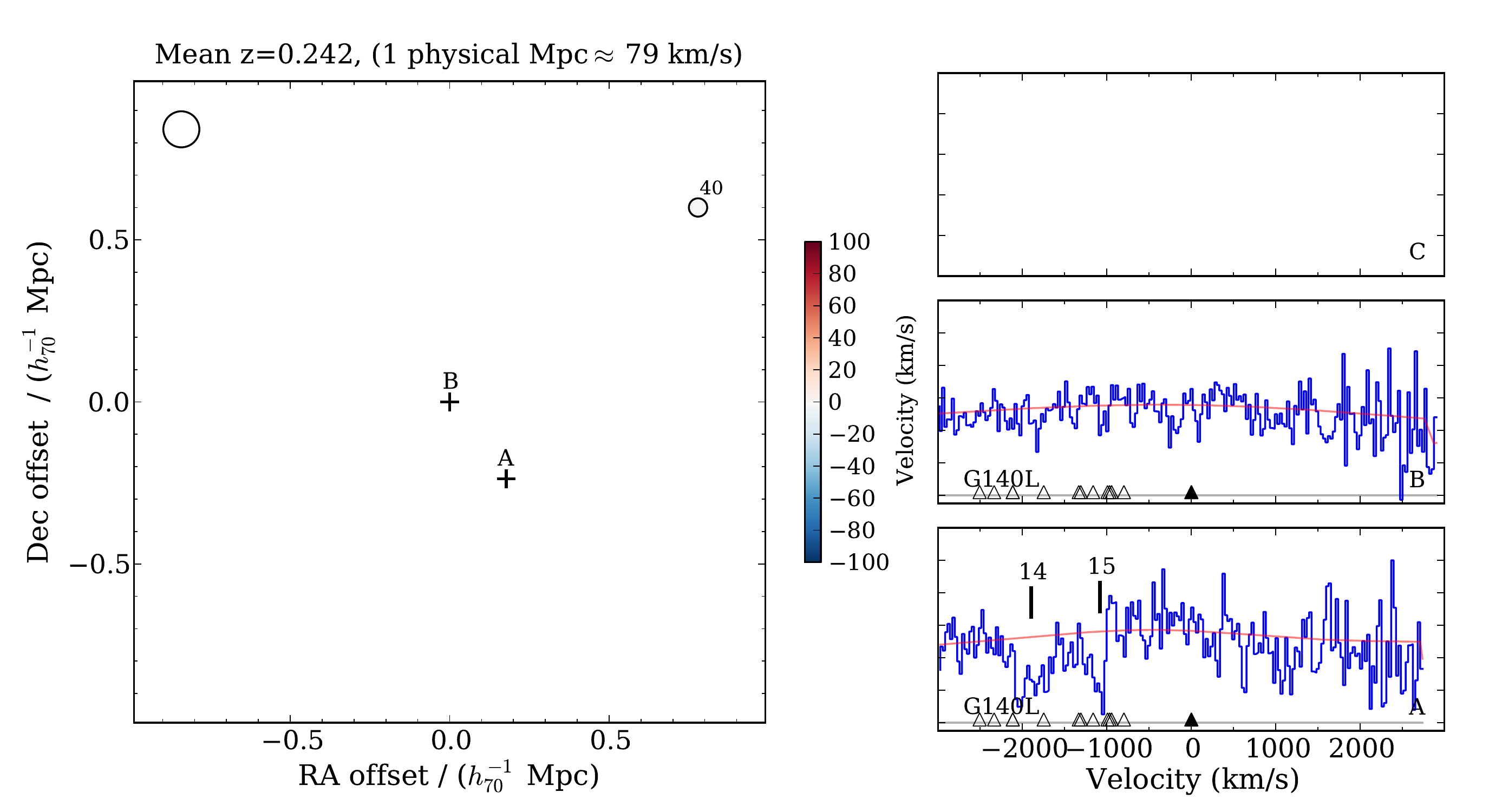}
Figure.~\ref{fig:galspec} --- continued.
\end{center}
\end{figure*}

\begin{figure*}
\begin{center}
\includegraphics[width=1.0\textwidth]{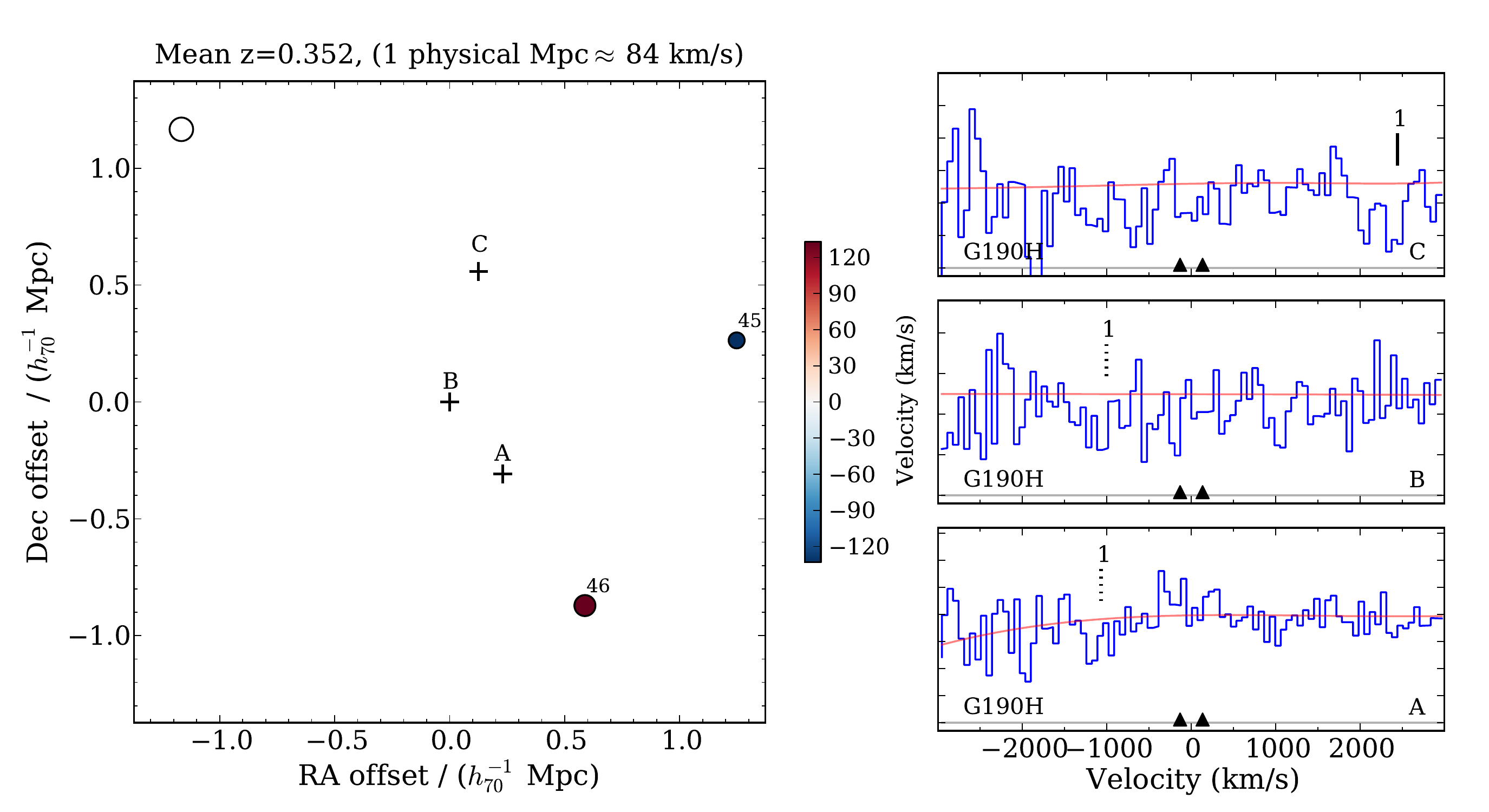}
Figure.~\ref{fig:galspec} --- continued.
\end{center}
\end{figure*}

\begin{figure*}
\begin{center}
\includegraphics[width=1.0\textwidth]{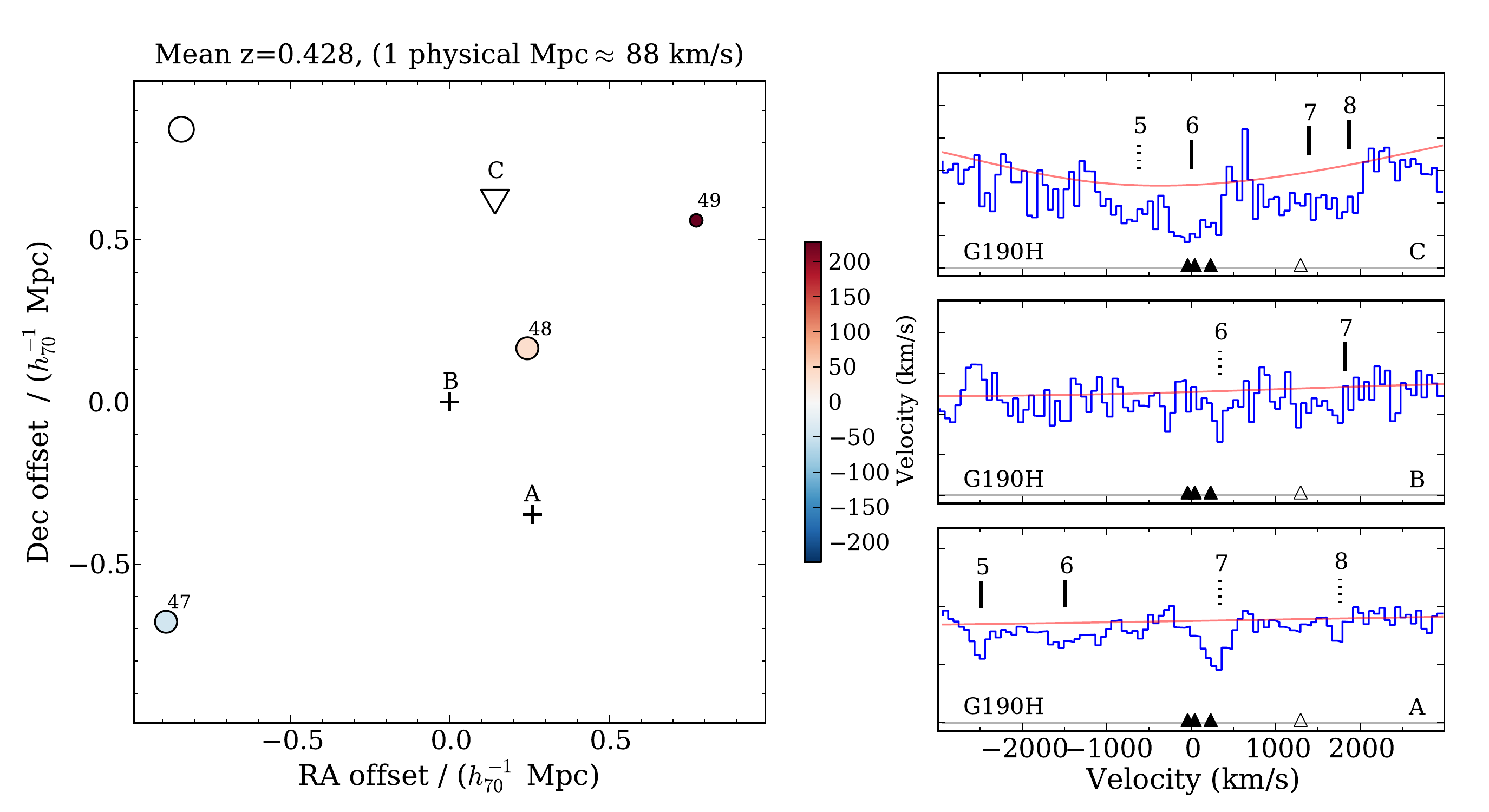}
Figure.~\ref{fig:galspec} --- continued.
\end{center}
\end{figure*}

\begin{figure*}
\begin{center}
\includegraphics[width=1.0\textwidth]{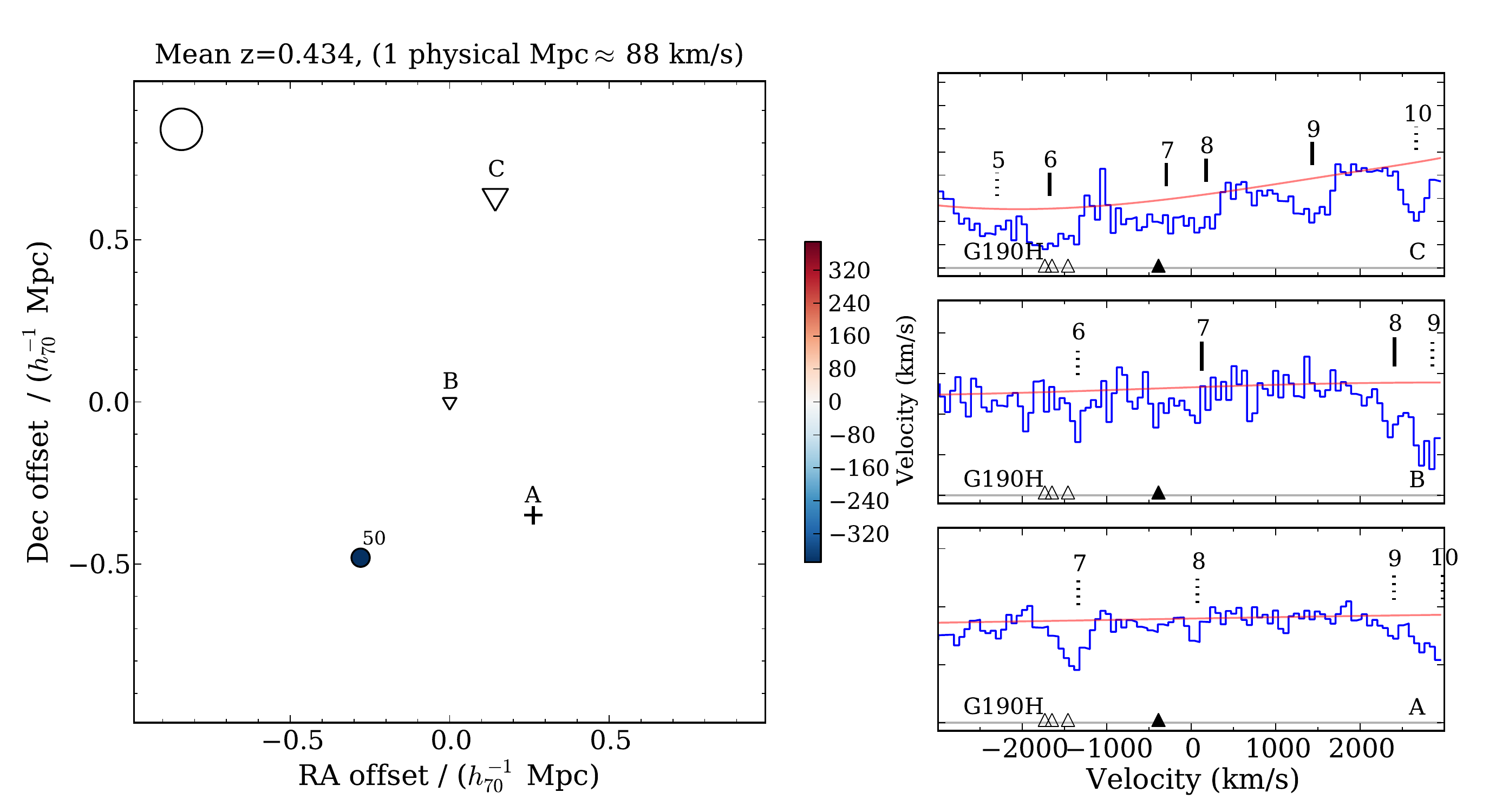}
Figure.~\ref{fig:galspec} --- continued.
\end{center}
\end{figure*}

\begin{figure*}
\begin{center}
\includegraphics[width=1.0\textwidth]{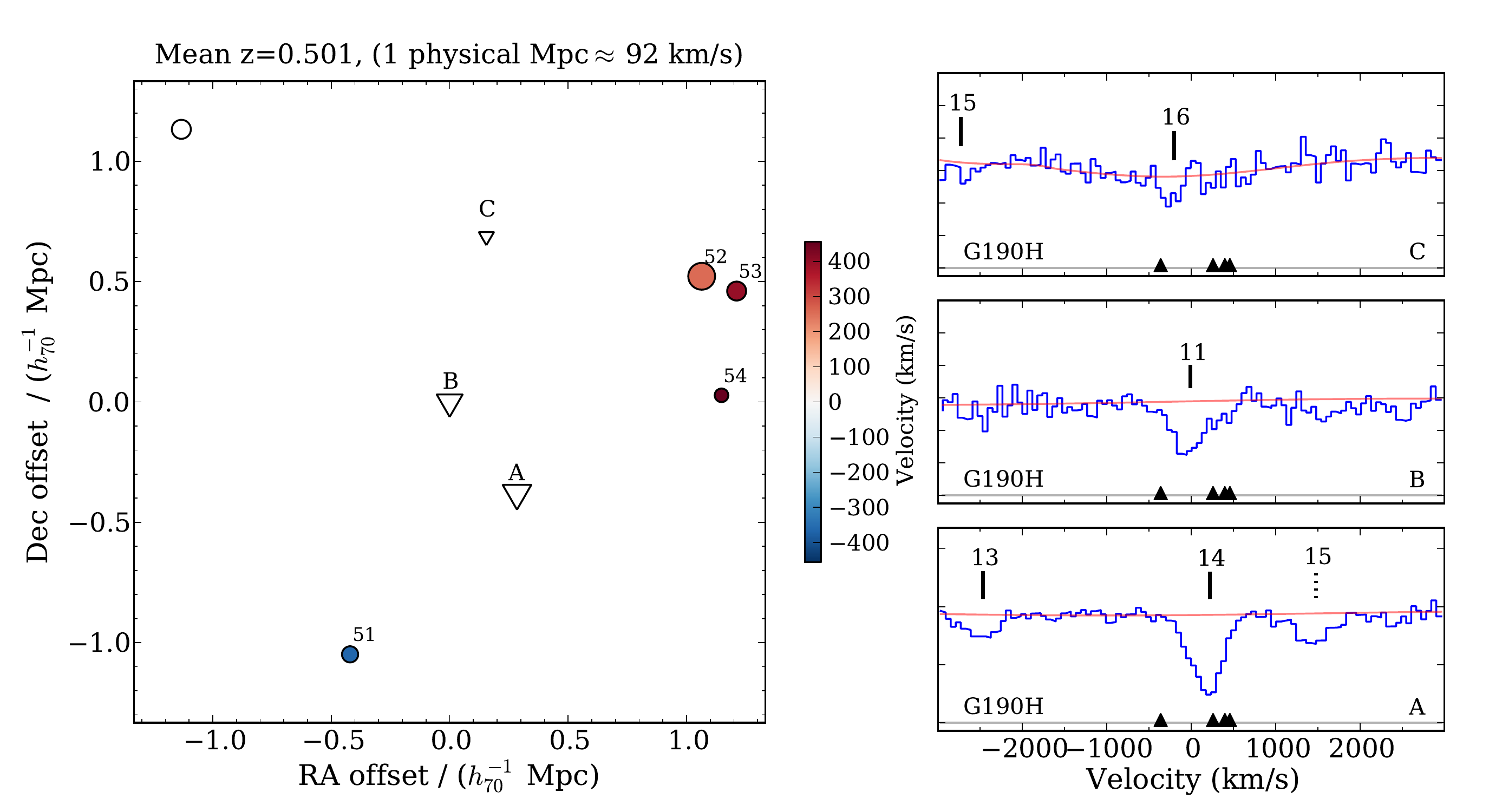}
Figure.~\ref{fig:galspec} --- continued.
\end{center}
\end{figure*}

\begin{figure*}
\begin{center}
\includegraphics[width=1.0\textwidth]{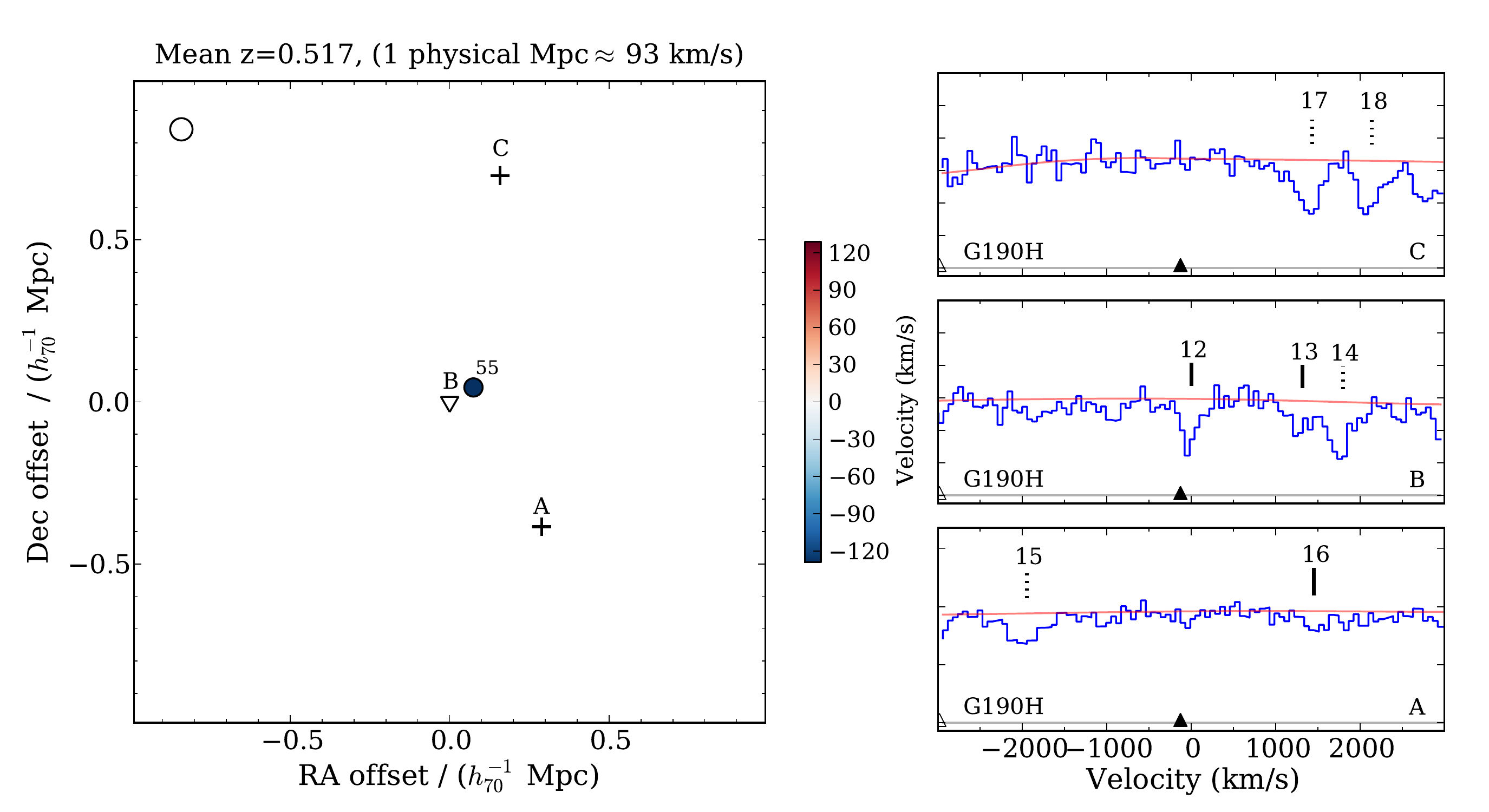}
Figure.~\ref{fig:galspec} --- continued.
\end{center}
\end{figure*}

\begin{figure*}
\begin{center}
\includegraphics[width=1.0\textwidth]{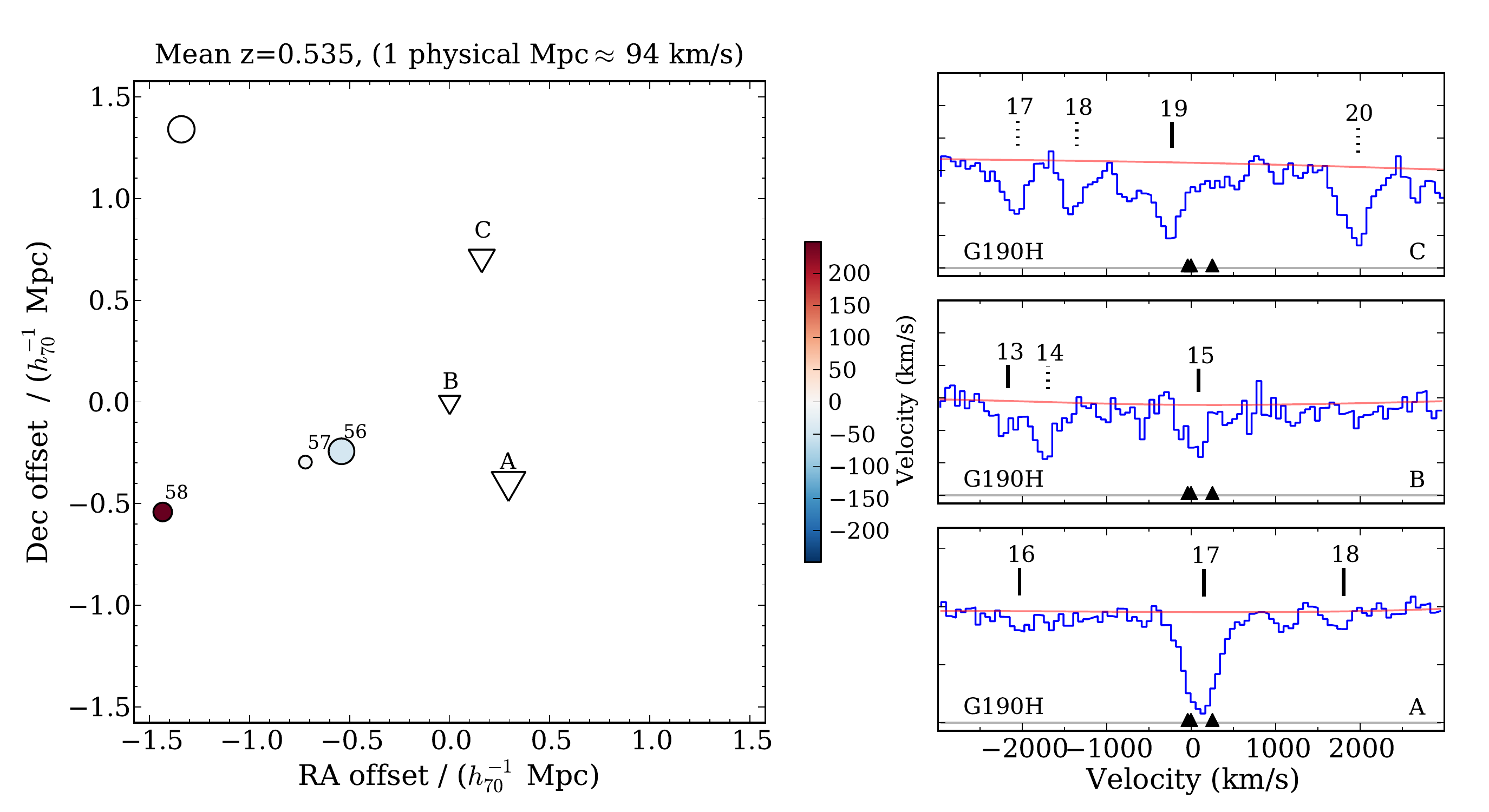}
Figure.~\ref{fig:galspec} --- continued.
\end{center}
\end{figure*}

\begin{figure*}
\begin{center}
\includegraphics[width=1.0\textwidth]{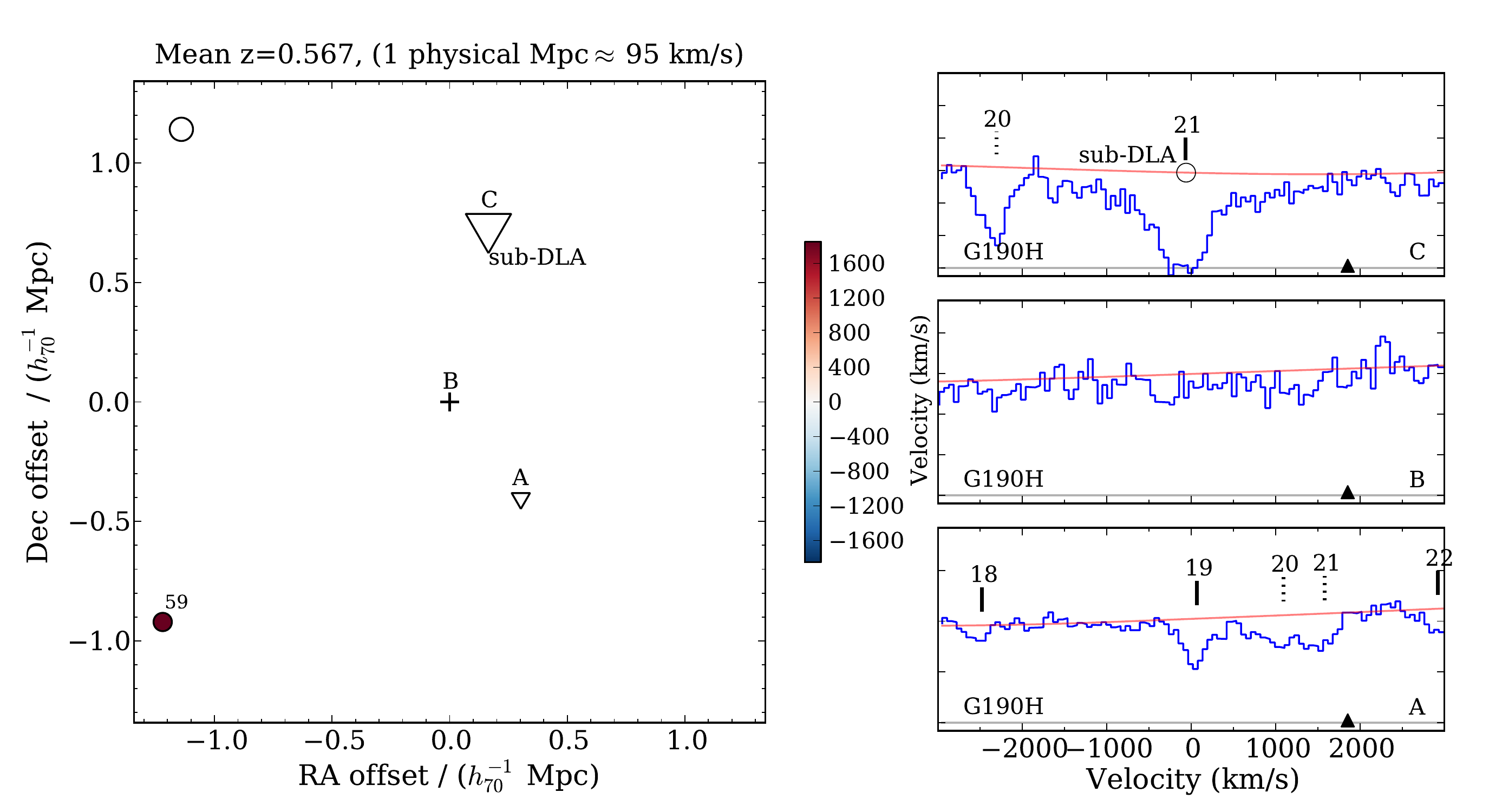}
Figure.~\ref{fig:galspec} --- continued. The position of the probable
sub-DLA in sightline C with strong low ionisation metal lines is
shown by a circle in the right-hand top panel.
\end{center}
\end{figure*}

\begin{figure*}
\begin{center}
\includegraphics[width=1.0\textwidth]{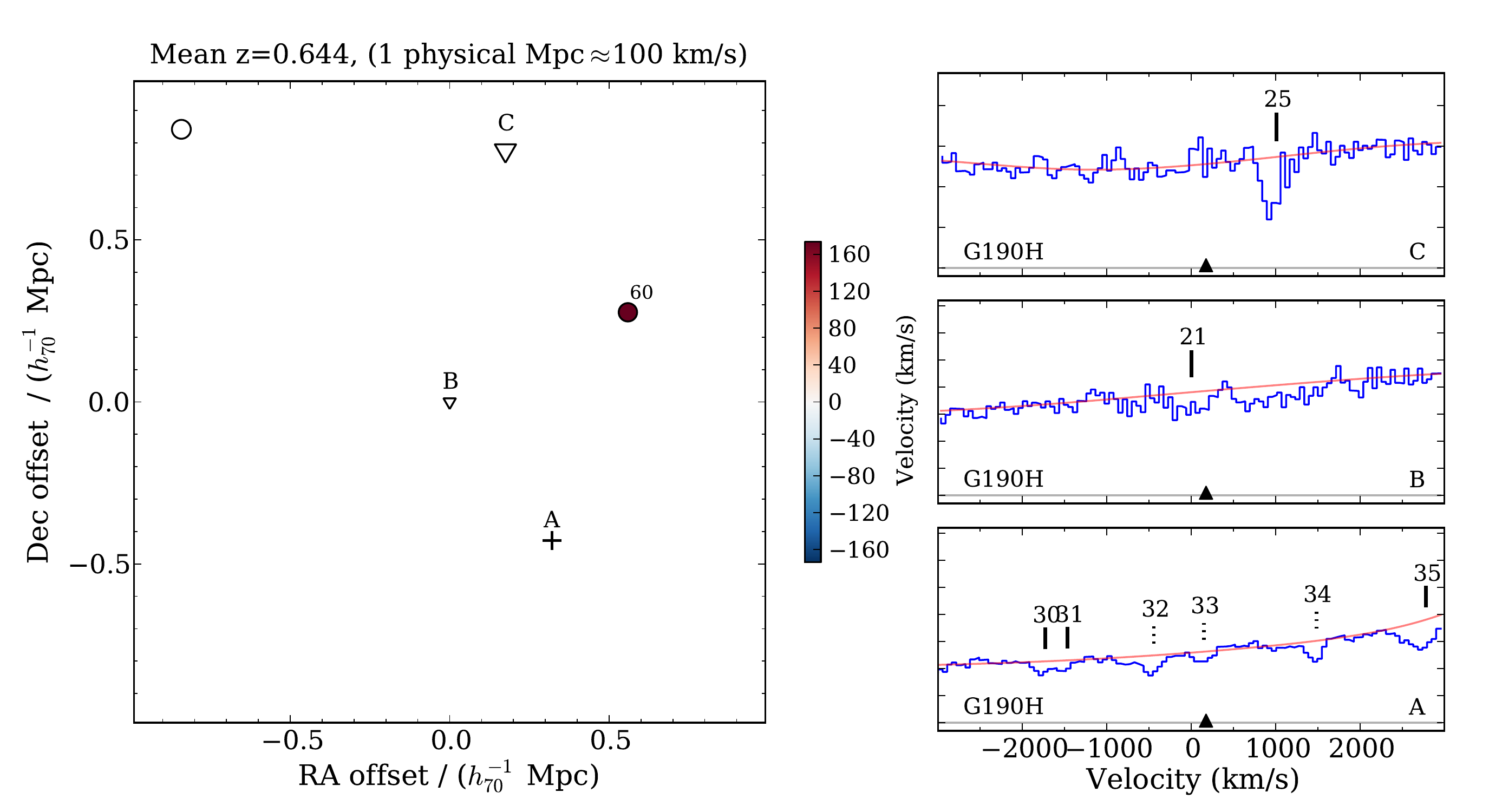}
Figure.~\ref{fig:galspec} --- continued.
\end{center}
\end{figure*}

\clearpage

\begin{figure*}
\begin{center}
\includegraphics[width=1.0\textwidth]{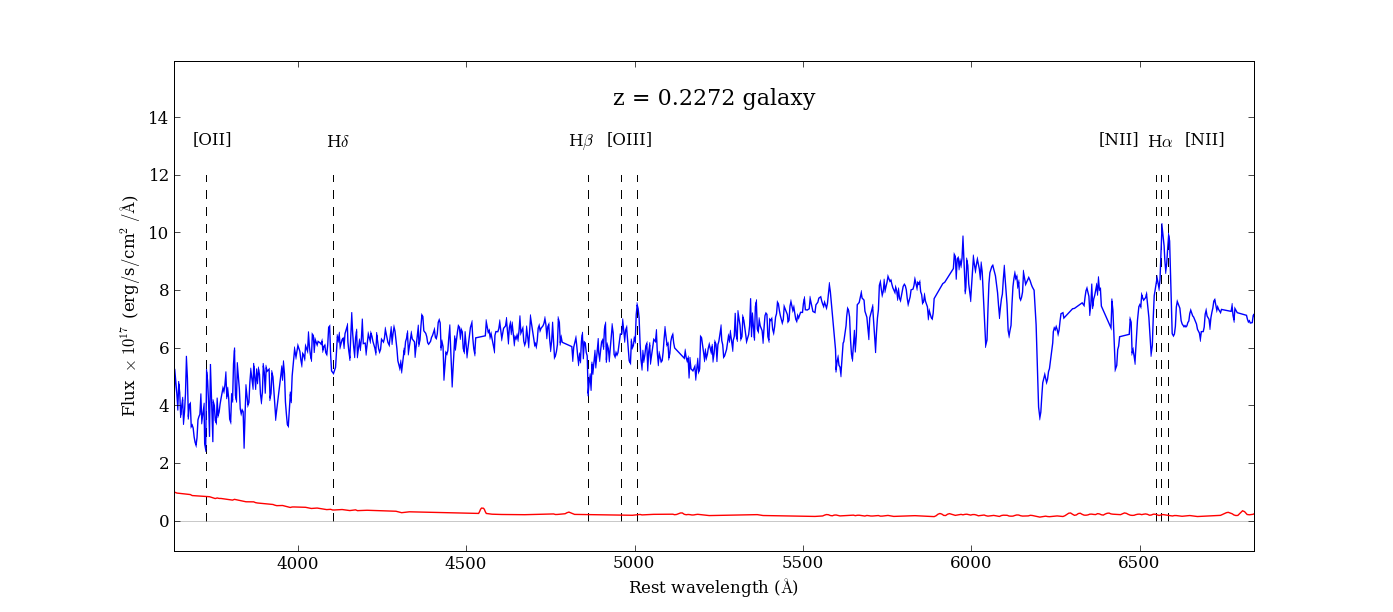}
\caption{\label{fig:closegalspec} \textsl{CFHT} MOS spectrum of the $z=0.2272$
  galaxy that has nearby metal absorption in QSO sightlines A and
  B. The flux is plotted as a function of rest wavelength, and the $1
  \sigma$ errors are shown. Features used in our analysis are
  labelled.}
\end{center}
\end{figure*}

\begin{figure*}
\begin{center}
\includegraphics[width=1.0\textwidth]{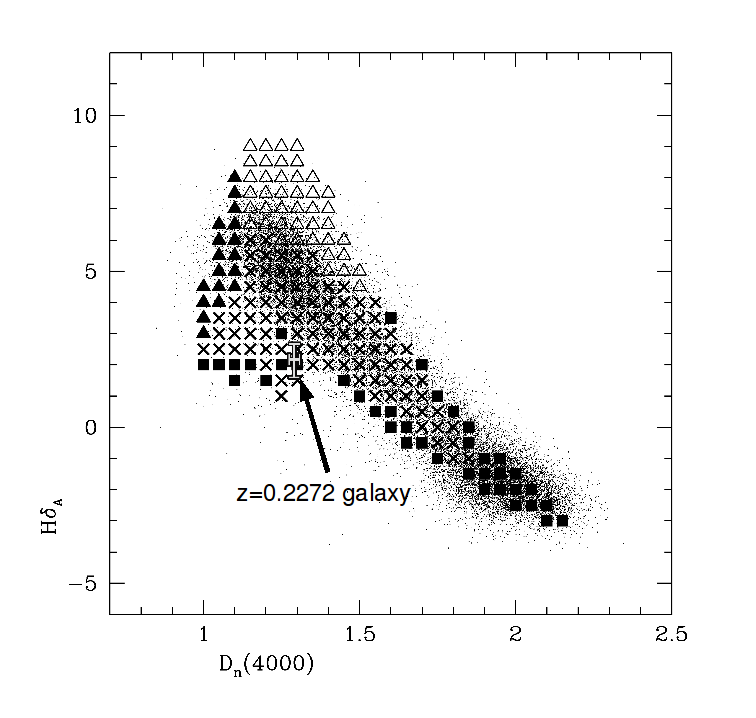}
\caption{\label{fig:kauf} \Hd$_A$ and D(4000) values for the
  $z=0.2272$ galaxy (grey cross) overlayed on a reproduction of
  Figure~6 from \citep{Kauffmann03}. Triangles represent regions where
  95\% of model galaxies have at least 5 percent of their stellar mass
  made up of stars produced in bursts of star formation over the last
  two Gyr. Open triangles represent galaxies where the most recent
  burst event began $<0.1$~Gyr ago; filled triangles correspond to
  galaxies where the most recent burst began $> 0.1$~Gyr ago. Squares
  show regions where 95\% of models have no stars produced in bursts
  of star formation over the last two Gyr. Crosses are other regions
  described by the model galaxies.}
\end{center}
\end{figure*}

\begin{figure}
\begin{center}
\includegraphics[width=0.49\textwidth]{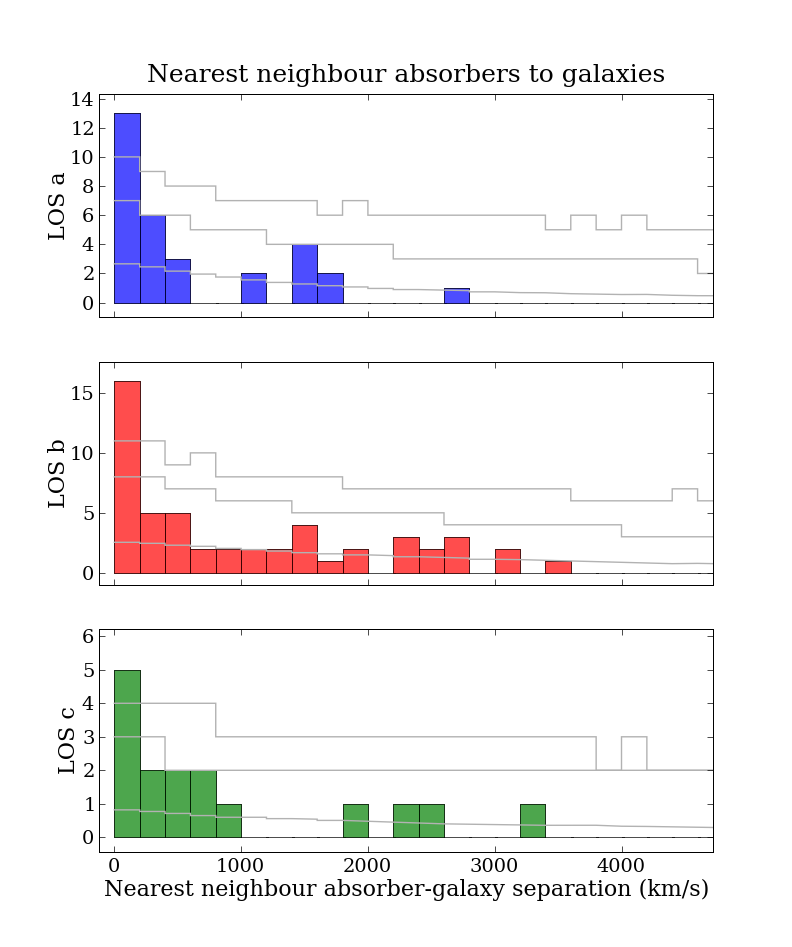}
\caption{\label{fig:NNag} The number of galaxy-absorber nearest
  neighbour pairs (one for each galaxy) as a function of pair velocity
  separation for each sightline. The velocity bins are 200~\kms\
  wide. In each panel the thin grey lines show the mean number of
  pairs constructed using the same galaxy sample with 5000 random sets
  of absorbers (bottom line), the 95\% confidence level from the
  random pairs (middle), and the 99\% confidence level (top).  There
  is a significant ($>99$\%) excess of pairs compared to the random
  pairs at velocity separations $<200$~\kms\ in all three sightlines.}
\end{center}
\end{figure}

\begin{figure}
\begin{center}
\includegraphics[width=0.49\textwidth]{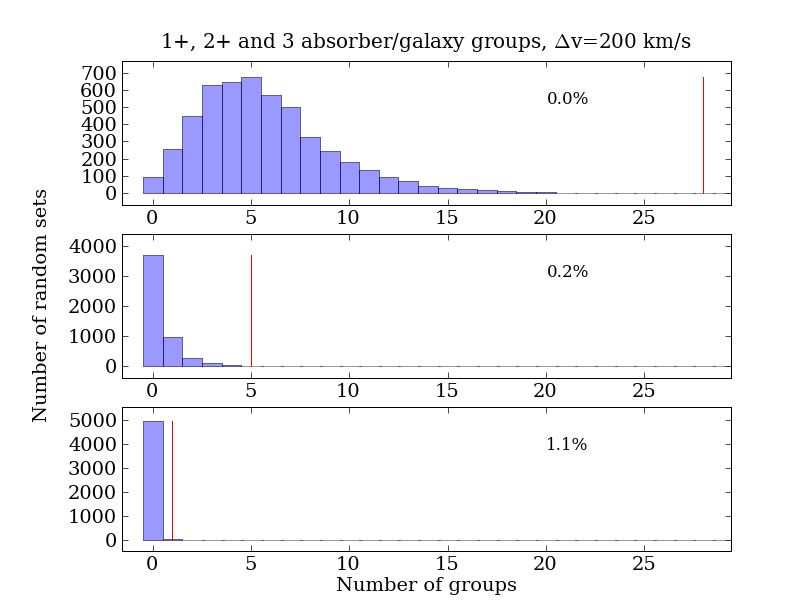}
\caption{\label{fig:gabs200} The number of galaxy-absorber `groups'
  (defined in Section~\ref{sec:stat-tests-absorb}) using the real
  galaxies and real absorbers and maximum velocity difference of
  200~\kms, compared to the number measured using the real galaxies
  and 5000 sets of random absorbers. The histogram shows the number of
  galaxies with associated absorbers in at least one (top), two
  (middle) and three (bottom) sightlines for random absorbers.  The
  vertical line shows the number of groups using real absorbers. The
  probability of the real number arising by chance is shown in each
  panel (0.0\% means none of the 5000 galaxy-random absorber sets had
  that many groups, i.e. $< 0.02$\%). There is a clear excess of
  galaxies associated with absorbers across one and two sightlines
  compared to random absorbers.}
\end{center}
\end{figure}

\begin{figure}
\begin{center}
\includegraphics[width=0.49\textwidth]{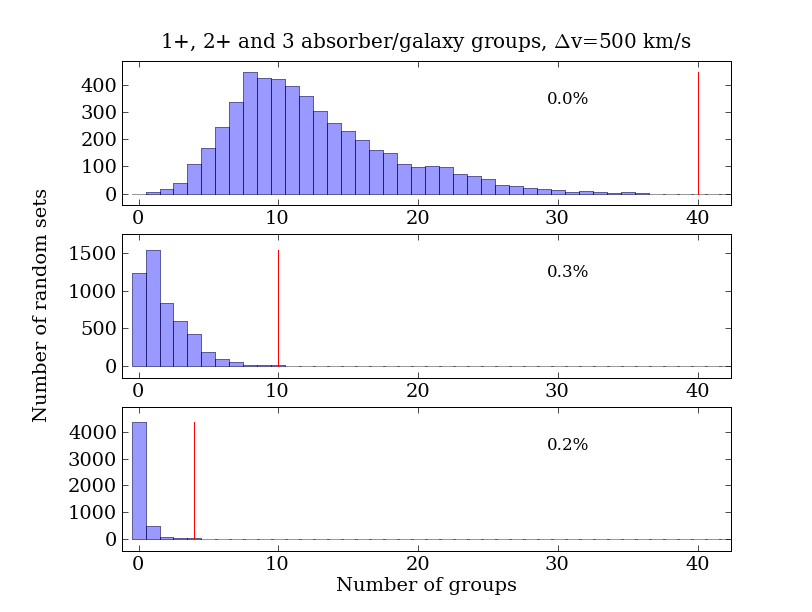}
\caption{\label{fig:gabs500} As for Figure~\ref{fig:gabs200}, but using a
  maximum velocity difference of 500~\kms. }
\end{center}
\end{figure}

\begin{figure}
\begin{center}
\includegraphics[width=0.49\textwidth]{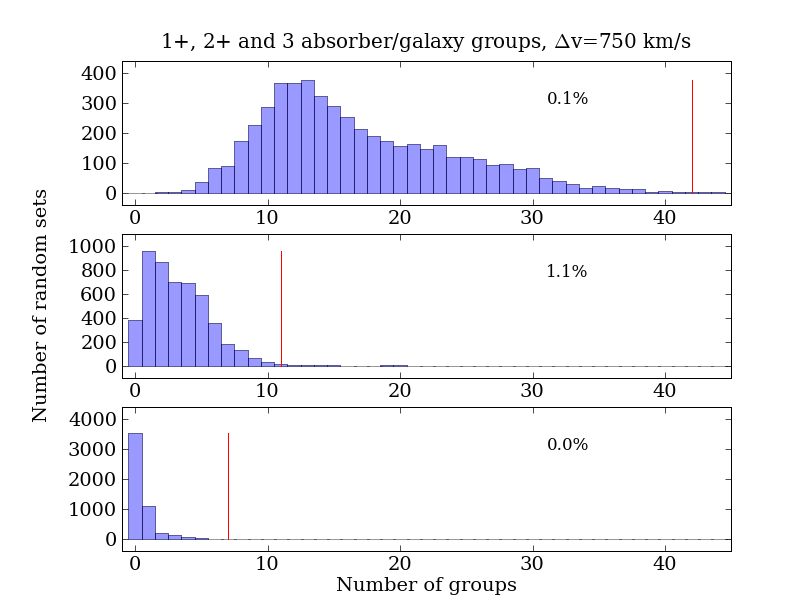}
\caption{\label{fig:gabs750} As for Figure~\ref{fig:gabs200}, but
  using a maximum velocity difference of 750~\kms. }
\end{center}
\end{figure}

\begin{figure}
\begin{center}
\includegraphics[width=0.49\textwidth]{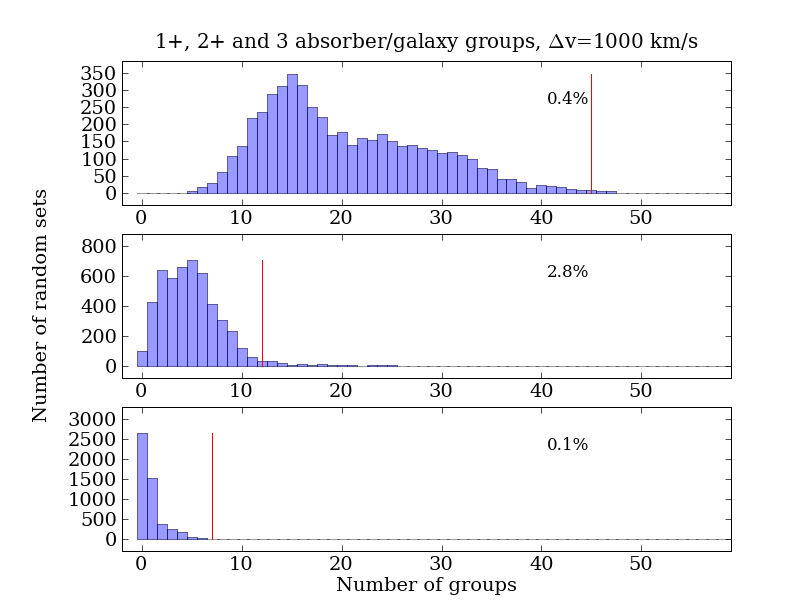}
\caption{\label{fig:gabs1000} As for Figure~\ref{fig:gabs200}, but
  using a maximum velocity difference of 1000~\kms.}
\end{center}
\end{figure}


\begin{figure*}
\begin{center}
\includegraphics[width=1.0\textwidth]{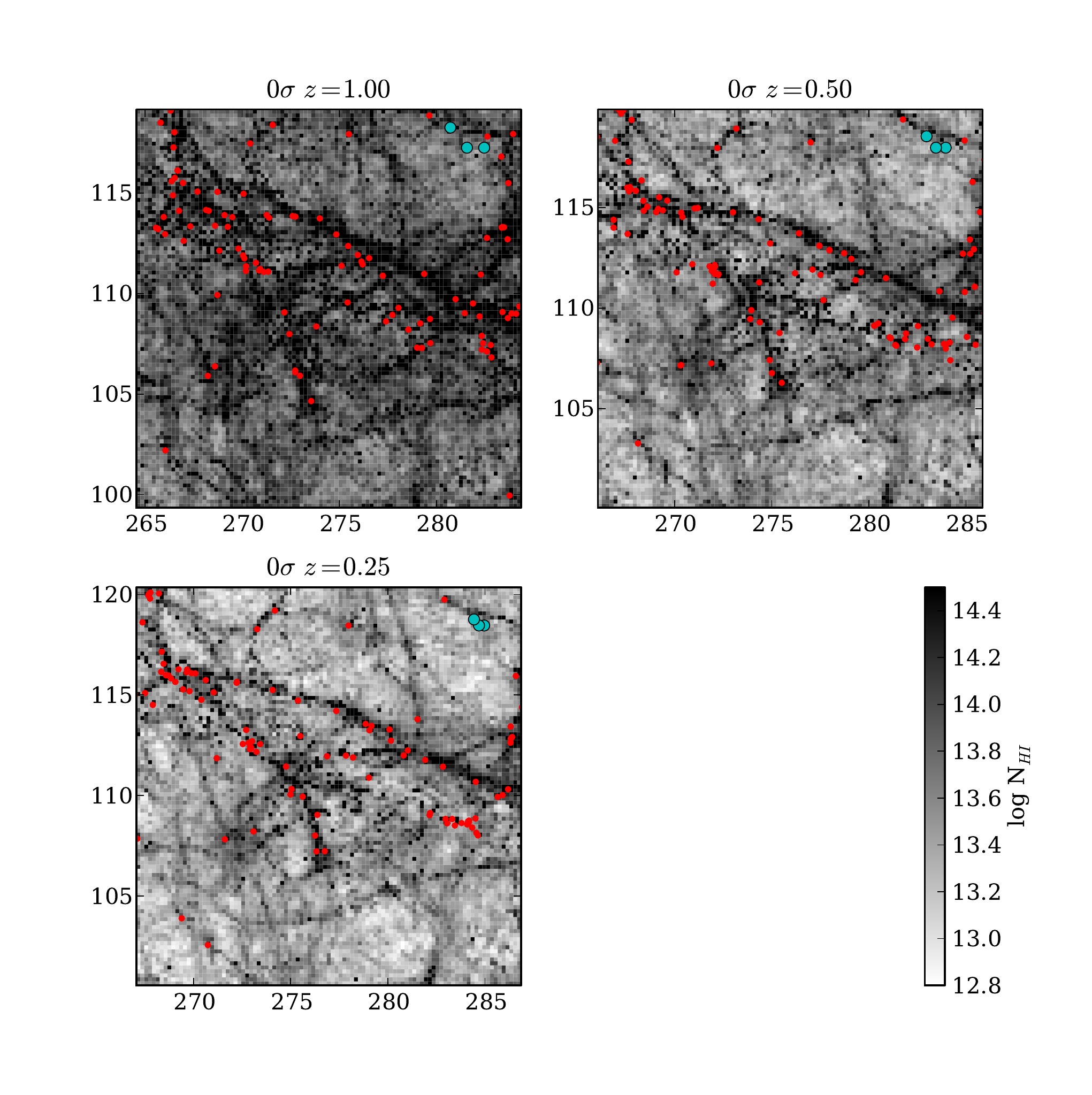}
\caption{\label{fig:sim_nhi} The X-Y positions of gas and galaxies in
  the mean density region of the \gimic\ simulation at redshifts 0.25,
  0.5 and 1. Coordinates are in $h^{-1}$ comoving Mpc and refer to the
  original Millennium positions. The dots show galaxies with stellar
  mass larger than $6\times10^{10}$ \msun. All galaxies in the
  20~$h^{-1}$~Mpc cubic region are plotted. Grayscale shows the total
  \NHI\ (\cmm) in $200 h^{-1}$ kpc $\times$ $200 h^{-1}$ kpc
  (comoving) bins, summed along the Z axis.  The group of three dots
  in the top right of each panel shows the configuration of triple
  sightlines at that redshift.}
\end{center}
\end{figure*}

\begin{figure*}
\begin{center}
\includegraphics[width=1.0\textwidth]{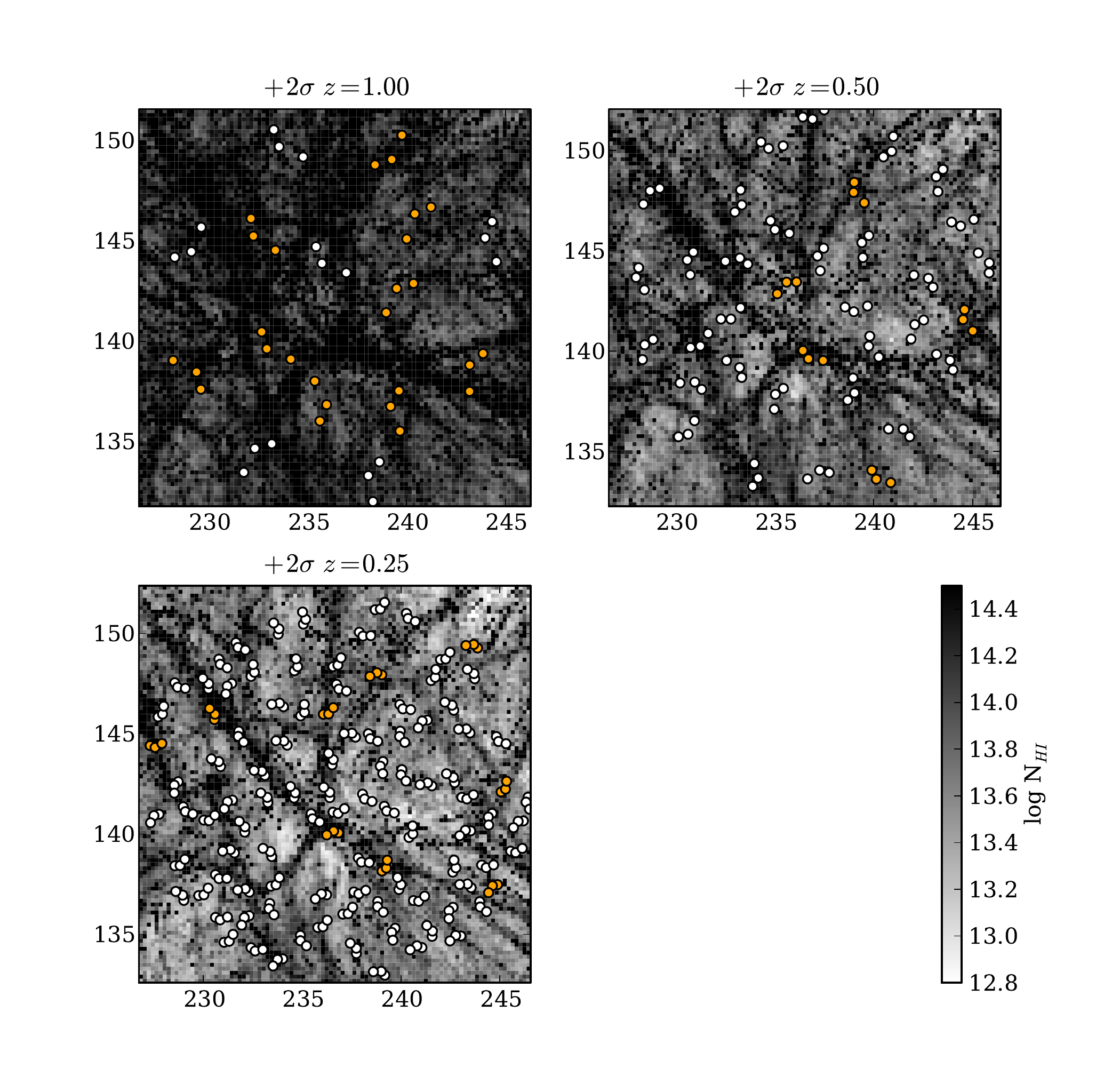}
\caption{\label{fig:sim_nhi_trip} Similar to Figure~\ref{fig:sim_nhi},
  but here dots show a sample of random triple sightlines through the
  $+2\sigma$ density region. The sightlines denoted by darker dots
  contain triple LOS-galaxy groups. At $z=1$, these groups fall inside
  and outside filamentary structures. At $z=0.5$ and $z=0.25$, as the
  comoving separation of the triplets and the characteristic column
  density of filamentary structures drops, the groups arise in knots
  and filaments.}
\end{center}
\end{figure*}

\begin{figure}
\begin{center}
\includegraphics[width=0.49\textwidth]{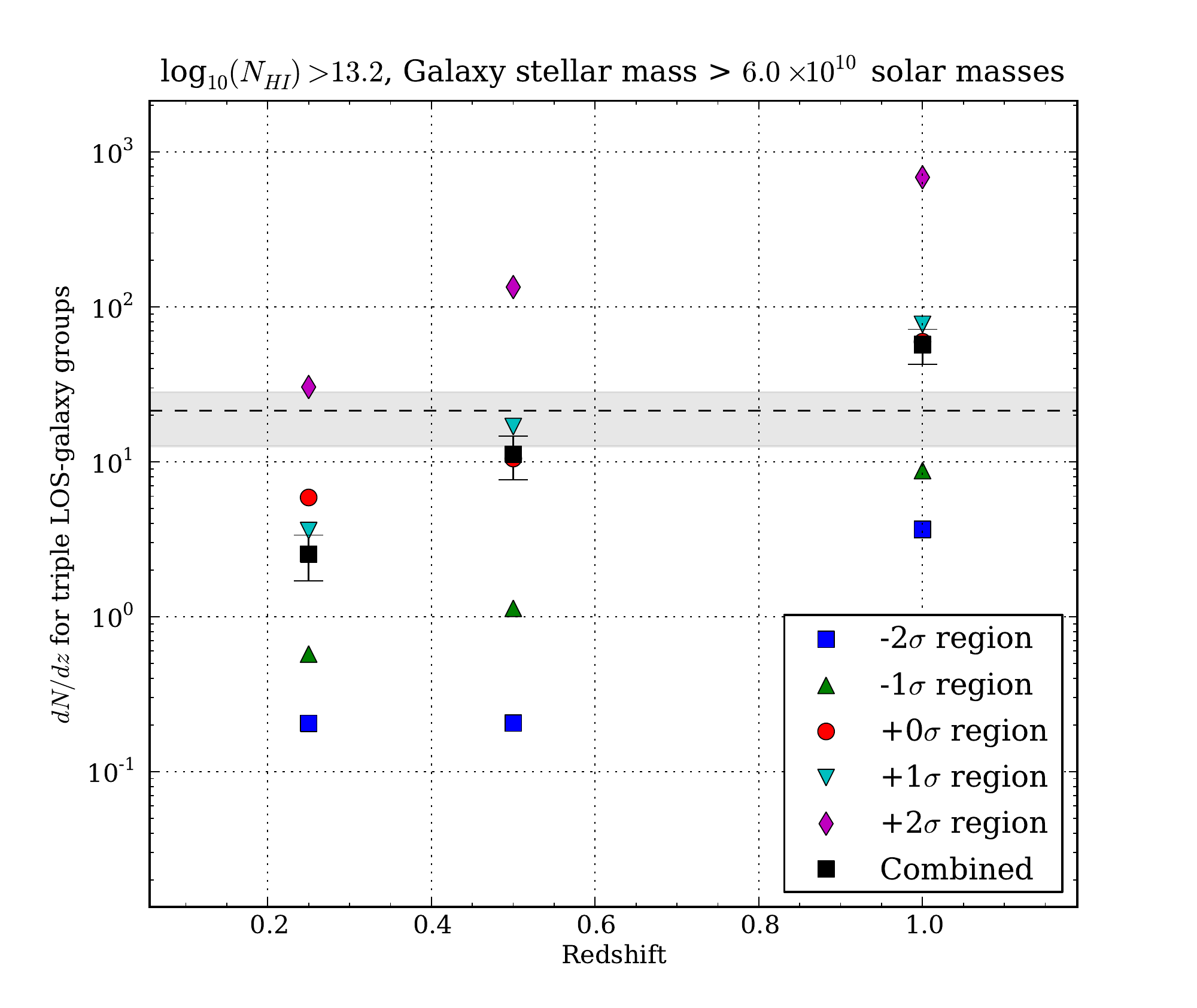}
\caption{\label{fig:dgdz} The number of galaxies with associated
  absorbers across all three sightlines (triple LOS-galaxy groups) per
  unit redshift in the simulations, as a function of redshift.  The
  numbers are shown for each region, and combined across all regions,
  weighting each region as described in
  Section~\ref{sec:comp-with-simul} (dark squares). The dotted line
  and shaded region show the observed incidence of such groups and
  $1\sigma$ Poisson error in the number of observed groups. The error
  bars show $1\sigma$ Poisson errors for the number of groups found in
  the simulations. The mean redshift of the observations is $\sim
  0.53$.}
\end{center}
\end{figure}

\appendix

\section{Method for generating random absorber sets}
\label{sec:meth-gener-rand}

We generate a single random set of \lya\ absorbers for one sightline
in the following way. We draw the number of random absorbers in a set
for a given sightline from a Poisson distribution with mean equal to
the number of real \lya\ absorbers in the sightline. We then generate
the individual random absorbers one at a time. Each absorber is given
a random wavelength between the minimum and maximum observed
wavelengths in that sightline, with a rest equivalent width drawn from
the exponential distribution observed in the FOS spectra in the
\textsl{HST} Quasar absorption line key project \citep{Weymann98}. The
minimum wavelengths were lower cutoff wavelengths for the GHRS
(sightlines A and B) or FOS (sightline C) wavelength settings. The
maximum wavelength was taken to be 3500~\kms\ bluewards of the QSO
emission redshift to match the cutoff applied to the observations. An
absorber is discarded if it falls at a wavelength where there is no
spectral coverage between the GHRS and FOS spectra, its observed
equivalent width is too low to be detected at its observed wavelength,
or it falls within 2~\AA\ of an existing absorber in the sightline.
We calculate the minimum detectable equivalent width as a function of
wavelength by fitting a low-order polynomial through the detection
significances of each line given in \citet{Petry06}.  We checked that
our combined spectra and errors were consistent with these detection
levels.  Any line with detection significance less than $3\sigma$ was
discarded. We discarded lines within 2~\AA\ of existing lines because
such coincidences would be fitted by a single absorption feature in
the real spectra. This process is repeated until we reach the number
of random absorbers for this set drawn from the Poisson distribution.

Using this method we build one thousand random sets for each
sightline. We note that instead of assigning lines a random
wavelength, we could have drawn their redshift from the observed line
distribution; $\ell(z) \propto (1+z)^{\gamma}$.  However, given the
weak dependence of this term on redshift at redshifts less than one
($\gamma \lesssim 0.4$), this will have a minor effect on the line
distribution. For the purposes of comparing absorbers with galaxies,
our method is conservative in the sense that we will see slightly more
random lines at lower redshifts compared to a $(1+z)^{0.4}$
distribution. At lower redshifts there is large overlap with the
galaxy distribution, thus correcting for this effect would result in
an increase in the significance of any correlation between \lya\ lines
and galaxies.

\section{Description of galaxy-absorber groups}
\label{sec:descr-galaxy-absorb}

This section describes the remaining galaxy-absorber groups shown in
Figure~\ref{fig:galspec}.

\subsubsection*{$z=0.053$}

This is an $\sim0.03$~L* galaxy very close to the A (25~kpc) and B
(90~kpc) sightlines. Two \lya\ lines in both sightlines are within
200~\kms\ of the galaxy velocity. The line at a smaller impact
parameter has a larger equivalent width. This is consistent with \HI\
gas surrounding the galaxy with a high covering factor out to
$\sim$65~kpc.

\subsubsection*{$z=0.115$}

A single bright galaxy 350~kpc~/~400~kpc from the A/B
sightlines. \lya\ lines appear within 200~\kms\ of the galaxy in both
sightlines.

\subsubsection*{$z=0.120$}

A single $0.7$~L* galaxy is 280~kpc from sightline B and 380~kpc from
sightline A. A weak \lya\ line within 100~\kms\ appears in sightline
B; no line is seen in sightline A.

\subsubsection*{$z=0.155$}

A single galaxy 780~kpc from sightline A and 970~kpc from sightline
B. \lya\ absorption is seen only in sightline B within 200~\kms\ of
the galaxy. The A spectrum is noisy, but consistent with weaker \lya\
at the same velocity as the line seen in the B spectrum.

\subsubsection*{$z=0.190$}

Three pairs of galaxies are seen close to this redshift. One is very
close ($130$~kpc) to the B sightline.  No significant \lya\
absorption is seen within 1000~\kms\ of any of these
galaxies. However, there is absorption in the B spectrum about
500~\kms\ from both the lower redshift pair and the higher redshift
pair, which has been identified as a transition other than \lya.  It
is possible that this is blended with \lya\ absorption.

\subsubsection*{$z=0.202$}

A group of 16 galaxies spread over $\sim 1000$~\kms, generally
distributed around sightline B. One is 175~kpc from sightline B and
several have a luminosity of a few times L*. A strong \lya\ line is
seen in spectrum B, but no line is seen in spectrum A. There is also a
feature in spectrum B at -700~\kms\ that we identified as \lyg\ from a
higher redshift absorber that may be blended with \lya.

\subsubsection*{$z=0.233$}

A group of five galaxies south of sightlines A and B.  Absorption is seen
within 500~\kms\ of the mean galaxy velocity in sightline B.  No
absorption is seen in sightline A.

\subsubsection*{$z=0.238$}

A group of seven galaxies close to sightline A.  The galaxy with the
smallest impact parameters is 280~kpc from sightline A and 500~kpc
from sightline B. Absorption is seen in sightline A, but not in
sightline B.

\subsubsection*{$z=0.242$}

A single galaxy 1.2~Mpc from sightlines A and B. No associated
absorption is seen within 1000~\kms.

\subsubsection*{$z=0.352$}

At this redshift \lya\ coverage begins for the spectrum of the third
sightline. There are two galaxies, the smallest separation for a QSO
sightline is 900~kpc from sightline A. No \lya\ absorption is seen
within 1000~\kms, but spectra B and C are noisy in this region.

\subsubsection*{$z=0.428$}

Three galaxies are seen, one within 730~kpc of sightlines A and C,
and 420~kpc of sightline B. Absorption is seen within 1000~\kms\ of
the galaxies in sightlines A and C, but not in sightline B.

\subsubsection*{$z=0.434$}

A single galaxy about 800~kpc from sightlines A and B. Absorption is
seen in the B and C spectra.  No \lya\ absorption is seen in spectrum
A, although there is an absorber identified with Ly-6 from a
higher-redshift system.

\subsubsection*{$z=0.501$}

A group of three galaxies 1.6~Mpc from all three sightlines, and a
fourth galaxy 1.5~Mpc from sightlines A and B. The galaxy with the
smallest impact parameter is 1.4~Mpc from sightline C. Strong
absorption is seen in sightline A; weaker \lya\ absorption is seen in
sightlines B and C. The absorption represents an absorber triplet (as
defined in Section~\ref{sec:lya-triple-coinc}), and this triplet forms
four triple LOS-galaxy groups (defined in
Section~\ref{sec:stat-tests-absorb}) with the four nearby galaxies.

\subsubsection*{$z=0.517$}

A single galaxy very close (130~kpc) to sightline B. \lya\ is seen in
sightline B only.

\subsubsection*{$z=0.535$}

Three galaxies, one of which is 900~kpc from sightline B.  Absorption
is seen within 200~km/s in all three sightlines. The absorption
represents a triplet (as defined in
Section~\ref{sec:lya-triple-coinc}) and it forms three triple
LOS-galaxy groups (defined in Section~\ref{sec:stat-tests-absorb})
with the three nearby galaxies.

\subsubsection*{$z=0.567$}

A single galaxy 2.4~Mpc from sightline B. No \lya\ absorption within
1000~\kms\ is seen in any of the spectra.  We do not observe a galaxy
linked with the sub-DLA absorption in sightline C at $z=0.5569$.

\subsubsection*{$z=0.644$}

A single galaxy 1~Mpc from sightlines B and C.  \lya\ is seen within
200~\kms\ of the galaxy position in sightlines B and A.

\end{document}